\documentclass[10pt,cleanfoot]{asme2ej}

\usepackage{overpic}  
\usepackage{tikz}     
\usetikzlibrary{arrows} 
\usepackage{epstopdf} 
\usepackage{stfloats}
\usepackage{adjustbox}
\usepackage{isomath}
\usepackage{amsmath}
\usepackage{booktabs}
\usepackage{bm}

\usepackage{lineno}
\usepackage{graphicx} 
\usepackage{epsfig}
\usepackage{fancyhdr}
\usepackage{setspace}
\usepackage{helvet}
\usepackage[hyphens]{url}

\usepackage{hyperref}   
\hypersetup{
	colorlinks=true,
	linkcolor=blue,
	citecolor=blue,
	urlcolor=blue,
}
\usepackage[square,numbers]{natbib}

\title{Boundary layer transition induced by surface roughness distributed over a low-pressure turbine blade}

\author{Xianwen Zhu
    \affiliation{
	HEDPS, CAPT\\
    School of Mechanics and Engineering Science\\
    Peking University\\
    Beijing 100871, China\\
    email: zxw@stu.pku.edu.cn
    }	
}

\author{Yuchen Ge
    \affiliation{ HEDPS, CAPT\\
    School of Mechanics and Engineering Science\\
    Peking University\\
    Beijing 100871, China\\
    email: yuchen.ge@stu.pku.edu.cn
    }
}

\author{Yaomin Zhao\thanks{Address all correspondence related to ASME style format and figures to this author.} 
    \affiliation{
	HEDPS, CAPT\\
    School of Mechanics and Engineering Science\\
    Peking University\\
    Beijing 100871, China\\
    email: yaomin.zhao@pku.edu.cn
    }
}

\author{Zuoli Xiao
    \affiliation{ HEDPS, CAPT\\
    School of Mechanics and Engineering Science\\
    Peking University\\
    Beijing 100871, China\\
    email: z.xiao@pku.edu.cn
    }
}

\author{Richard D. Sandberg
    \affiliation{Department of Mechanical Engineering\\
	University of Melbourne\\
	Melbourne, VIC3010, Australia\\
	Email: richard.sandberg@unimelb.edu.au
	}
}

\begin{document}
\definecolor{Qgreen}{RGB}{15,98,12}
\definecolor{mycolor}{RGB}{132,36,33}
\definecolor{mycolor1}{RGB}{139,0,18}
\definecolor{mycolor2}{RGB}{0,114,189}
\definecolor{mycolor3}{RGB}{235,145,132}

\maketitle    
\doublespacing

\begin{abstract}
{\it 
Direct numerical simulations of a low-pressure turbine with roughness elements distributed over the blade surface have been performed.
A series of fifteen cases with varying roughness heights and streamwise wavenumbers are introduced to present a systematic study of the effect of roughness on the various transition phenomena in the suction-side boundary layer.
For cases with large roughness heights, the boundary layer is violently disturbed by the wake of roughness elements in the leading edge (LE) region, and maintains a turbulent state over the whole blade suction-side.
For cases with small roughness heights, however, the disturbances induced by the LE roughness are suppressed by the favorable pressure gradient in the downstream boundary layer, and the relaminarized flow does not undergo transition until the separation near the blade trailing edge (TE).
Furthermore, the streamwise wavenumber of the distributed roughness plays an important role in cases with intermediate roughness height.
Specifically, cases with larger streamwise slope show earlier transition induced by strong shear layer instability, which manages to suppress the mean flow separation near the TE region.
Overall, the combined effect of several factors, including the geometric effect at the blade LE and TE, the complex pressure gradient distribution across the turbine vane, and the various roughness configurations, is responsible for the intriguing boundary layer behaviors in the present study. 
}
\end{abstract}

\begin{nomenclature}
\entry{$Re$}{exit Reynolds number}
\entry{$Ma_{e}$}{exit Mach number}
\entry{$\xi$}{blade tangential distance}
\entry{$\eta$}{blade normal distance}
\entry{$\eta_{wall}$}{local roughness height}
\entry{$L_{\xi}$}{blade surface length}
\entry{$L_{z}$}{spanwise domain size}
\entry{$x$}{axial coordinates}
\entry{$y$}{pitchwise coordinates}
\entry{$z$}{spanwise coordinates}
\entry{$k$}{peak value of roughness height}
\entry{$\alpha$}{wavenumber}
\entry{$\lambda$}{wavelength}
\entry{$ES$}{roughness effective slope}
\entry{$C$}{chord length}
\entry{$C_{ax}$}{axial chord length}
\entry{Tu}{turbulence intensity}
\entry{$U_{\infty}$}{reference velocity}
\entry{$\Delta \xi^+$}{wall-tangential grid spacing}
\entry{$\Delta \eta^+$}{wall-normal grid spacing}
\entry{$\Delta z^+$}{spanwise grid spacing}
\entry{$C_p$}{pressure coefficient}
\entry{$\tau_{w}$}{wall total drag}
\entry{$C_d$}{drag coefficient}

\subsection*{Superscripts and Subscripts}
\entry{$\left\langle \right\rangle$}{time-averaged quantity}
\entry{$\bar{}$}{time- and spanwise-averaged quantity}
\entry{$\tilde{}$}{dispersive fluctuating component}
\entry{$ ^{\prime}$}{turbulent fluctuating component}
\entry{$ _{dis}$}{instantaneous dispersive velocity}
\entry{$_{\xi}$}{streamwise component}
\entry{$_{\eta}$}{normal component}
\entry{$_{z}$}{spanwise component}
\entry{$\infty$}{reference value}
\end{nomenclature}


\section{Introduction}\label{sec:Introduction}

In turbomachinery applications, specifically gas turbines, the blade surfaces inevitably develop distributed roughness over extended operation. This roughness significantly impacts the blade boundary layer, which is characterized by complex pressure gradients, laminar-turbulent transition, and flow separation, thereby affecting both the efficiency and operational safety of the machine \cite{Bons2010review}. 
However, existing work regarding roughness effects in turbomachinery applications has been largely restricted to overall performance metrics, particularly empirical models for drag and heat transfer. 
The fundamental mechanisms governing roughness effects on the blade boundary layer remain less understood, and many important questions regarding the detailed flow physics are yet to be answered. 
In the following, we will briefly summarize existing roughness studies on gas turbine flows. 


The influence of roughness on turbomachinery flows has been extensively investigated through experiments, initially focusing on mean flow characteristics and efficiency. 
Early studies by Bammert and Milsch \cite{Bammert1972} and Bammert and Sandstede \cite{Bammert1980} established that roughness increases kinetic losses, alters flow turning angles, and significantly thickens the boundary layer, especially in decelerating regions. 
Subsequent work by Kind et al. \cite{Kind1998} highlighted the dominance of suction-side roughness on loss generation. 
The sensitivity to operating conditions was further examined by Bogard et al.~\cite{Bogard1998} and Boyle and Senyitko~\cite{Boyle2003}, who demonstrated the critical roles of freestream turbulence and Reynolds number in augmenting heat transfer and aerodynamic losses. 
Addressing the complexity of real surfaces, Bons~\cite{Bons2002} showed that irregular topography results in skin friction and heat transfer levels significantly higher than standard sand-grain predictions. 
Furthermore, Roberts and Yaras~\cite{Roberts2005} linked earlier transition inception to specific geometric parameters such as spacing and skewness. 
More recently, in the context of low-pressure turbines, Montis et al.~\cite{Montis2010,Montis2011} and Lorenz et al.~\cite{Lorenz2012} revealed that while roughness can suppress laminar separation bubbles by promoting early transition, it ultimately increases profile losses due to the extended turbulent wetted area and enhanced wake mixing. 
Nonetheless, limited by measurement techniques, these studies predominantly focused on time-averaged quantities, leaving the intricate flow dynamics within the boundary layer largely unexplored.

Compared to experiments, numerical simulations usually provide more details of the flow fields and thus are desired for deeper understanding on the rough-wall boundary layer flows in turbo-machines.
Early numerical simulations of turbomachinery flows with surface roughness, however, mainly employed Reynolds-averaged Navier–Stokes (RANS) calculations, in which the accuracy was highly dependent on the turbulence models~\cite{Stripf2009,Dassler2012,Ge2015,Wei2017,Liu2020}. 
Zeng et al.~\cite{Zeng2022} compared experimental data with RANS simulations for a high-lift low-pressure turbine blade, revealing that while RANS models can capture qualitative trends, they exhibit significant quantitative discrepancies, notably overpredicting the roughness-induced profile losses. 

Only recently, the development of supercomputers and algorithms has made high-fidelity numerical simulations of rough-wall turbomachinery flows possible. 
Joo et al.~\cite{Joo2016} analyzed the flow over a roughened turbine blade using LES and RANS and found that LES successfully predicted the roughness-induced turbulent separation, while RANS roughness models failed. 
Hammer et al.~\cite{Hammer2018} performed LES with immersed boundary methods, showing that the distinct roughness peaks located on the blade surface produced velocity streaks, significantly impacting the transition locations. 
Focusing on in-depth analysis of flow dynamics, Wang et al.~\cite{Wang2021} conducted LES of compressor blades with trigonometric-function roughness, revealing the critical role of roughness-induced spanwise velocity components in governing streak merging and shear layer destabilization. 
By scaling up computational resources, Jelly et al.~\cite{Jelly2023} conducted the first high-fidelity roughness-resolved LES study on high-pressure turbine blades at engine-relevant Reynolds number, revealing that surface roughness amplifies total pressure loss and heat flux through premature transition onset, boundary layer thickening, and intensified turbulent mixing. 
To enhance computational fidelity, Nardini et al.~\cite{Nardini2023,Nardini2023_2} pioneered the integration of a three-dimensional boundary data immersion method (BDIM) with DNS for resolving multiscale rough surfaces on high-pressure turbine (HPT) blades, revealing roughness-induced mechanisms governing boundary layer transition modulation, shockwave structural alterations, and Reynolds analogy breakdown. 
Introducing actual roughness configurations from in-service blades, Jelly et al.~\cite{Jelly2025} performed DNS of HPT vane covered by localized non-Gaussian roughness. 
By varying the roughness height in a systematic way, they demonstrated a strong sensitivity of suction-side skin friction and heat transfer to the location of the roughness.
However, the above studies mostly considered the influence of roughness height, suggesting that the geometric effects of roughness on turbomachinery flows warrant further study. 

In the present study, we investigate the roughness effects on LPT blade boundary layer flows, leveraging the ability of direct numerical simulations to resolve the details of flow structures.
By varying the height and streamwise wavenumber of roughness elements, a systematic investigation is enabled with high-fidelity flow fields.
Specifically, the boundary layer covered with distributed roughness elements is expected to be affected by complex factors such as blade geometry and strong pressure gradient, which result in more intriguing transition behaviors when compared with canonical flows.
In particular, in addition to the roughness height, the effects of the roughness wavenumber (thus the streamwise slope) are also investigated, highlighting the geometric effects of roughness on the blade boundary layer transition. 

The outline of this paper is as follows. 
An introduction to the numerical simulations, along with validation of the results, is given in section~\ref{sec:Methodology}. 
Then, an overview of the flow fields obtained from the roughened LPT simulations is given in section~\ref{sec:Overview}. 
Detailed analysis on the mechanisms for the complex boundary layer behaviors affected by various roughness parameters, including transition, relaminarization, and separation, \emph{etc}, are discussed in section~\ref{sec:Results}. 
Finally, conclusions are drawn in section~\ref{sec:Conclusion}.

\section{Methodology}\label{sec:Methodology}

\subsection{Case set-up}
\begin{figure*}[t!]
	\centering
	\begin{overpic}[width=0.8\textwidth]{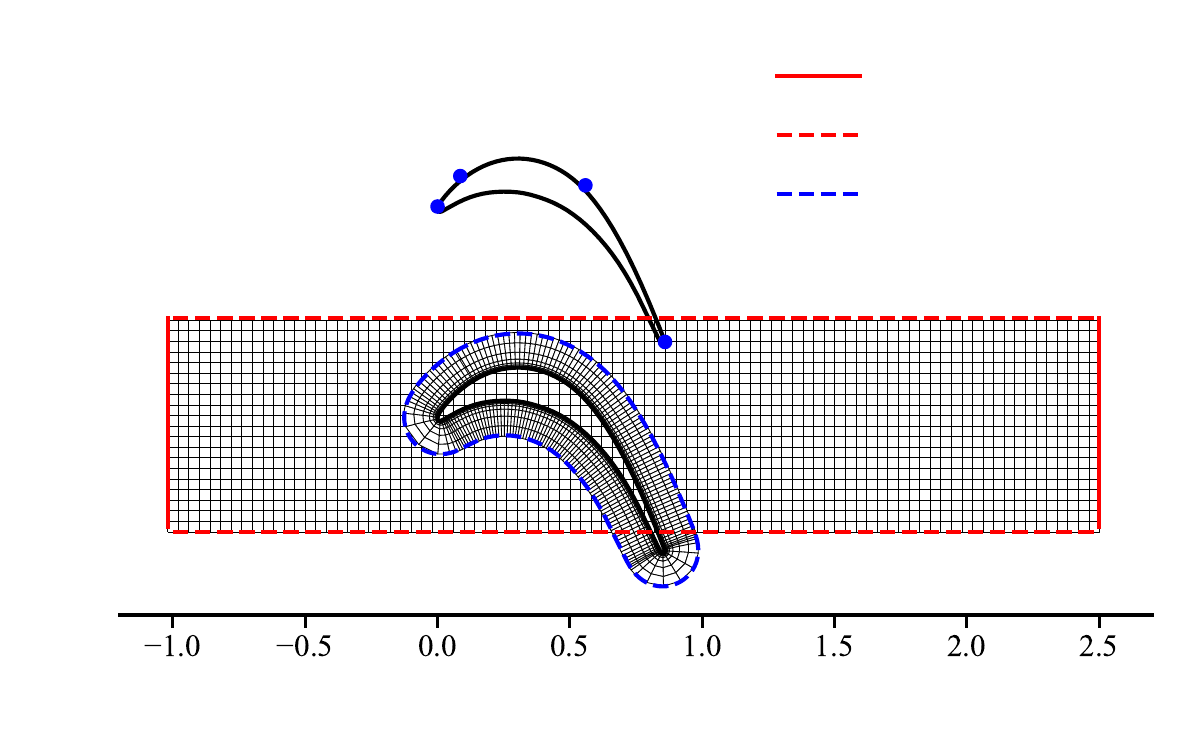}
            \put(0,55){(\textit{a})}
            \begin{tikzpicture}[overlay, x=0.0592\textwidth, y=0.0592\textwidth]
                \draw[white] (0,0)--(0,1);
                \draw[white] (0,0)--(1,0);
                \draw [line width =1pt][-latex](0,1.5)--(1,1.5) node (A) at (0.8,1.7) {$x$};
                \draw [line width =1pt][-latex](0,1.5)--(0,2.5) node (B) at (0.2,2.3) {$y$};
                \draw [line width =1pt][densely dashed][-stealth](1,2.4)--+(46.5:1);
                \draw [line width =1pt][densely dashed][-stealth](1,3)--+(46.5:1);
                \draw [line width =1pt][densely dashed][-stealth](1,3.6)--+(46.5:1);
                \node (C) at (1,4.2) {$U_{in}$};
                \node (D) at (4.5,6) {LE};
                \node (E) at (5.2,6.8) {$x/C_{ax}=0.1$};
                \node (F) at (7.3,6.6) {$x/C_{ax}=0.65$};
                \node (G) at (7.6,5) {TE};
                \draw [line width =1pt][densely dotted][<->](4.9,1.65)--+(2.62,0);
                \node (H) at (6.195,1.9) {$C_{ax}$};
                \node (I) at (5.8,5.7)[align=center] {Pressure \\ side};
                \node (J) at (6.4,7.2)[align=center] {Suction \\ side};
                \node (k) at (7.1,0.7) {$x/C$};
                \node (L) at (9.8,7.6)[anchor=west]{Inlet/outlet};
                \node (M) at (9.8,6.94)[anchor=west]{Periodic};
                \node (N) at (9.8,6.28)[anchor=west]{Overset interface};
            \end{tikzpicture}
	\end{overpic}
    \hspace{0.02\textwidth}
	\begin{overpic}[width=0.392\textwidth]{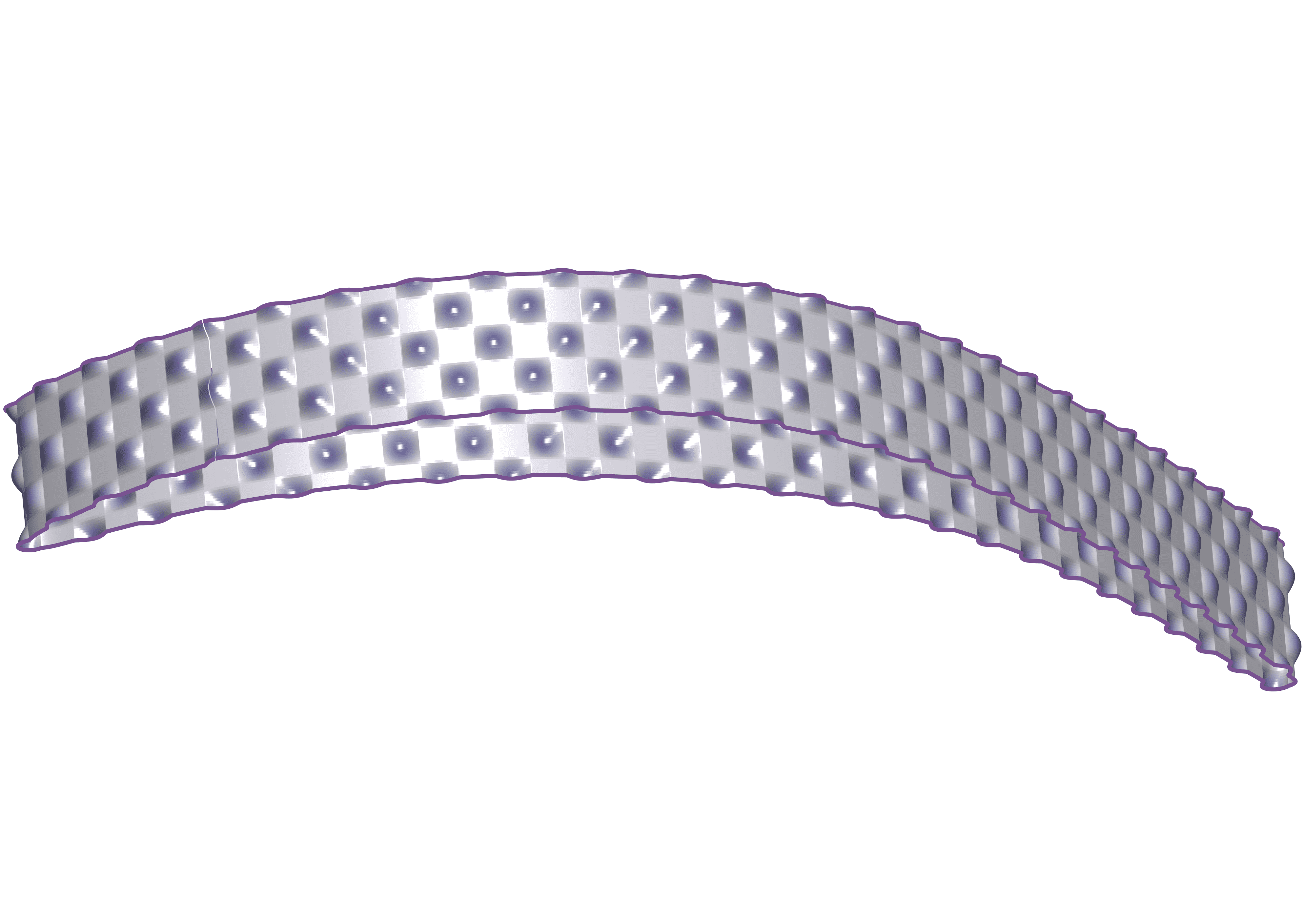}
            \put(0,71){(\textit{b})}
	\end{overpic}
    \hspace{0.02\textwidth}
    \begin{overpic}[width=0.36\textwidth]{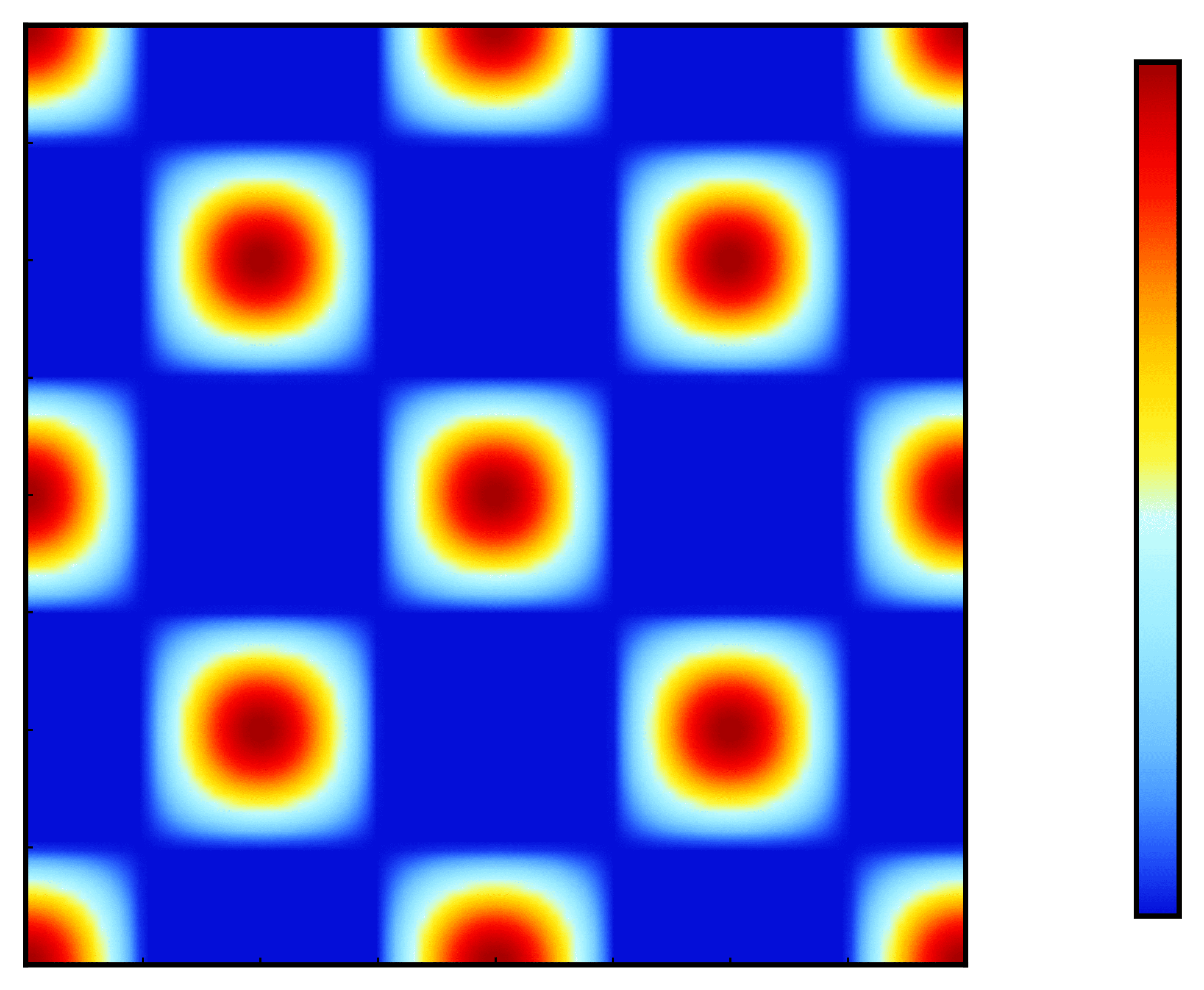}
        \put(-8,78){(\textit{c})}
        \put(-5,40){\scalebox{1.0}{$z$}}
        \put(40,-4){\scalebox{1.0}{$\xi$}}
        \begin{tikzpicture}[overlay, x=0.0592\textwidth, y=0.0592\textwidth]
            \draw[white] (0,0)--(0,1);
            \draw[white] (0,0)--(1,0);
            \draw[line width =1pt] (2.5,4.95)--+(0,0.3);
            \draw[line width =1pt] (4.85,4.95)--+(0,0.3);
            \draw[line width =1pt][<->] (2.5,5.1)--(4.85,5.1);
            \node (A) at (3.675,5.3) {$\lambda_{\xi}$};
            \draw[line width =1pt][|<->|] (5.1,4.85)--(5.1,2.45);
            \node (B) at (5.35,3.65) {$\lambda_{z}$};
            \node (C) at (6.1,0.35) {0};
            \node (D) at (6.1,4.65) {$k$};
        \end{tikzpicture}
	\end{overpic}\\[1em]
	\caption{Schematic for LPT case set-up and roughness configuration. (\textit{a}) The computational grid is showing every twentieth line in each direction. Blade boundary layer divided by critical points marked by the blue circles (LE, leading edge; TE, trailing edge). $C_{ax}$ means axial chord. (\textit{b}) A roughened blade surface profile. (\textit{c}) Contour of blade surface height.}
    \label{fig:schematic}
\end{figure*}

A schematic for the configuration of the LPT simulations is shown in Fig.~\ref{fig:schematic}(\textit{a}). 
The baseline LPT is a T106A cascade, and the computational domain is bounded by the red lines highlighted on the axial and pitchwise ($x$–$y$) plane intersection. 
The simulations are performed at an exit Reynolds number of $Re = 60,000$ and an exit Mach number $Ma_e = 0.405$, which are in agreement with experimental investigations reported by Stadtm\"uller~\cite{Stadtmuller2001} and numerical simulations by Michelassi et al.~\cite{Michelassi2015}. 

In order to investigate the effects of distributed roughness on the LPT flow, the whole blade surface in the present simulations is covered by roughness elements.
With $\xi$ and $\eta$ denoting the coordinates in the wall-tangential and wall-normal directions along the blade surface, respectively, the height of roughness elements $\eta_{wall}$ is defined as follows:
\begin{align}
\eta_{wall} & =\max \left\{k \cos \left(\frac{2 \pi}{\lambda_\xi} \xi\right) \cos \left(\frac{2 \pi}{\lambda_z} z\right), 0\right\} \\
& = \max \left\{k \cos \left(\frac{2 \pi \alpha_\xi}{L_\xi} \xi\right) \cos \left(\frac{2 \pi \alpha_z}{L_z} z\right), 0\right\}.
\end{align}
An example of the rough wall blade is shown in Fig.~\ref{fig:schematic}(\textit{b,c}).
Here, $\xi$ is the coordinate around the blade surface, with $L_\xi$ representing arc length around the blade, and $z$ is the spanwise coordinate, with $L_z$ denoting the spanwise width of the computational domain. 
Moreover, $k$ is the peak value of the roughness height, while $\alpha$ and $\lambda$ denote the wavenumber and wavelength of the roughness elements, with the subscripts $_\xi$ and $_z$ are representing the tangential and spanwise components, respectively. 
Thereafter, the roughness effective slope is defined as :
\begin{equation}
ES_{\xi}=\frac{1}{L_{\xi} L_z} \int_0^{L_z} \int_0^{L_{\xi}}\left|\frac{\partial \eta_{wall}(\xi, z)}{\partial \xi}\right| \mathrm{d} \xi \mathrm{~d} z.
\end{equation}
The trigonometric form of the roughness is chosen for two reasons: one is that this roughness is simple to generate and has been widely studied in canonical flows \cite{Chan2018,Ma2022}, and the other is that different trigonometric functions can be superimposed to produce irregular roughness \cite{Napoli2008,Demarchis2015}. 
Moreover, following the setup in Vadlamani et al.~\cite{Vadlamani2018}, the 'peaks-only' component is taken to simplify the mesh generation. 

A list of the cases with different surface roughness is shown in Table~\ref{tab:roughness_parameters}.
In the present study, five roughness heights were considered, together with three streamwise roughness wavenumbers, resulting in a total of $15$ rough cases.
Throughout this paper, the roughness cases are identified by the following code
\begin{equation}
\underbrace{k48}_{k/C \times 10^4}\underbrace{\alpha50}_{\alpha_\xi}.
\end{equation}
Note that for typical turbine blades with the chord length around $C=100mm$ \cite{Ciorciari2014}, the average roughness heights in the present study vary in the range of $32.4\mu m$ to $162.1\mu m$, representative of rough surfaces observed in used turbine blades \cite{Tarada1993}.
Particularly, nondimensionalized with the viscous friction length scale at $x/C_{ax}=0.4$ in the suction-side boundary layer of the smooth case, the dimensionless roughness heights are $k^+ = 5.3, 10.7, 16, 21.3$ and $26.5$, respectively. 
The minimum roughness height $k16$ cases are considered hydraulically smooth for most of the blade boundary layer because $k^+ \approx 5$. 
For other cases with increasing roughness heights from $k/C=3.2 \times 10^{-3}$ to $k/C=8.0 \times 10^{-3}$, the flow structures vary significantly.
More details will be discussed in sections~\ref{sec:Overview} and \ref{sec:Results}. 

\begin{table}[t]
\caption{Surface roughness parameters, including: the peak height $k/C$, streamwise wavelength $\lambda_\xi / C$, dimensionless height $k^+$, streamwise effective slope $ES_\xi$, and spanwise effective slope $ES_z$.}\label{tab:roughness_parameters}
\centering{%
\begin{tabular}{l c c c c c}
\toprule
ID code & $k/C$ & $\lambda_{\xi}/C$ & $k^{+}$ & $ES_{\xi}$ & $ES_{z}$ \\
\midrule
smooth         & $0$                  & $--$    & $0$    & $0$     & $0$     \\
$k16\alpha50$  & $1.6 \times 10^{-3}$ & $0.05$  & $5.3$  & $0.041$ & $0.041$ \\
$k16\alpha100$ & $1.6 \times 10^{-3}$ & $0.025$ & $5.3$  & $0.082$ & $0.041$ \\
$k16\alpha150$ & $1.6 \times 10^{-3}$ & $0.017$ & $5.3$  & $0.123$ & $0.041$ \\
$k32\alpha50$  & $3.2 \times 10^{-3}$ & $0.05$  & $10.7$ & $0.082$ & $0.081$ \\
$k32\alpha100$ & $3.2 \times 10^{-3}$ & $0.025$ & $10.7$ & $0.164$ & $0.081$ \\
$k32\alpha150$ & $3.2 \times 10^{-3}$ & $0.017$ & $10.7$ & $0.246$ & $0.081$ \\
$k48\alpha50$  & $4.8 \times 10^{-3}$ & $0.05$  & $16.0$ & $0.123$ & $0.122$ \\
$k48\alpha100$ & $4.8 \times 10^{-3}$ & $0.025$ & $16.0$ & $0.246$ & $0.122$ \\
$k48\alpha150$ & $4.8 \times 10^{-3}$ & $0.017$ & $16.0$ & $0.368$ & $0.122$ \\
$k64\alpha50$  & $6.4 \times 10^{-3}$ & $0.05$  & $21.3$ & $0.164$ & $0.163$ \\
$k64\alpha100$ & $6.4 \times 10^{-3}$ & $0.025$ & $21.3$ & $0.327$ & $0.163$ \\
$k64\alpha150$ & $6.4 \times 10^{-3}$ & $0.017$ & $21.3$ & $0.491$ & $0.163$ \\
$k80\alpha50$  & $8.0 \times 10^{-3}$ & $0.05$  & $26.5$ & $0.205$ & $0.204$ \\
$k80\alpha100$ & $8.0 \times 10^{-3}$ & $0.025$ & $26.5$ & $0.409$ & $0.204$ \\
$k80\alpha150$ & $8.0 \times 10^{-3}$ & $0.017$ & $26.5$ & $0.614$ & $0.204$ \\
\bottomrule
\end{tabular}
}%
\end{table}

\subsection{Numerical methods}
The non-dimensionalized three-dimensional compressible Navier–Stokes equations are solved using the multi-block structured curvilinear solver HiPSTAR \cite{Sandberg2015}, which has been successfully employed in a number of numerical studies of both low-pressure and high-pressure turbine flows \cite{Michelassi2015,Sandberg2015,Zhao2020,Zhao2021}. 
A fourth-order finite-difference scheme is applied for spatial discretization, and the ultra-low storage frequency optimized explicit Runge–Kutta method \cite{Kennedy2000} is used for time integration. 
Furthermore, the overset method \cite{Deuse2020a} is applied in the present LPT simulations. 
The computational domain employs a similar overset mesh configuration as in \cite{Zhao2021} and \cite{Jelly2023}, consisting of an O-type grid wrapped around the LPT blade and an embedded background H-type grid as shown in Fig.~\ref{fig:schematic}.  
The H-type and O-type grids overlap with each other, and continuity conditions are imposed at the overlapping boundaries, with variables interpolated using a fourth-order Lagrangian method between the blocks. 

At the inlet, freestream turbulence (FST) is introduced by a digital filter method \cite{Klein2003}, in which the generated fields can reproduce first- and second-order one-point statistics as well as a given autocorrelation function efficiently. 
The incoming turbulence intensity $T_u/U_{\infty}=3.2\%$, and the integral turbulence length scale is $5\%C$ for all cases. 
At the outlet, the zonal characteristic boundary condition \cite{Sandberg2006} is applied to reduce reflections due to passing vortices from turbulent flow or wakes. 
Furthermore, no-slip isothermal wall conditions are applied at the blade surface. 
In particular, the complex geometries of roughened blade surfaces have been resolved by a second-order boundary data immersion method (BDIM) \cite{Schlanderer2017}, which has been extensively tested in compressible simulations, including the recent high-fidelity simulations of high-pressure turbines \cite{Jelly2023}. 
Moreover, the HiPSTAR–BDIM framework has been successfully employed in several high-fidelity numerical studies of roughened high-pressure turbine flows \cite{Nardini2023,Jelly2025}. 
Further details of the BDIM formulation can be found in Schlanderer et al. \cite{Schlanderer2017}.

\subsection{Validation}
To extensively validate the numerical setups for the present simulations, a series of test cases with different meshes have been performed, and the mesh parameters are listed in Table~\ref{tab:grid_convergence}. 
\begin{table}[t!]
\caption{Parameters for different meshes.\label{tab:grid_convergence}}
\centering
\begin{tabular}{l c c c}
\toprule
Symbol & $N_\xi$ & $N_\eta$ & $N_z$ \\
\midrule
Mesh -- C      & 1544 & 129 & 60 \\
Mesh -- P      & 3707 & 149 & 60 \\
Mesh -- $\xi$  & 5352 & 149 & 60 \\
Mesh -- $\eta$ & 3707 & 208 & 60 \\
Mesh -- $z$    & 3707 & 149 & 100\\
\bottomrule
\end{tabular}
\end{table}

\begin{figure}[t!]
	\centering
	\begin{overpic}[width=0.46\textwidth]{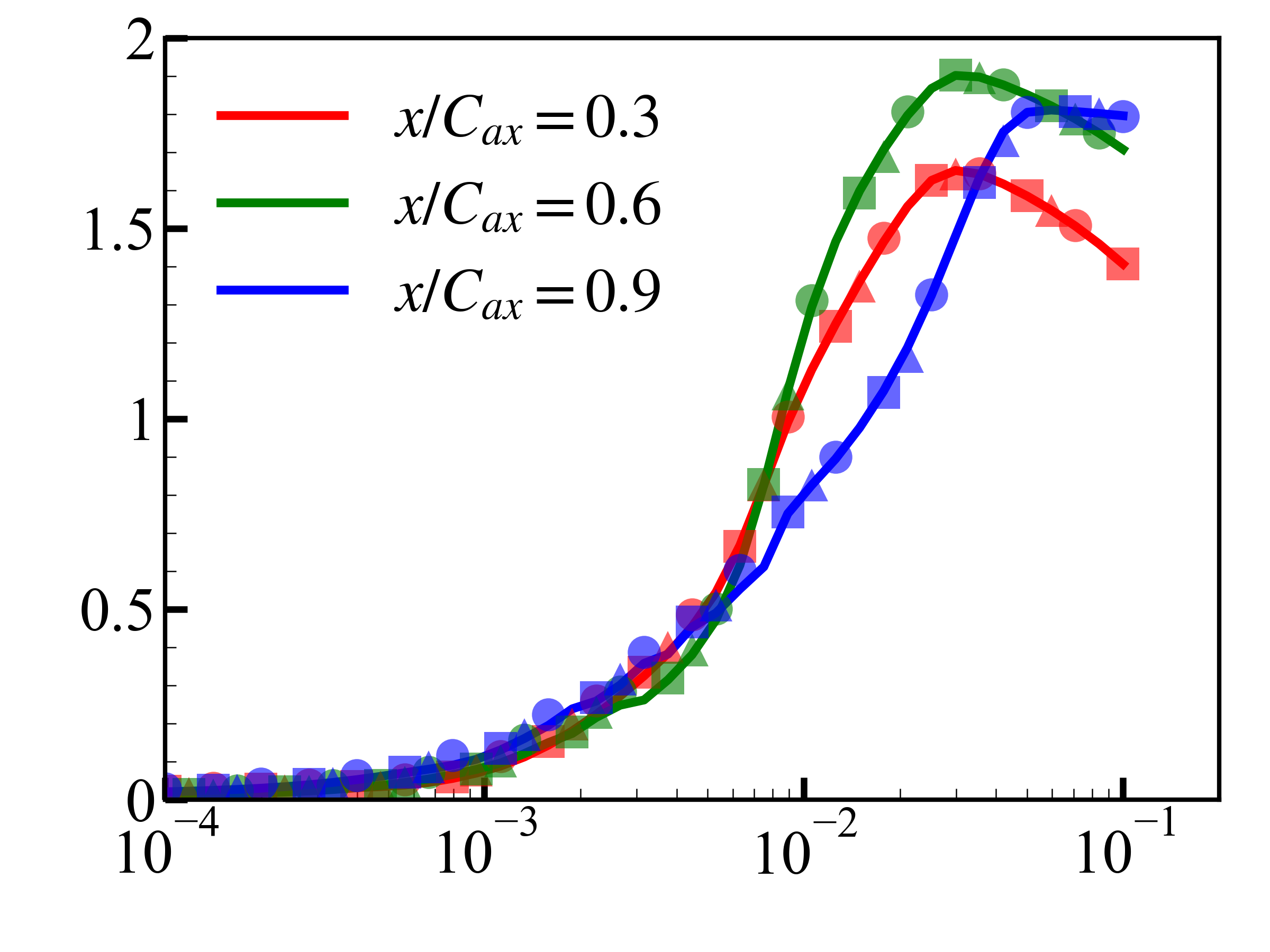}
		\put(-4,71){(\textit{a})}
            \put(0,40){\scalebox{1.0}{$U_{\xi}$}}
            \put(45,0){\scalebox{1.0}{$\eta/C_{ax}$}}
	\end{overpic}\\
    \hspace{0.02\textwidth}
	\begin{overpic}[width=0.46\textwidth]{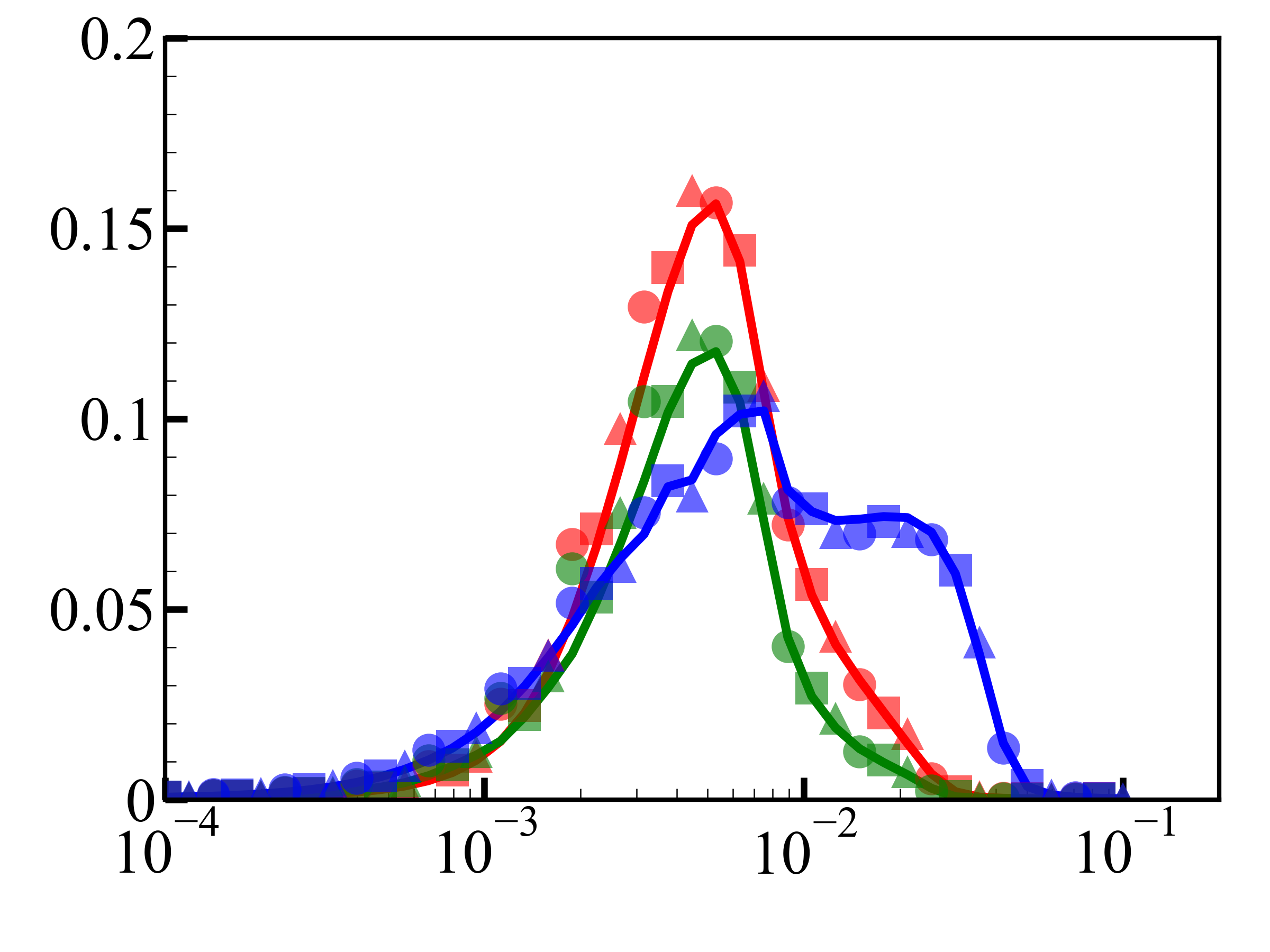}
		\put(-4,71){(\textit{b})}
            \put(45,0){\scalebox{1.0}{$\eta/C_{ax}$}}
            \put(-4,40){\scalebox{1.0}{\rotatebox[origin=c]{90}{$\overline{u^{\prime}_{\xi} u^{\prime}_{\xi}}$}}}
	\end{overpic}\\
	\caption{Validation of grid convergence in the $k64\alpha50$ case: (\textit{a}) wall normal profiles of mean tangential velocity; (\textit{b}) wall normal profiles of tangential Reynolds normal stress. 
    }
        \label{fig:grid_convergence}
\end{figure}

The Mesh-C was used exclusively for the smooth-blade case, and the corresponding results are in very good agreement with the results reported by Sandberg et al.~\cite{Sandberg2015}. 
Although the Mesh-C is fine enough for the smooth case, the rough cases obviously require finer grid resolution~\cite{Jelly2023,Nardini2023_2,Nardini2023}. 
Therefore, to further validate the grid independence, the $k64\alpha50$ case has been tested by  
a series of progressively refined meshes, which are summarized in Table~\ref{tab:grid_convergence}. 
The results presented in Fig.~\ref{fig:grid_convergence} show that the Mesh-P is able to accurately predict the mean velocity and Reynolds normal stress profiles at diverse streamwise locations.  
Therefore, for the rough cases simulated in the present study, the Mesh-P is used for production.

In addition, 
the grid spacings in proximity to the blade surface in the tangential, wall-normal and spanwise directions are expressed in non-dimensional terms via the local viscous length scale, denoted as $\Delta \xi^+$, $\Delta \eta^+$ and $\Delta z^+$, respectively. 
Although there exists a variation in grid spacing over the extent of the blade surface, efforts have been made to restrict these spacings to comparatively minimal magnitudes. 
On the suction side, the grid spacings in the wall-tangential, wall-normal, and spanwise directions are maintained below $\Delta \xi^+ < 3$, $\Delta \eta^+<1$ and $\Delta z^+ < 6$, respectively.
In particular, the present mesh for the rough-wall LPT flow is much finer in the wall-tangential direction due to the need to resolve the roughness elements, compared to previously conducted smooth-wall simulations ($\Delta \xi^+ < 11$, $\Delta \eta^+<1$ and $\Delta z^+ < 9$) \cite{Sandberg2015}.

\section{Overview of the flow field}\label{sec:Overview}

\begin{figure*}
    \centering
    \begin{overpic}[width=0.36\textwidth]{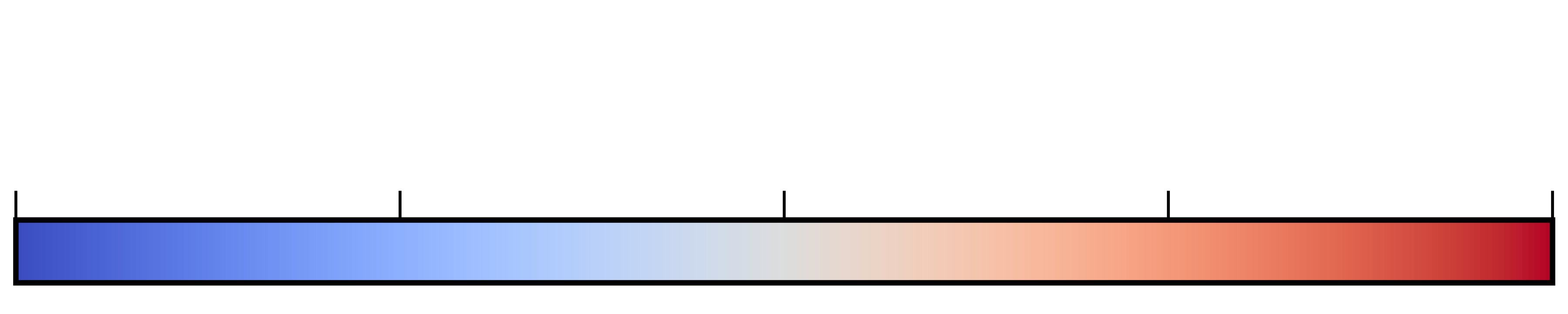}
        \put(-5,10){\scalebox{1.0}{$-1.8$}}
        \put(17,10){\scalebox{1.0}{$-0.9$}}
        \put(48,10){\scalebox{1.0}{$0$}}
        \put(70,10){\scalebox{1.0}{$0.9$}}
        \put(95,10){\scalebox{1.0}{$1.8$}}
        \put(45,18){\scalebox{1.0}{$u_{\xi}$}}
    \end{overpic}\\[1em]
    \begin{overpic}[width=0.32\textwidth]{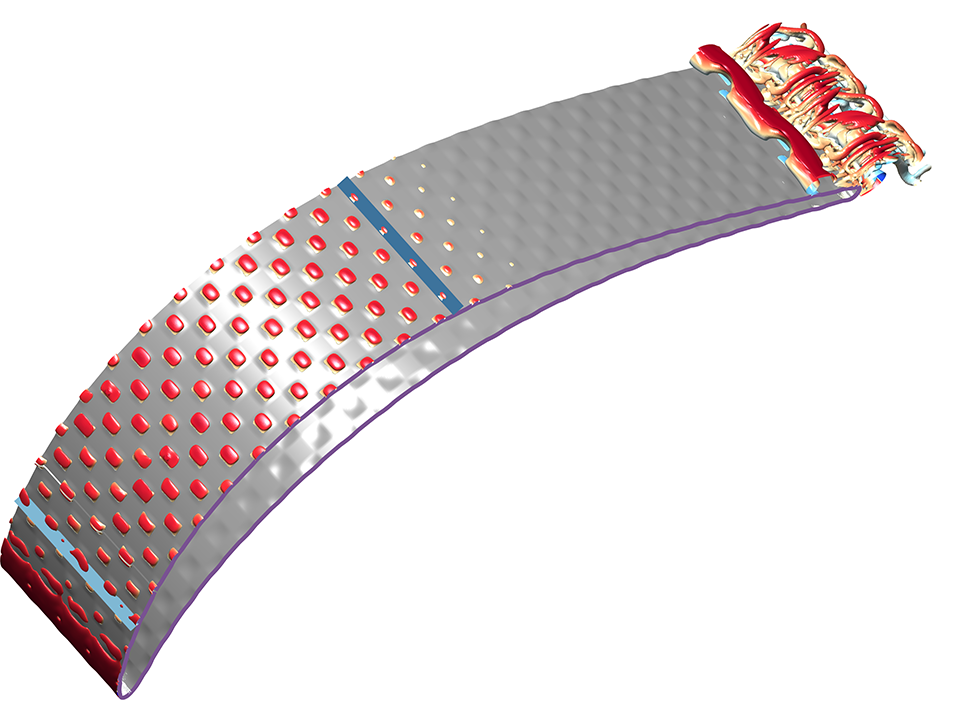}
	\put(-2,42){(\textit{a})}
    \end{overpic}
    \begin{overpic}[width=0.32\textwidth]{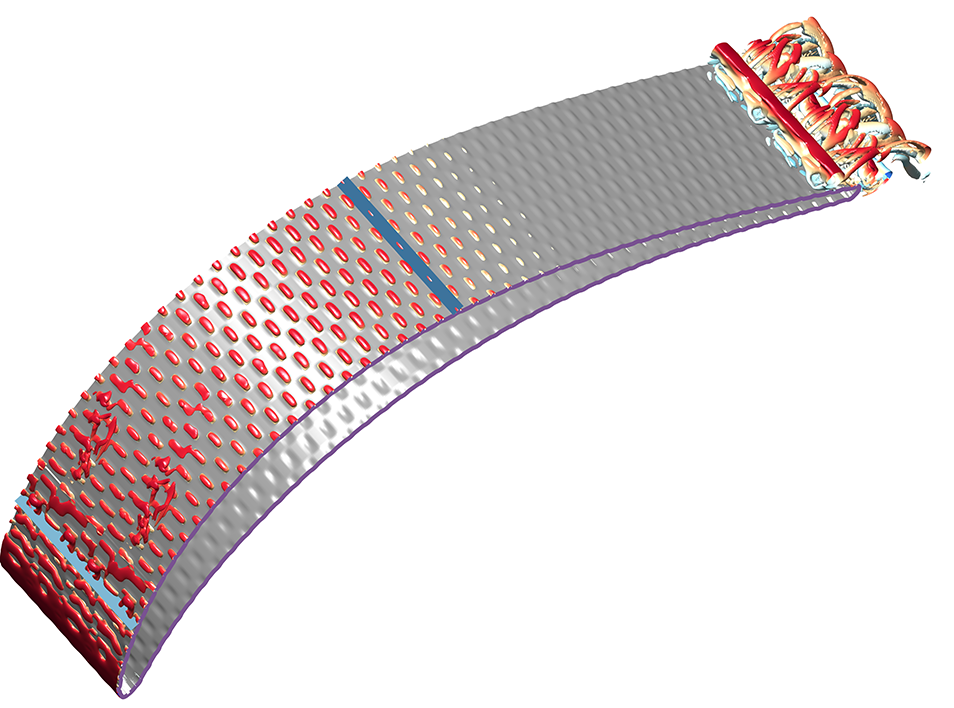}
	\put(-2,42){(\textit{b})}
    \end{overpic}
    \begin{overpic}[width=0.32\textwidth]{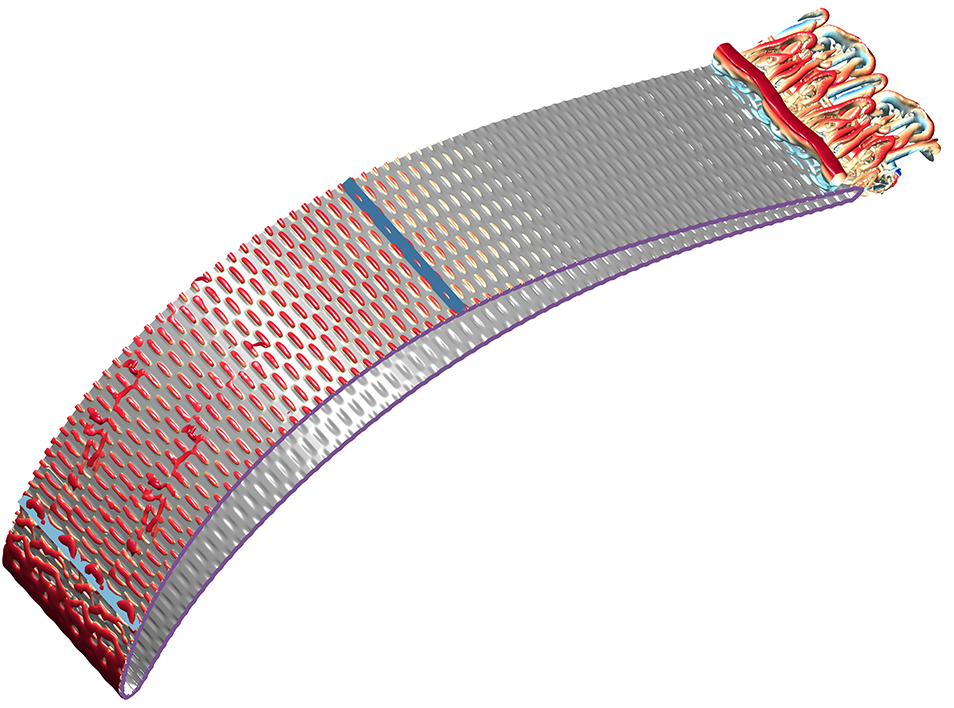}
	\put(-2,42){(\textit{c})}
    \end{overpic}\\
    \begin{overpic}[width=0.32\textwidth]{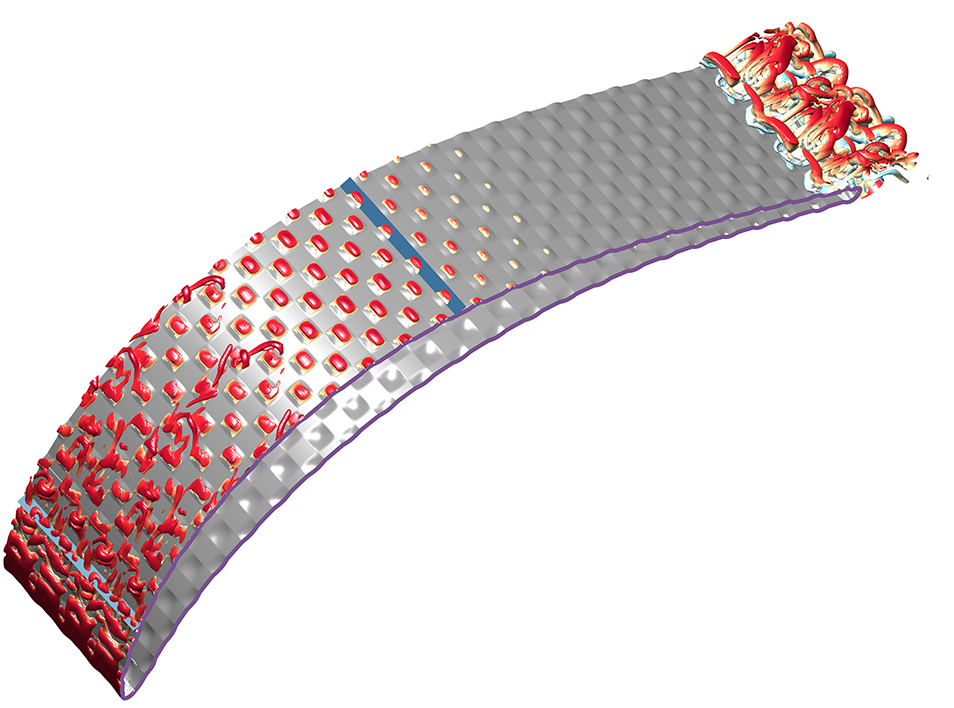}
	\put(-2,42){(\textit{d})}
    \end{overpic}
    \begin{overpic}[width=0.32\textwidth]{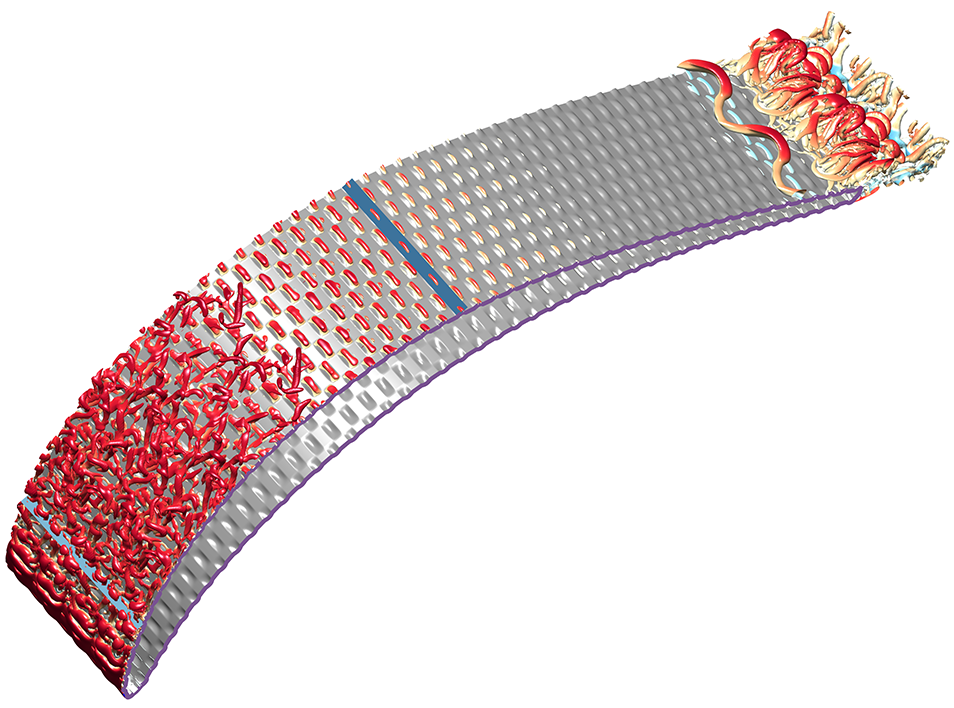}
	\put(-2,42){(\textit{e})}
    \end{overpic}
    \begin{overpic}[width=0.32\textwidth]{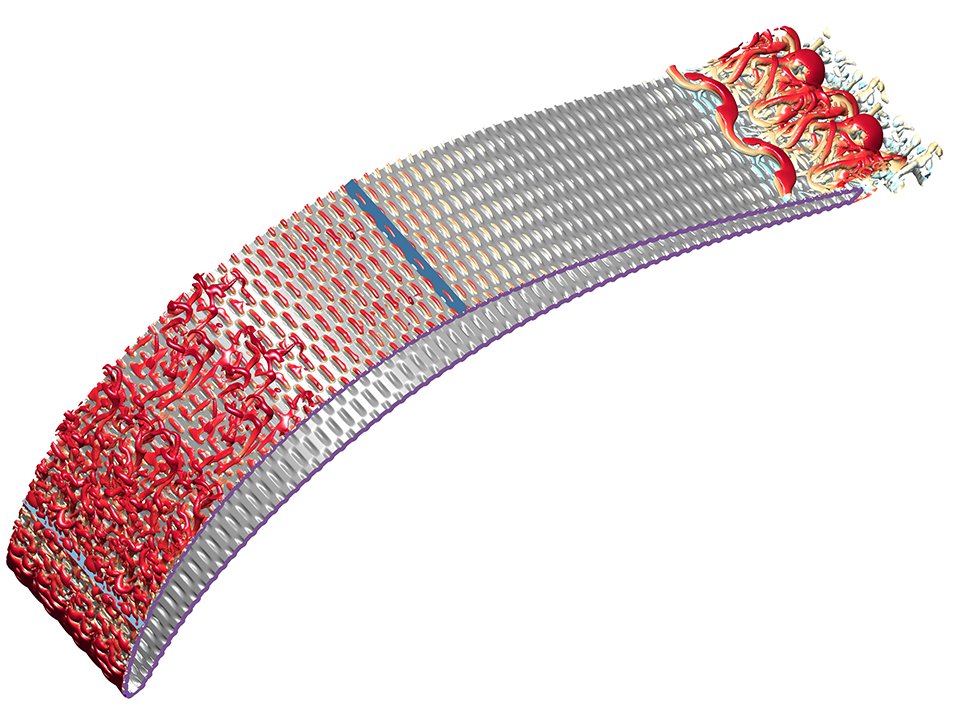}
	\put(-2,42){(\textit{f})}
    \end{overpic}\\
    \begin{overpic}[width=0.32\textwidth]{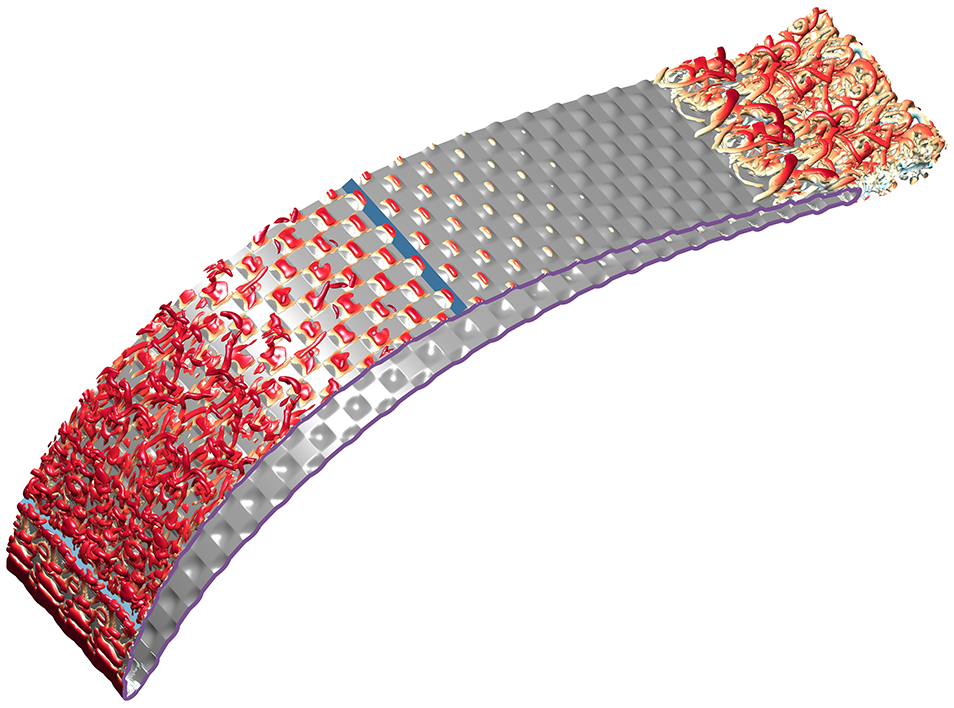}
	\put(-2,42){(\textit{g})}
    \end{overpic}
    \begin{overpic}[width=0.32\textwidth]{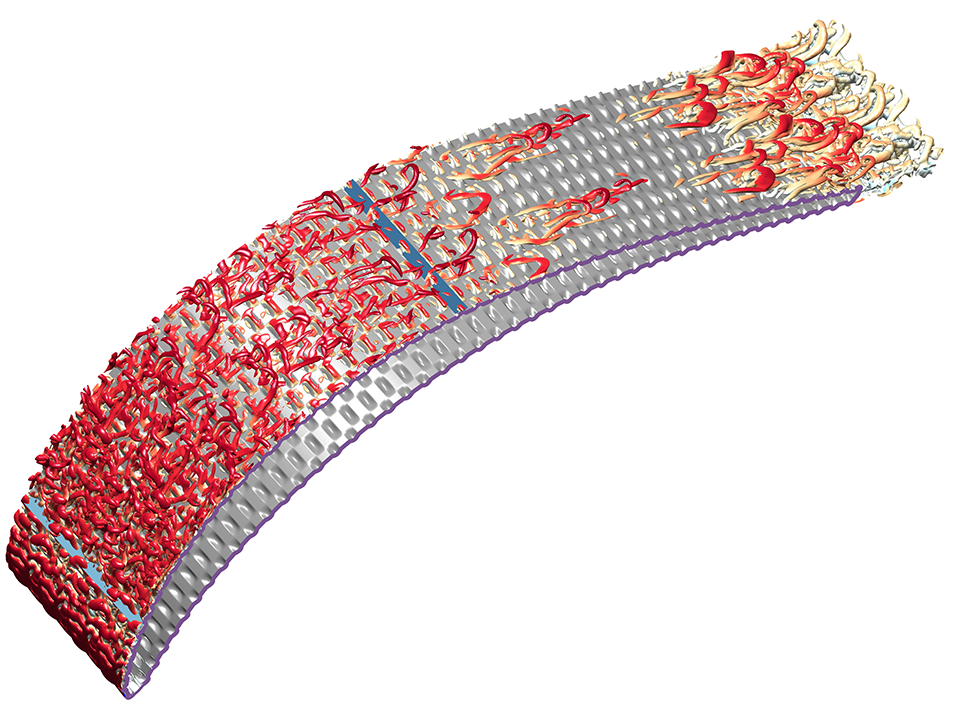}
	\put(-2,42){(\textit{h})}
        \begin{tikzpicture}[overlay, x=0.074\textwidth, y=0.074\textwidth]
            \draw[white] (0,0)--(0,1);
            \draw[white] (0,0)--(1,0);
            \draw [black][line width = 1.5pt][densely dashed](2.4,2.3) ellipse (0.8 and 0.5);
        \end{tikzpicture}
    \end{overpic}
    \begin{overpic}[width=0.32\textwidth]{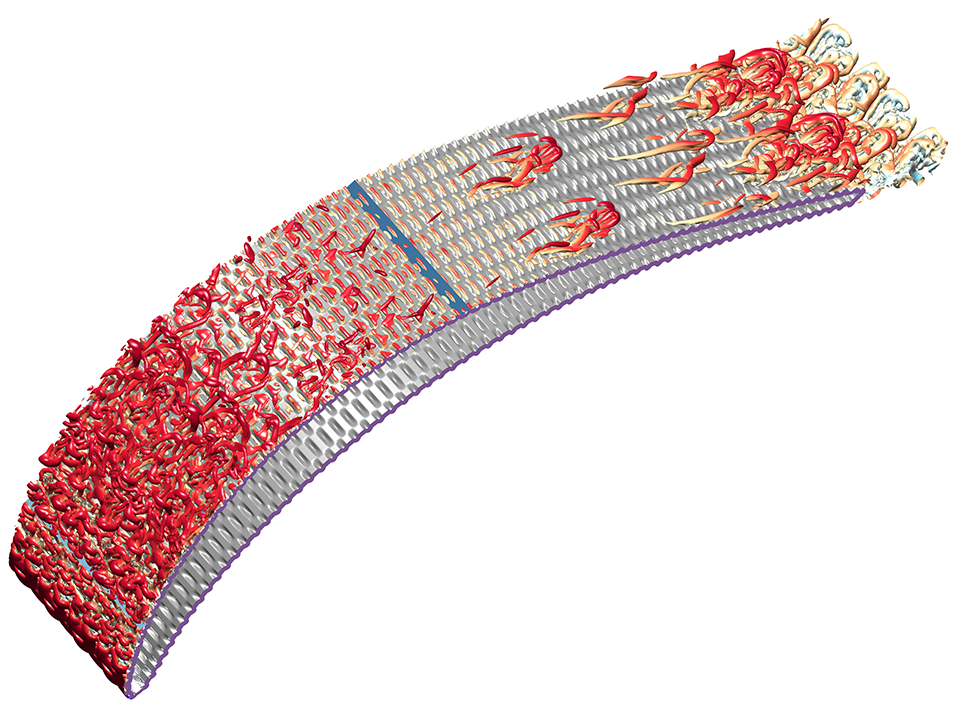}
	\put(-2,42){(\textit{i})}
    \end{overpic}\\
    \begin{overpic}[width=0.32\textwidth]{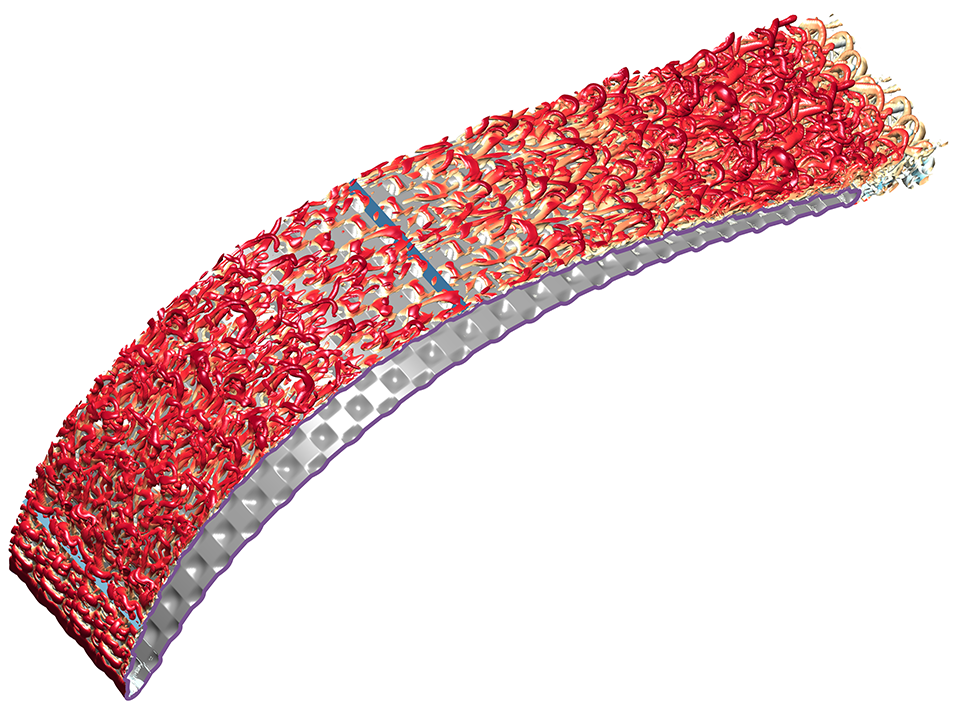}
	\put(-2,42){(\textit{j})}
    \end{overpic}
    \begin{overpic}[width=0.32\textwidth]{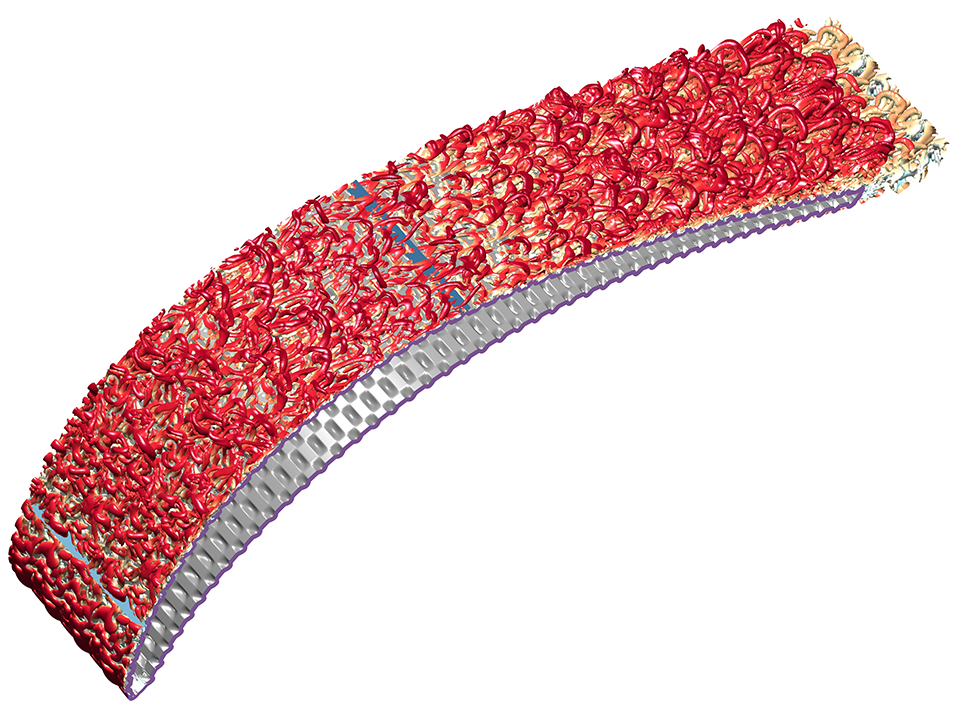}
	\put(-2,42){(\textit{k})}
    \end{overpic}
    \begin{overpic}[width=0.32\textwidth]{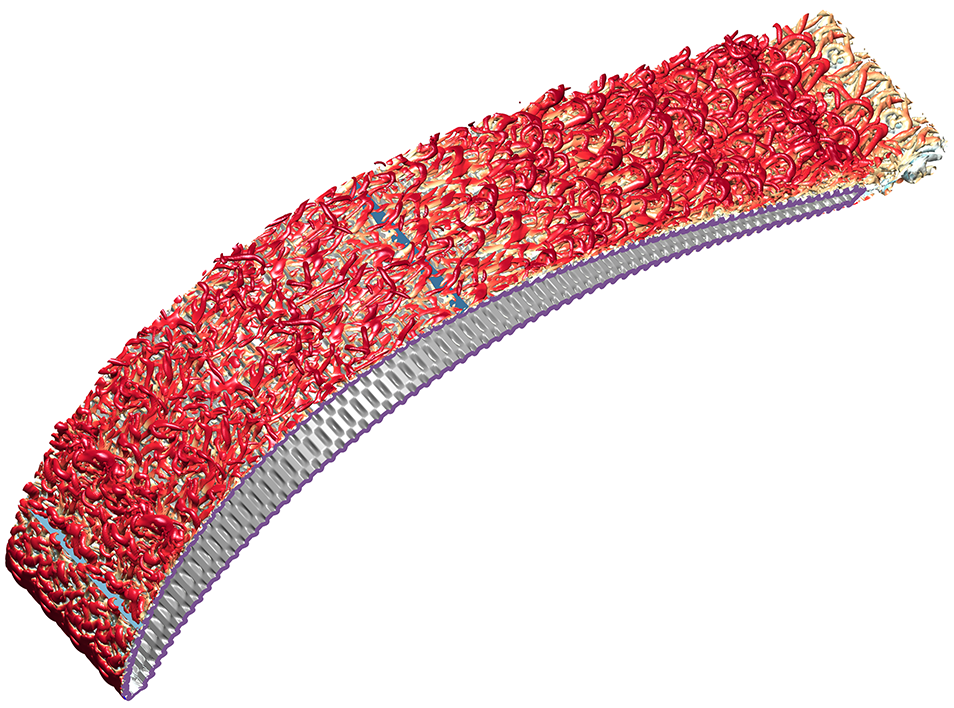}
	\put(-2,42){(\textit{l})}
    \end{overpic}\\
    \begin{overpic}[width=0.32\textwidth]{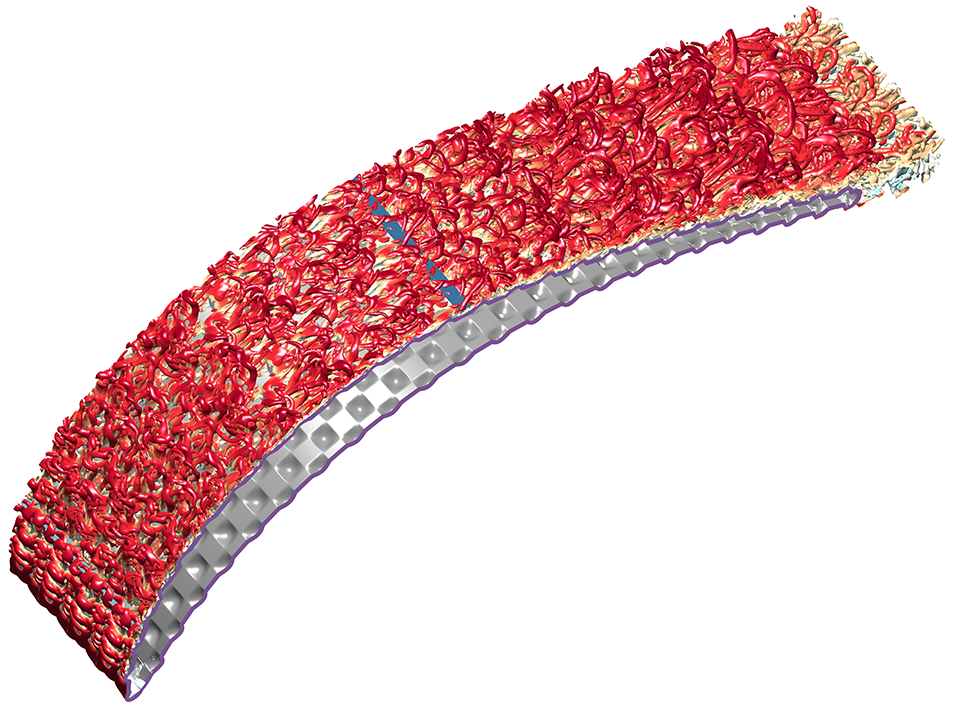}
	\put(-2,42){(\textit{m})}
        \begin{tikzpicture}[overlay, x=0.074\textwidth, y=0.074\textwidth]
            \draw[white] (0,0)--(1,0);
            \node[Qgreen](A) at (-0.7,6.3) {$k$};
            \draw[Qgreen][line width = 1pt][->](-0.4,6.18)--+(90:0.3);
            \node[Qgreen](B) at (0.45,16.2) {$\alpha$};
            \draw[Qgreen][line width = 1pt][->](0.75,16.08)--+(90:0.3);
            \draw[Qgreen][line width =1.5pt][dashed][-latex](-0.5,16)--+(0:2);
            \draw[Qgreen][line width =1.5pt][dashed][-latex](-1,7.3)--+(-90:2);
        \end{tikzpicture}
    \end{overpic}
    \begin{overpic}[width=0.32\textwidth]{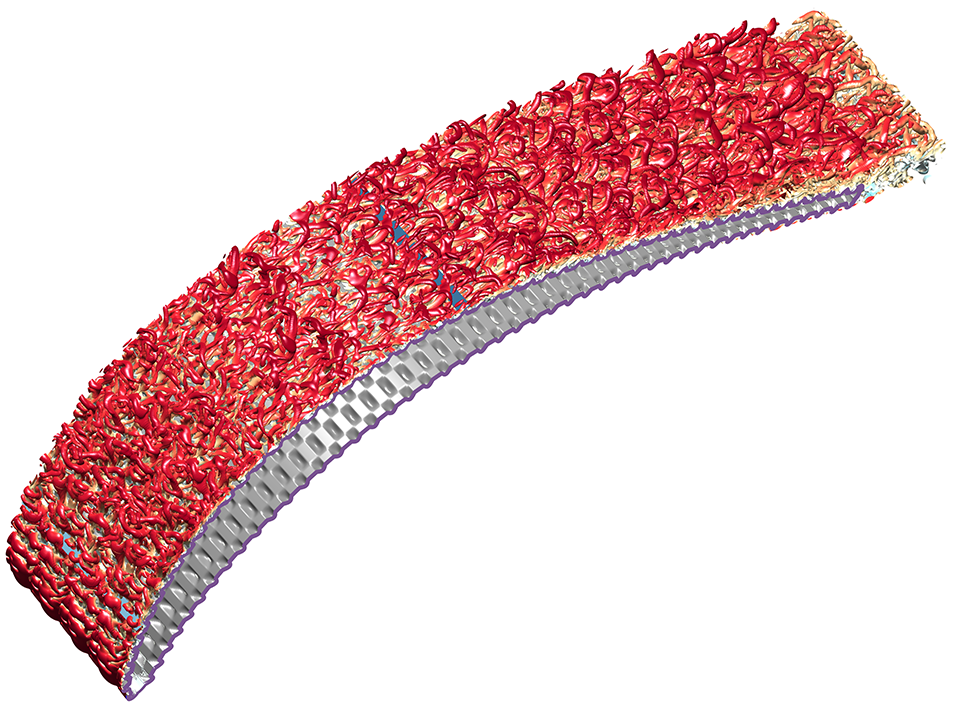}
	\put(-2,42){(\textit{n})}
    \end{overpic}
    \begin{overpic}[width=0.32\textwidth]{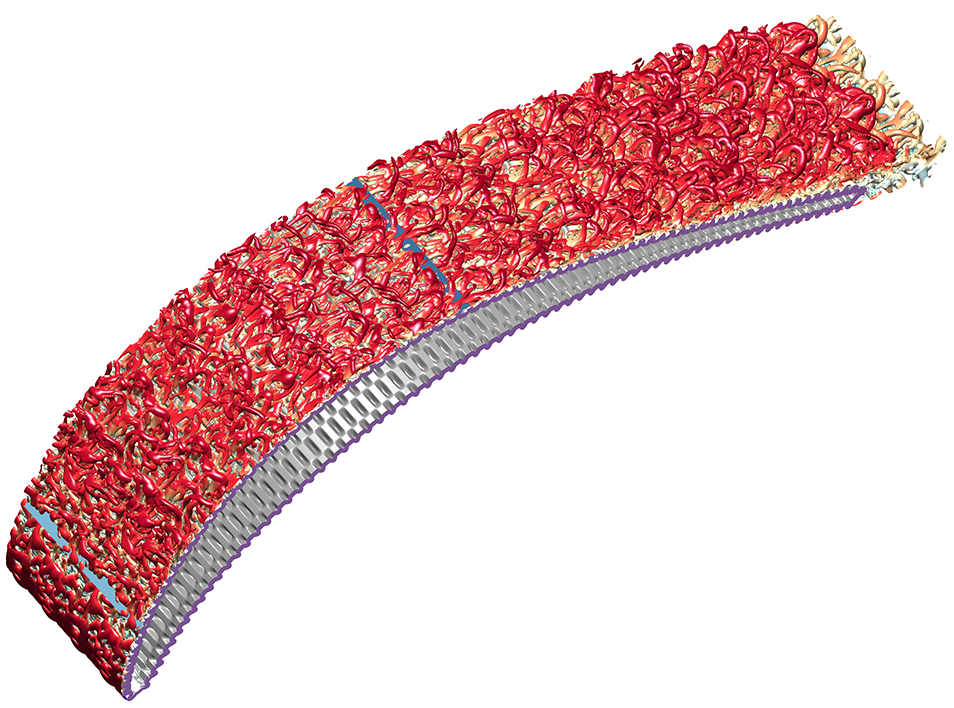}
	\put(-2,42){(\textit{o})}
    \end{overpic}
    \caption{The vortical structures on the suction-side boundary layer: (\textit{a}) $k16\alpha50$; (\textit{b}) $k16\alpha100$; (\textit{c}) $k16\alpha150$; (\textit{d}) $k32\alpha50$; (\textit{e}) $k32\alpha100$; (\textit{f}) $k32\alpha150$; (\textit{g}) $k48\alpha50$; (\textit{h}) $k48\alpha100$; (\textit{i}) $k48\alpha150$; (\textit{j}) $k64\alpha50$; (\textit{k}) $k64\alpha100$; (\textit{l}) $k64\alpha150$; (\textit{m}) $k80\alpha50$; (\textit{n}) $k80\alpha100$; (\textit{o}) $k80\alpha150$. Instantaneous iso-surfaces of Q=1000 are presented, colored by mean tangential velocity. The blue lines represent $x/C_{ax}=0.1$ and $x/C_{ax}=0.65$, respectively.}
    \label{fig:Q_vel}
\end{figure*}

An overview of the suction-side boundary layer is given to present the complex flow phenomena affected by wall roughness. 
Specifically, Fig.~\ref{fig:Q_vel} shows the instantaneous vortical structures identified by iso-surfaces of the Q-criterion \cite{Hunt1988}, which are colored by the mean wall-tangential velocity. 
Obviously, the surface roughness has a significant impact on the suction-side boundary layer, and cases with different surface roughness show varying flow structures. 
We first focus on the effect of roughness height.
In cases with relatively low roughness height, such as the $k16$ and $k32$ cases, the roughness-induced disturbances are mainly limited to the proximity of roughness elements.
In particular, the disturbances at the leading edge are suppressed in the FPG region, until turbulent vortical structures re-occur near the trailing edge. 
In cases with increasingly higher roughness amplitude, however, pronounced vortical structures emerge. 
Specifically, across the FPG region, escalated roughness heights intensify boundary layer disturbances, thereby sustaining vortical structures emanating from the leading edge in the $k64$ to $k80$ cases. 
Furthermore, we can analyze the effect of the streamwise wavenumber of the surface roughness, focusing on the $k48$ cases shown in Fig.~\ref{fig:Q_vel}(\textit{g,h,i}). 
For case $k48\alpha50$ in Fig.~\ref{fig:Q_vel}(\textit{g}), no strong vortical structures are observed in the APG region, except for the region near the trailing edge. 
As a comparison, cases $k48\alpha100$ and $k48\alpha150$ in Fig.~\ref{fig:Q_vel}(\textit{h,i}) present intermittent transitional structures in the APG region, which finally develop to turbulence near the blade trailing edge. 
This suggests that the roughness wavenumber may have a significant impact on the transition process on the suction-side boundary layer.  

\begin{figure*}[t!]
    \centering
	\begin{overpic}[width=0.32\textwidth]{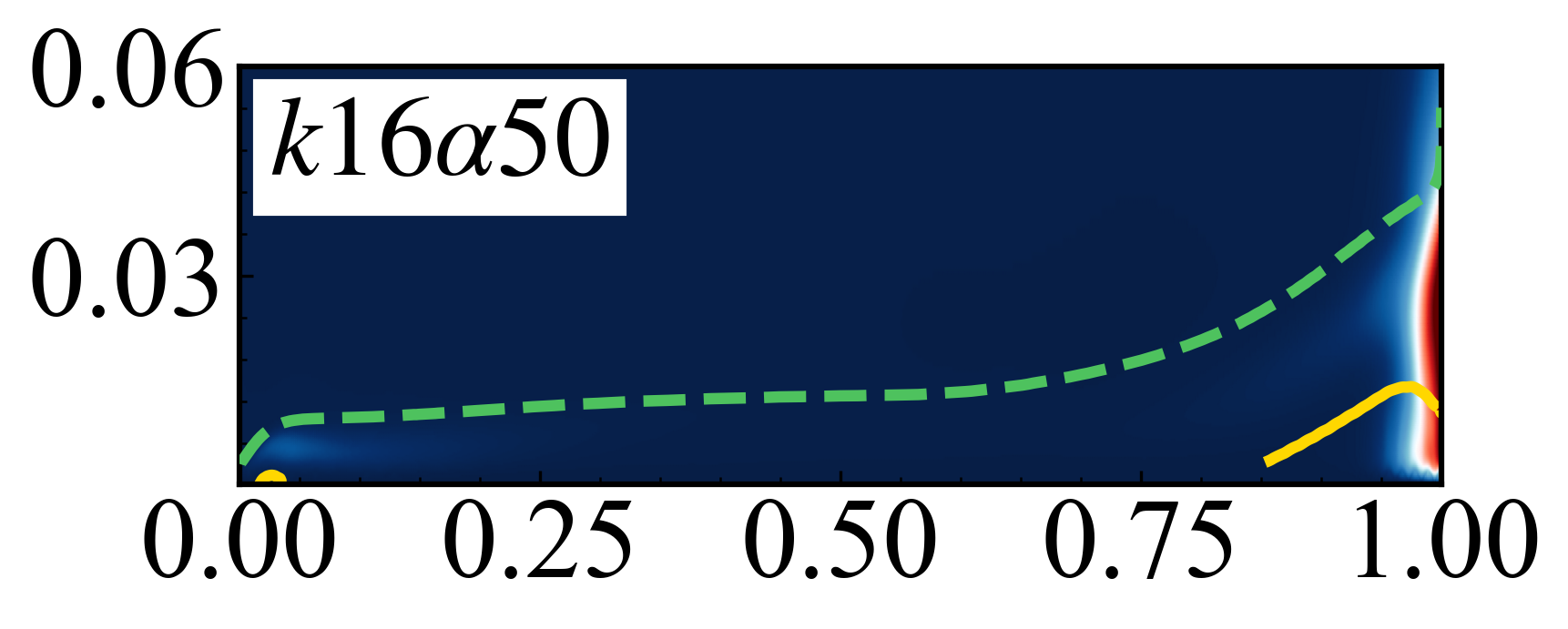}
		\put(-7,35){(\textit{a})}
		\put(-6,20){\scalebox{1}{\rotatebox[origin=c]{90}{$\eta/C_{ax}$}}}
	\end{overpic}
	\begin{overpic}[width=0.32\textwidth]{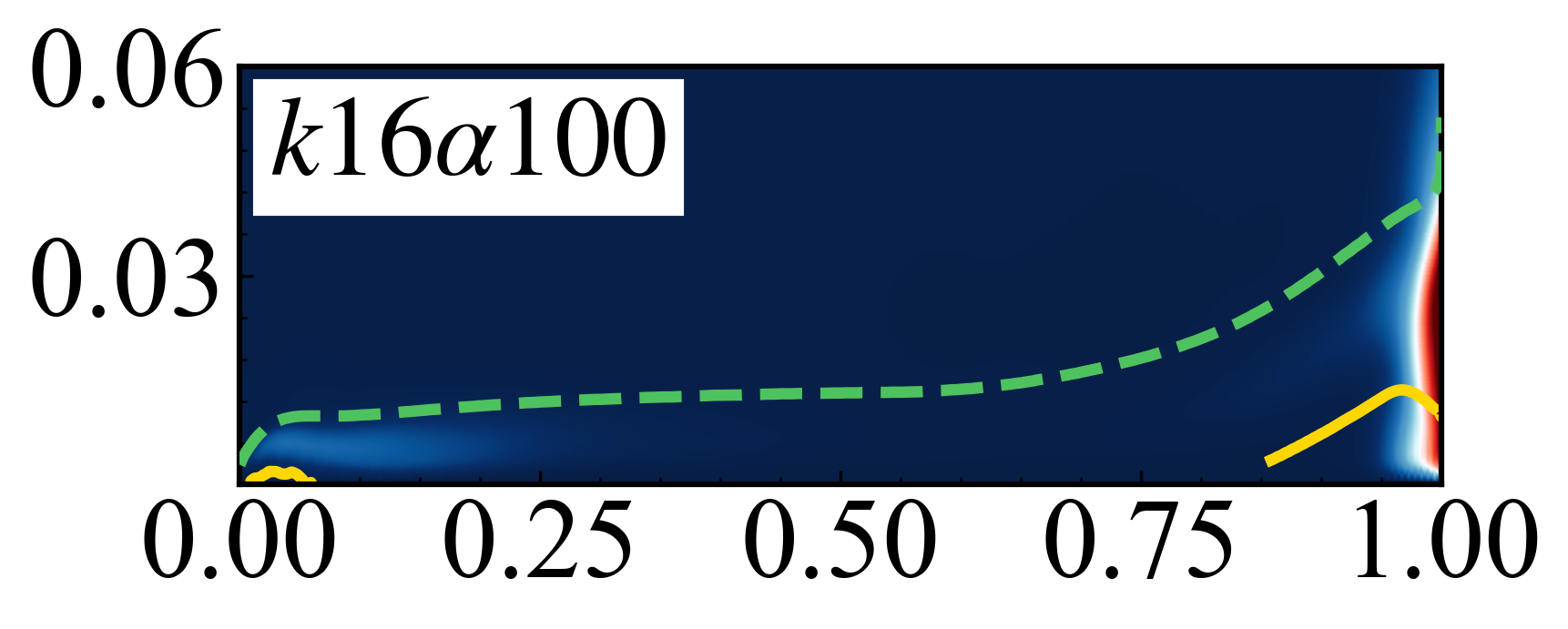}
		\put(-7,35){(\textit{b})}
	\end{overpic}
	\begin{overpic}[width=0.32\textwidth]{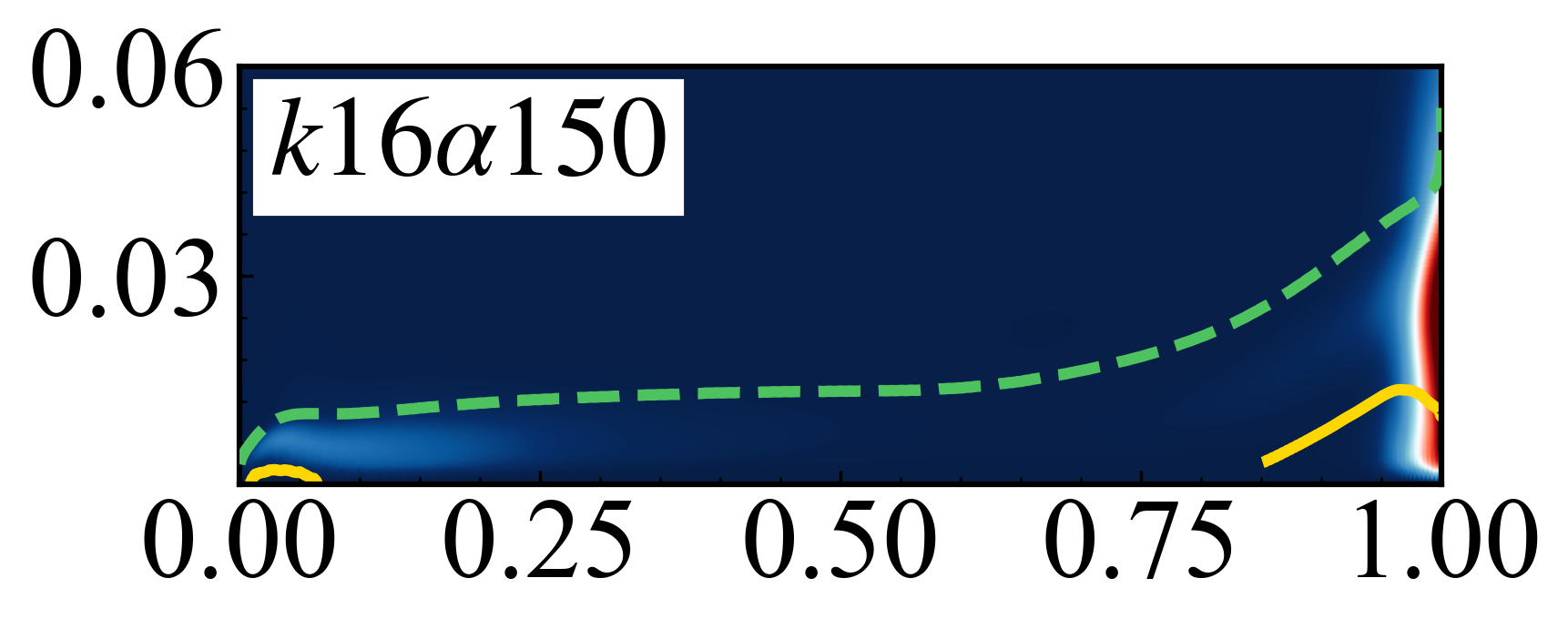}
		\put(-7,35){(\textit{c})}
	\end{overpic}\\[0.8em]
        \begin{overpic}[width=0.32\textwidth]{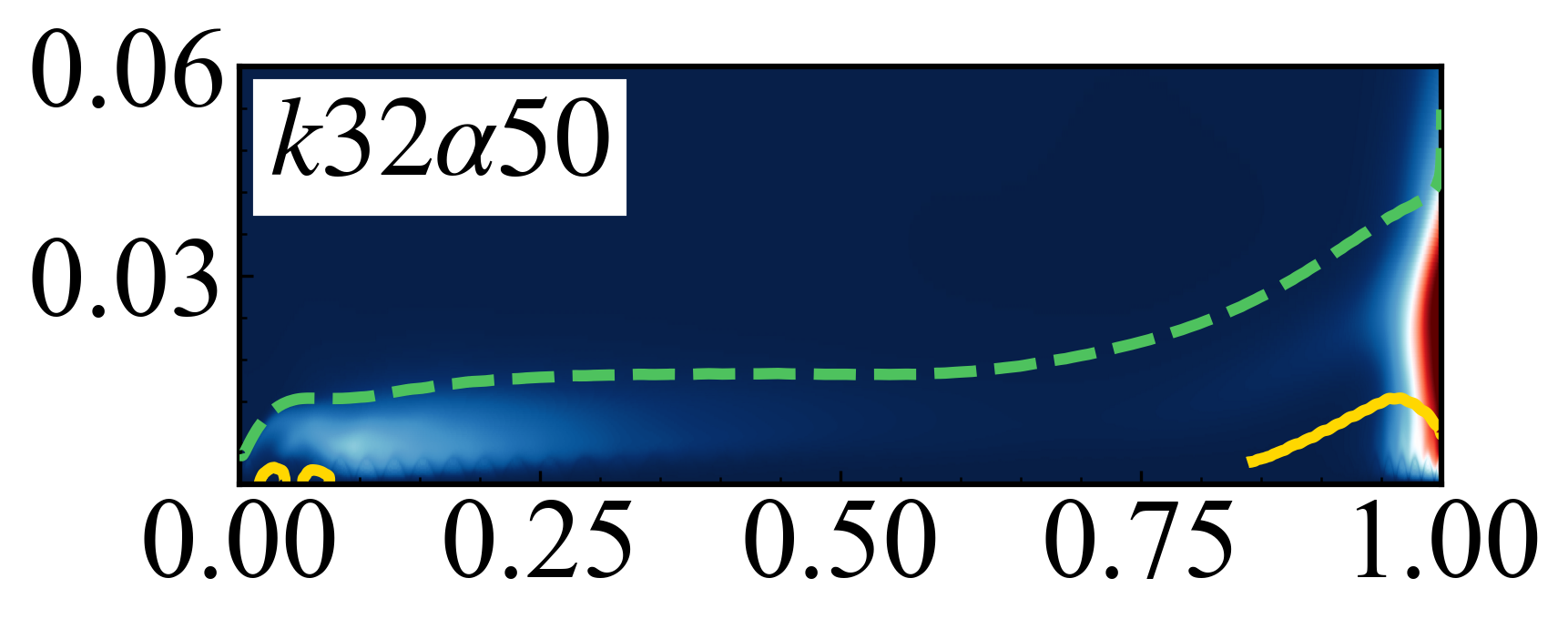}
		\put(-7,35){(\textit{d})}
		\put(-6,20){\scalebox{1}{\rotatebox[origin=c]{90}{$\eta/C_{ax}$}}}
	\end{overpic}
        \begin{overpic}[width=0.32\textwidth]{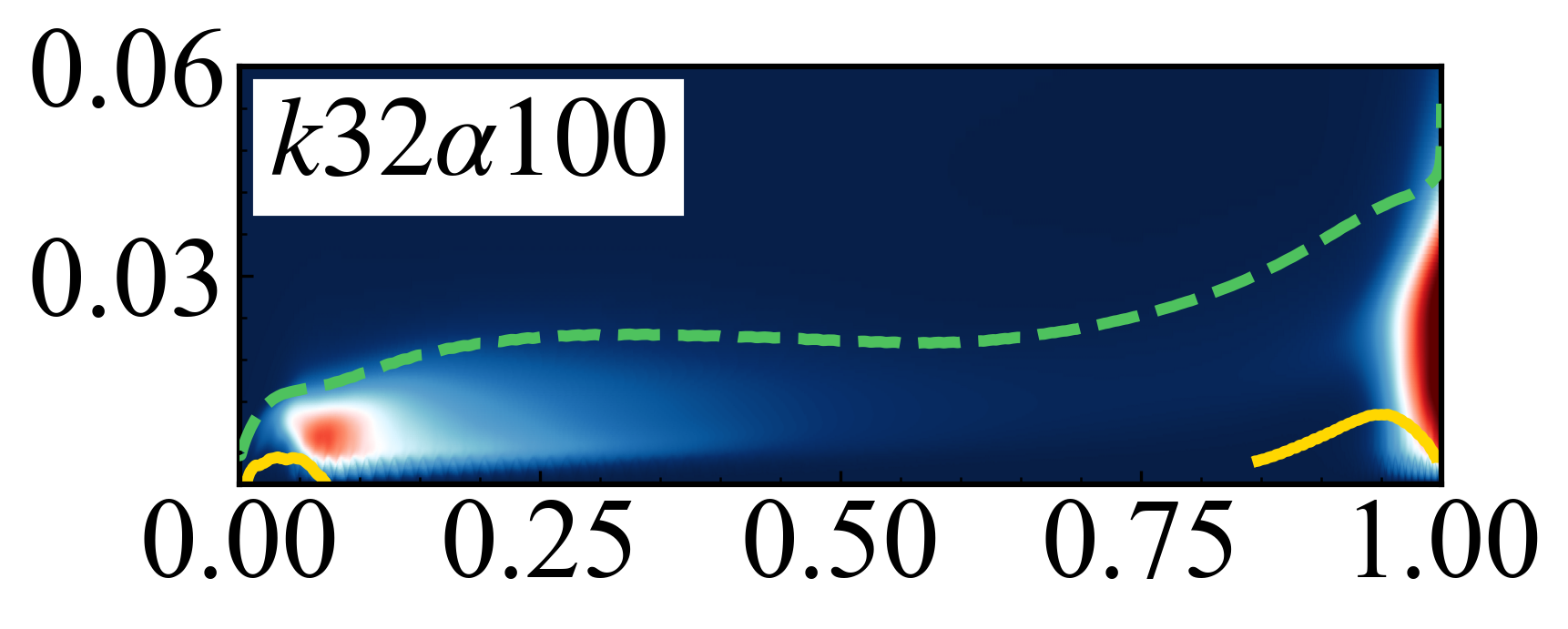}
		\put(-7,35){(\textit{e})}
	\end{overpic}
        \begin{overpic}[width=0.32\textwidth]{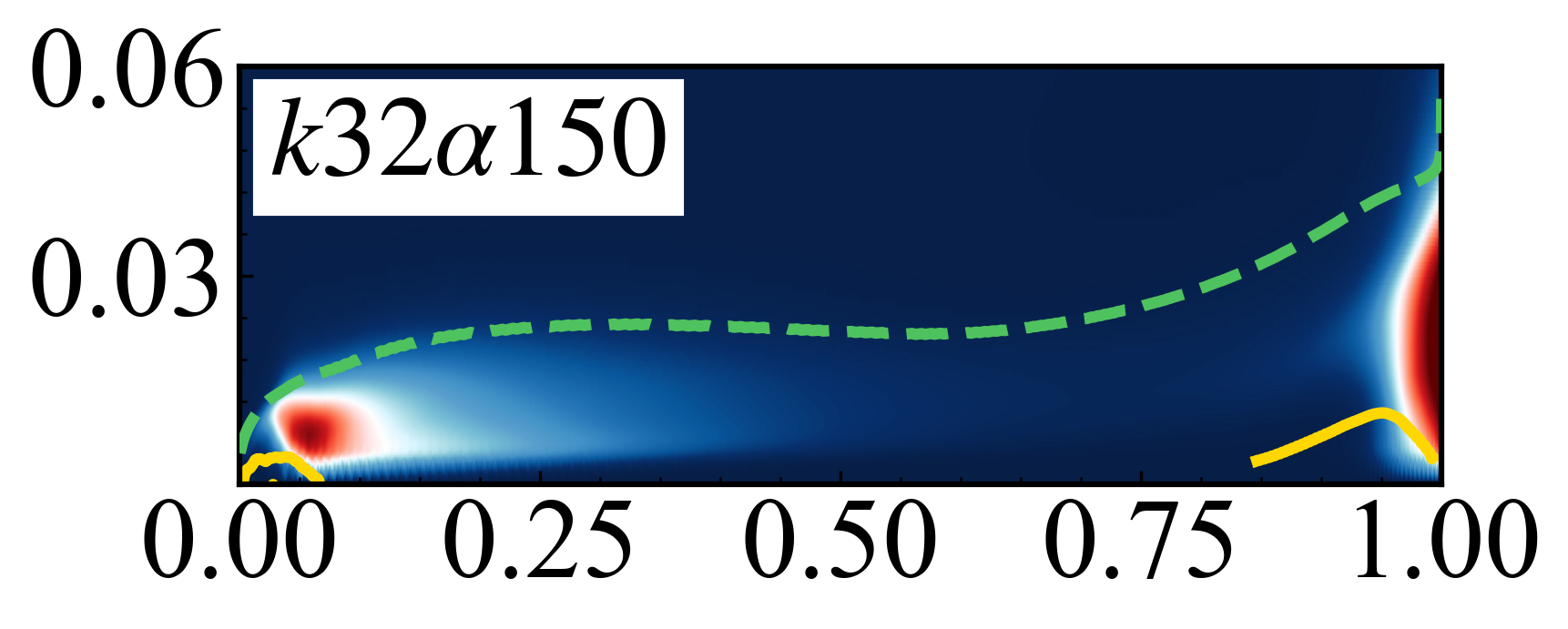}
		\put(-7,35){(\textit{f})}
	\end{overpic}\\[0.8em]
        \begin{overpic}[width=0.32\textwidth]{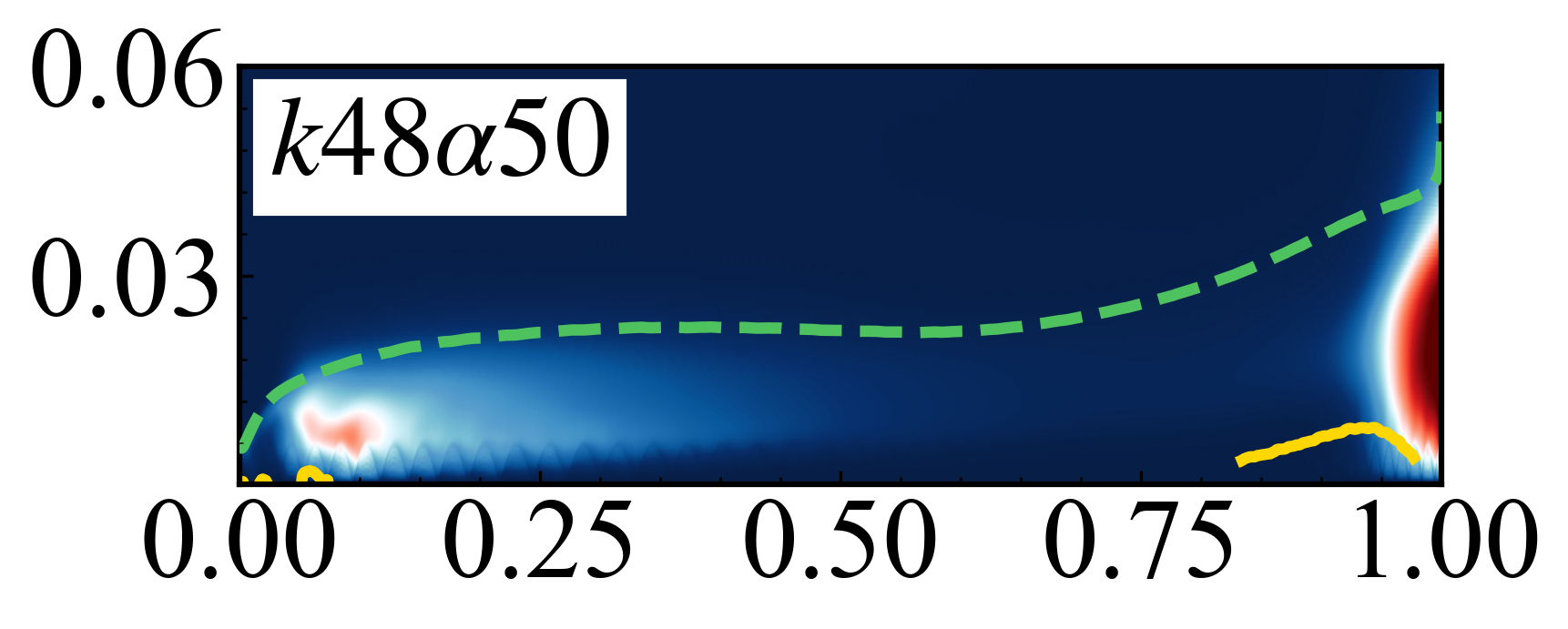}
		\put(-7,35){(\textit{g})}
		\put(-6,20){\scalebox{1}{\rotatebox[origin=c]{90}{$\eta/C_{ax}$}}}
	\end{overpic}
        \begin{overpic}[width=0.32\textwidth]{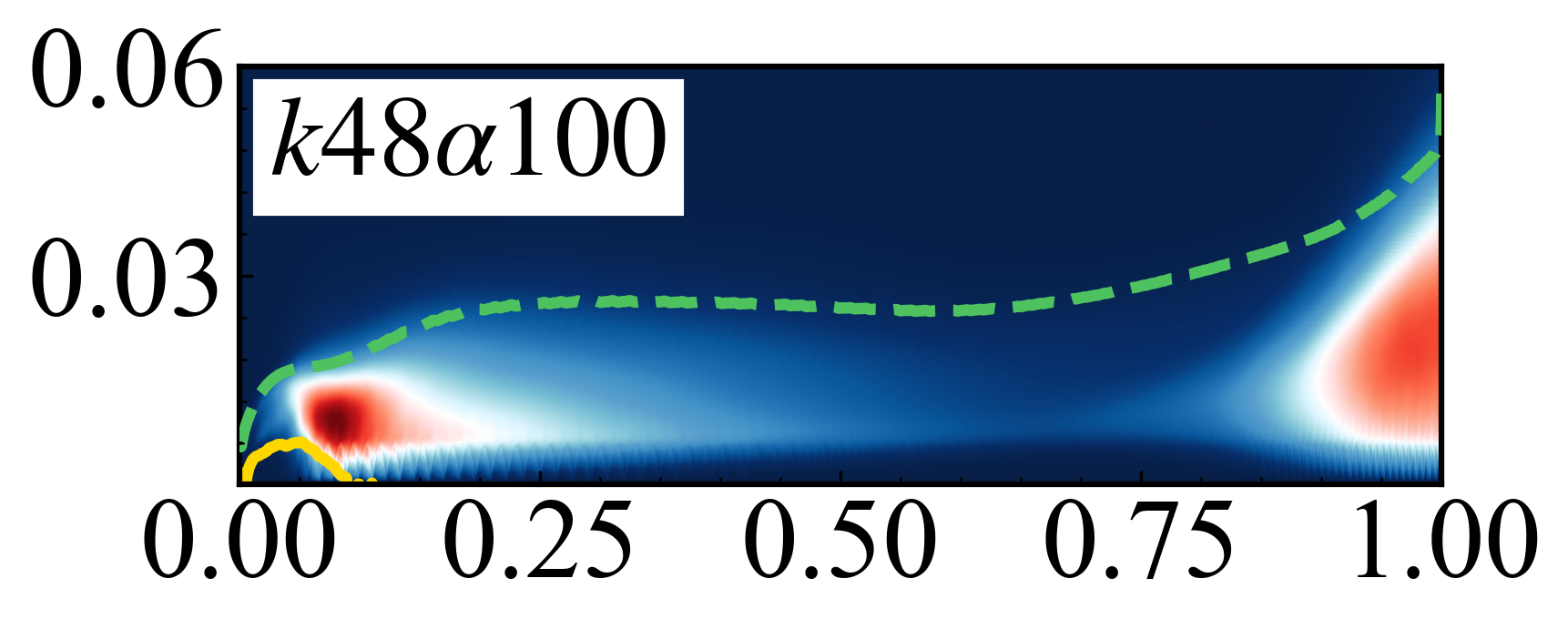}
		\put(-7,35){(\textit{h})}
	\end{overpic}
        \begin{overpic}[width=0.32\textwidth]{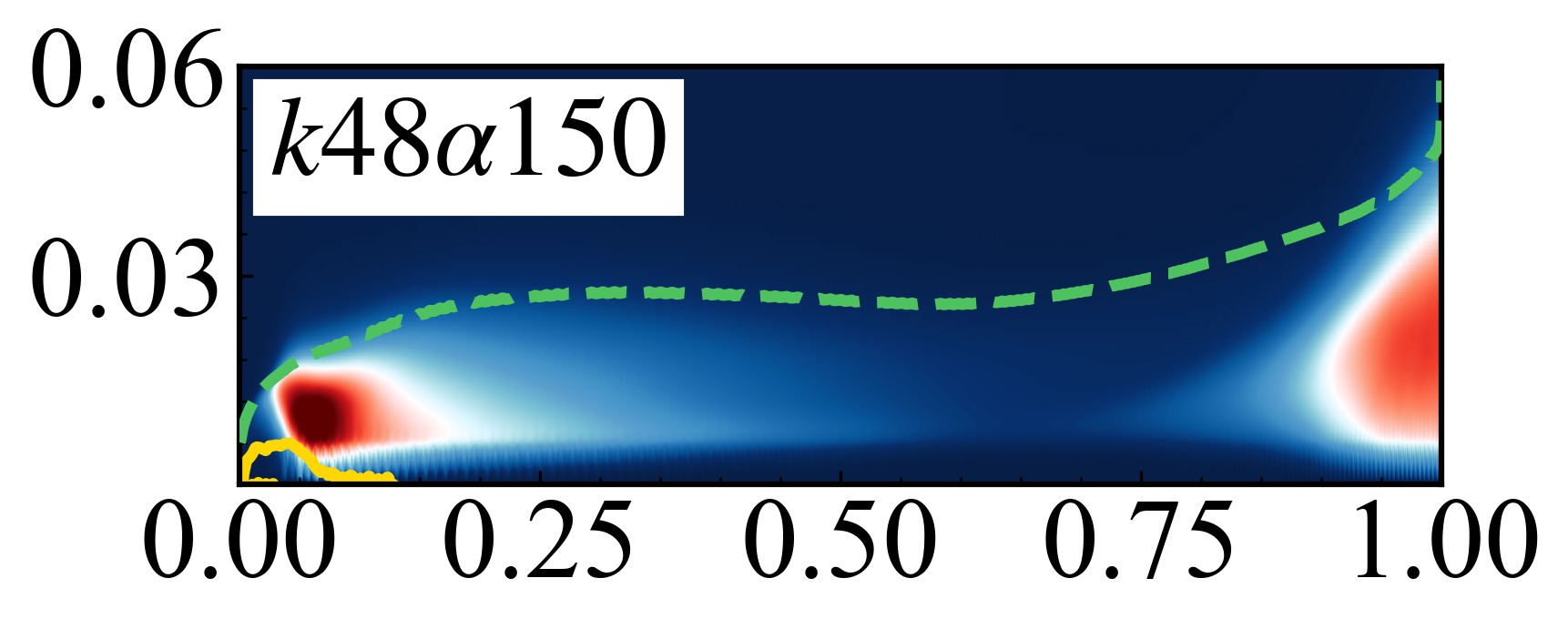}
		\put(-7,35){(\textit{i})}
	\end{overpic}\\[0.8em]
        \begin{overpic}[width=0.32\textwidth]{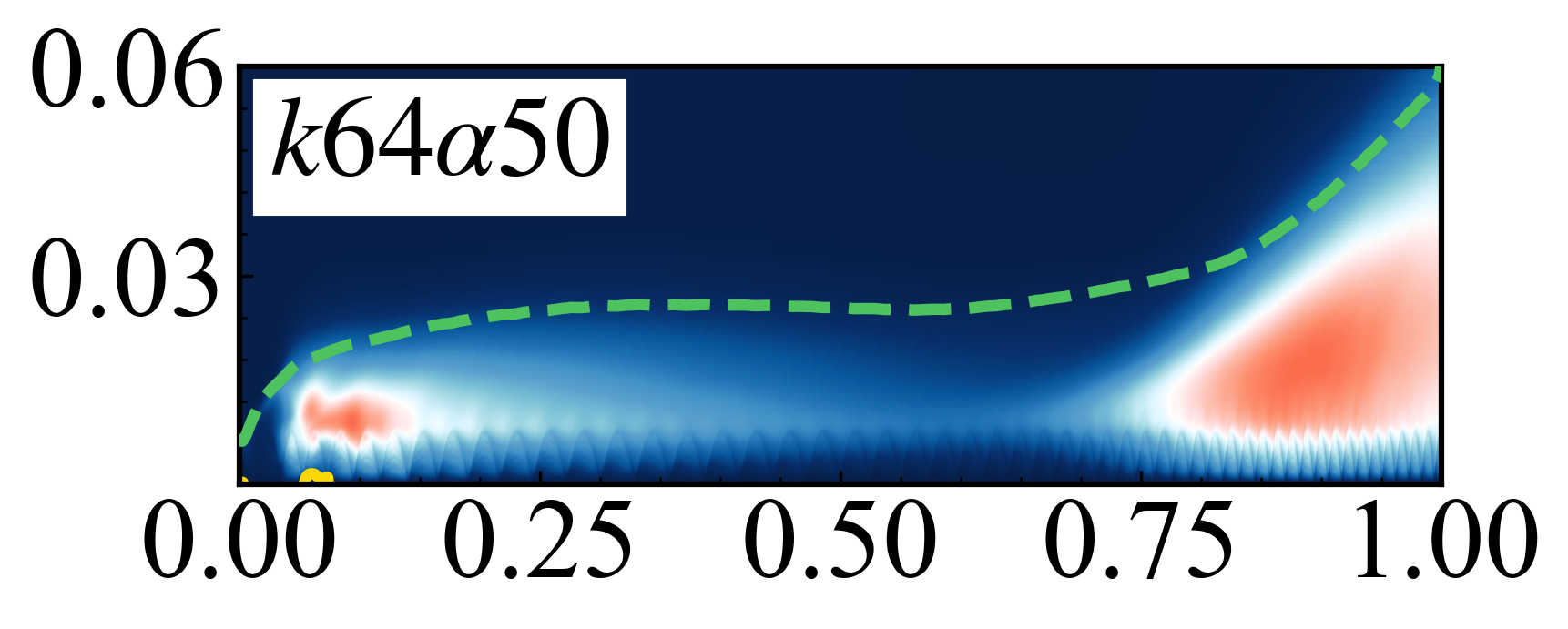}
		\put(-7,35){(\textit{j})}
		\put(-6,20){\scalebox{1}{\rotatebox[origin=c]{90}{$\eta/C_{ax}$}}}
	\end{overpic}
        \begin{overpic}[width=0.32\textwidth]{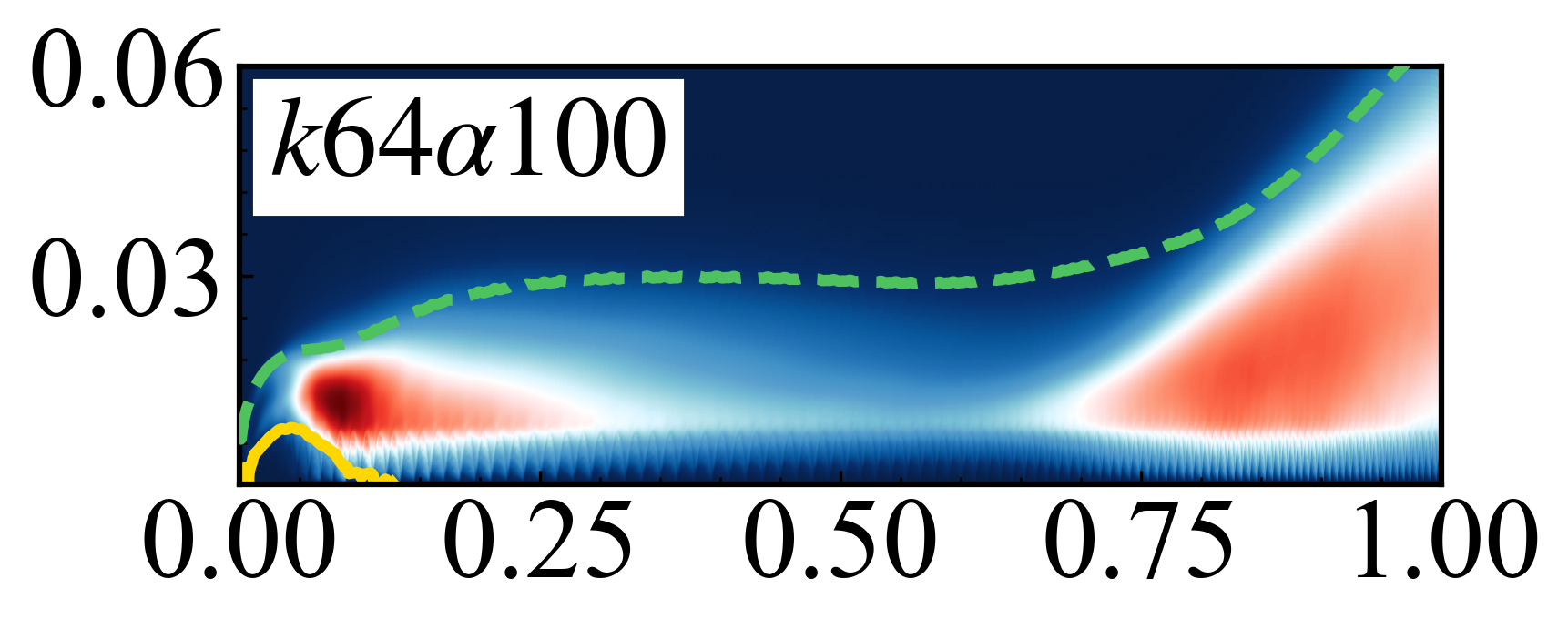}
		\put(-7,35){(\textit{k})}
	\end{overpic}
        \begin{overpic}[width=0.32\textwidth]{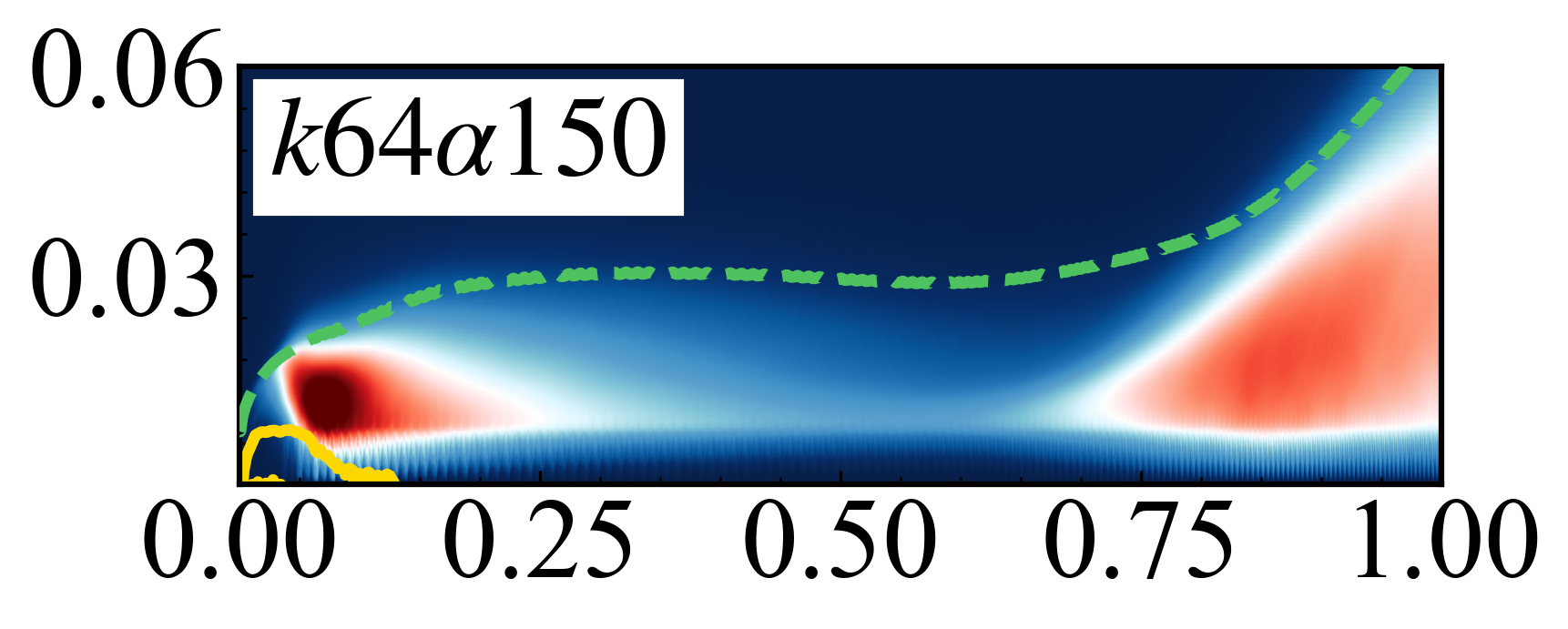}
		\put(-7,35){(\textit{l})}
	\end{overpic}\\[0.8em]
        \begin{overpic}[width=0.32\textwidth]{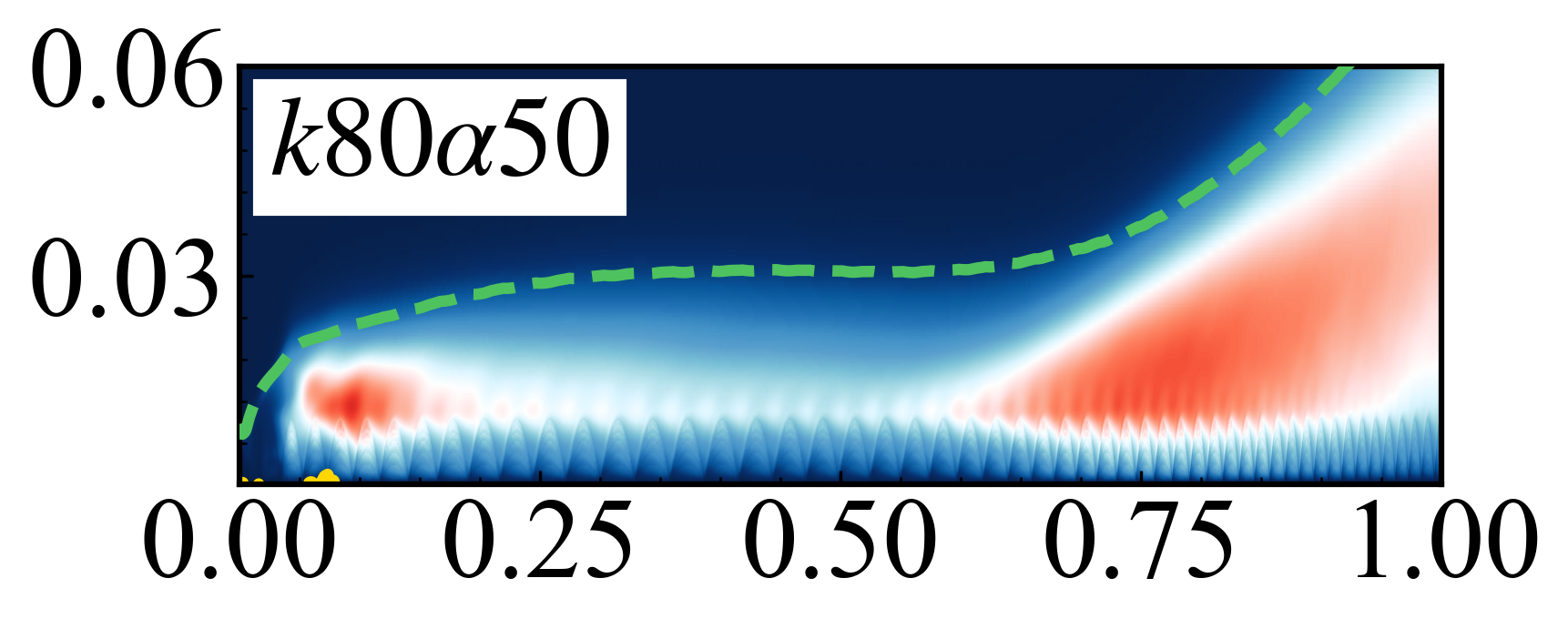}
		\put(-8,35){(\textit{m})}
		\put(-6,20){\scalebox{1}{\rotatebox[origin=c]{90}{$\eta/C_{ax}$}}}
		\put(45,-4){\scalebox{1}{$x/C_{ax}$}}
	\end{overpic}
        \begin{overpic}[width=0.32\textwidth]{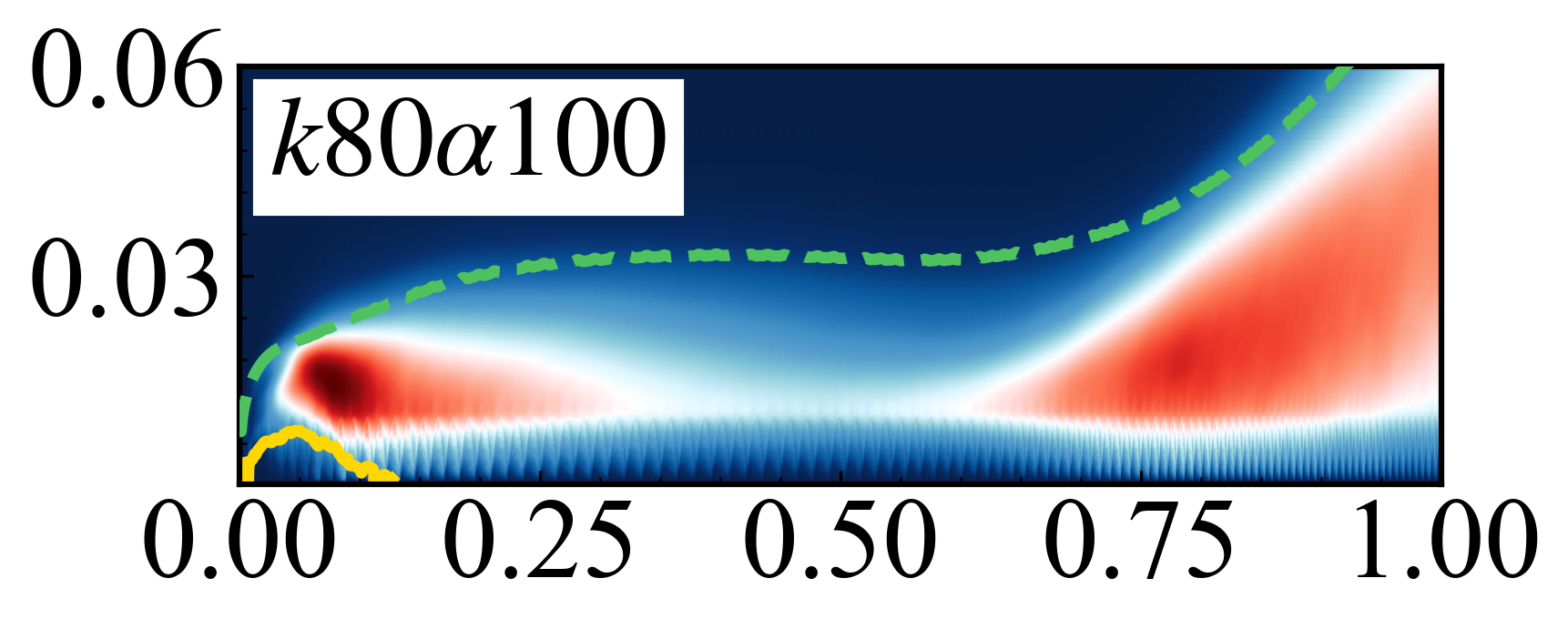}
		\put(-7,35){(\textit{n})}
		\put(45,-4){\scalebox{1}{$x/C_{ax}$}}
	\end{overpic}
        \begin{overpic}[width=0.32\textwidth]{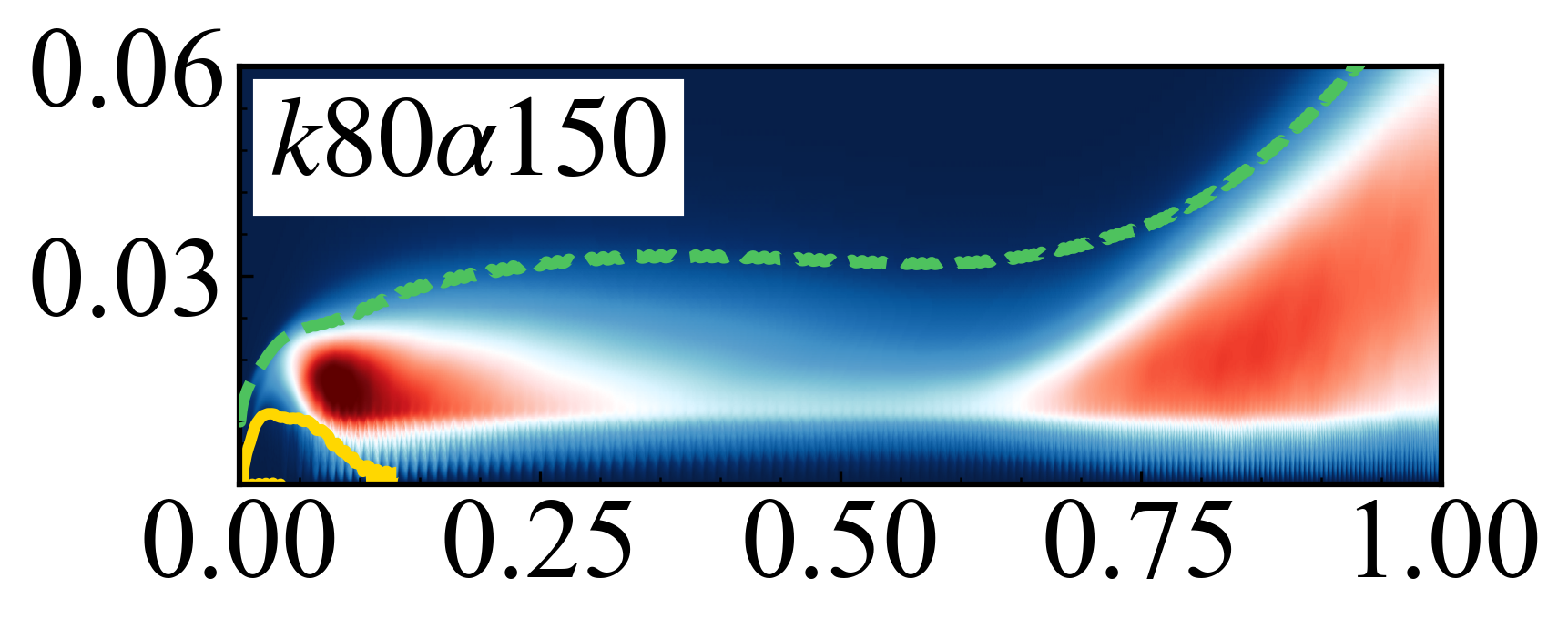}
		\put(-7,35){(\textit{o})}
		\put(45,-4){\scalebox{1}{$x/C_{ax}$}}
	\end{overpic}\\[2em]
	\begin{overpic}[width=0.36\textwidth]{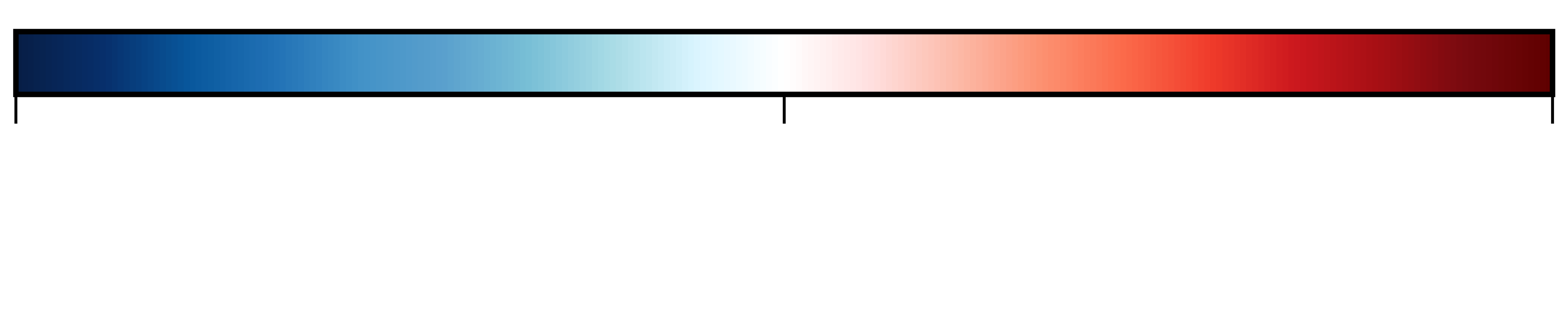}
            \put(0,6){\scalebox{1.0}{$0$}}
            \put(45,6){\scalebox{1.0}{$0.05$}}
            \put(95,6){\scalebox{1.0}{$0.1$}}
            \put(45,-1){\scalebox{1.0}{$TKE$}}
	\end{overpic}
	\caption{Contours of spanwise- and time-averaged turbulent kinetic energy in the suction-side boundary layer. The green dashed lines are the edge of the boundary layer, and the yellow solid lines indicate the separation bubbles.}
	\label{fig:tke_fluc_linear_contour}
\end{figure*}

To further shed light on the roughness effects, the contours of the turbulent kinetic energy (TKE) in the suction-side boundary layer are shown in Fig.~\ref{fig:tke_fluc_linear_contour}. 
The TKE is computed based on the triple decomposition method \cite{Reynolds_Hussain_1972}, as 
\begin{align}
u_i&=\left\langle u_i\right\rangle+u_i^{\prime}=\bar{u}_i+\tilde{u}_i+u_i^{\prime},\\
\overline{u_i u_j}&=\overline{(\bar{u}_i+\tilde{u}_i+u_i^{\prime})(\bar{u}_j+\tilde{u}_j+u_j^{\prime})}=\bar{u}_i \bar{u}_j+\overline{\tilde{u}_i \tilde{u}_j}+\overline{u_i^{\prime} u_j^{\prime}},\\
TKE&=\frac{1}{2}\overline{u_i^{\prime} u_i^{\prime}}; \quad E_{dis}=\frac{1}{2}\overline{\tilde{u}_i \tilde{u}_i}.
\end{align}
Here, $\bar{}$ denotes the time- and spanwise- averaged quantity, while $\left\langle \right\rangle$ denotes the time-averaged quantity.
Accordingly, $u_i^{\prime}$ denotes the turbulent fluctuating velocity, while $\tilde{u}_i$ is the dispersive fluctuating velocity that accounts for spanwise non-homogeneity induced by roughness elements distributions.
Moreover, TKE and $E_{dis}$ represent the turbulent and dispersive fluctuating energies, respectively. 

It is not surprising to see in Fig.~\ref{fig:tke_fluc_linear_contour} that the TKE distribution is significantly impacted by the roughness height.
In cases with low roughness, such as the $k16$ and $k32$ cases, the high TKE region is mainly near the blade trailing edge, which is presumably caused by the boundary layer separation due to the APG as indicated by the zero velocity iso-lines in Figs.~\ref{fig:tke_fluc_linear_contour}(\textit{a$\sim$f}).
In cases with higher roughness, however, the TKE increases violently in the APG region of the suction-side boundary layer ($x/C_{ax}>0.65$), and the trailing-edge separation is suppressed accordingly.
The other interesting observation is the significant impact of the roughness streamwise wavenumber $\alpha$, as shown by the $k48$ cases in Figs.~\ref{fig:tke_fluc_linear_contour}(\textit{g,h,i}).
Though the roughness height is the same for these three cases, the cases $k48\alpha100$ and $k48\alpha150$ in Figs.~\ref{fig:tke_fluc_linear_contour}(\textit{h,i}) show earlier increase of the TKE in the APG region and suppression of the trailing-edge separation, in contrast to the  $k48\alpha50$ case in Fig.~\ref{fig:tke_fluc_linear_contour}(\textit{g}).
This agrees with the observation about the vortical structures in Fig.~\ref{fig:Q_vel}.
In addition to the TKE distribution in the APG region, the streamwise wavenumber of roughness also has a significant effect on the leading-edge region.
It is noted that cases with higher roughness wavenumber (and thus higher effective slope), like cases $k48\alpha100$ and $k48\alpha150$, induce a leading-edge separation, and the boundary layer in that region is thus highly disturbed.
Although the leading-edge disturbances seem to be suppressed in the following region with strong FPG, whether they have direct impact on the APG transiton behavior requires further investigation in the following sections.

\begin{figure*}[t!]
	\centering
	\begin{overpic}[width=0.46\textwidth]{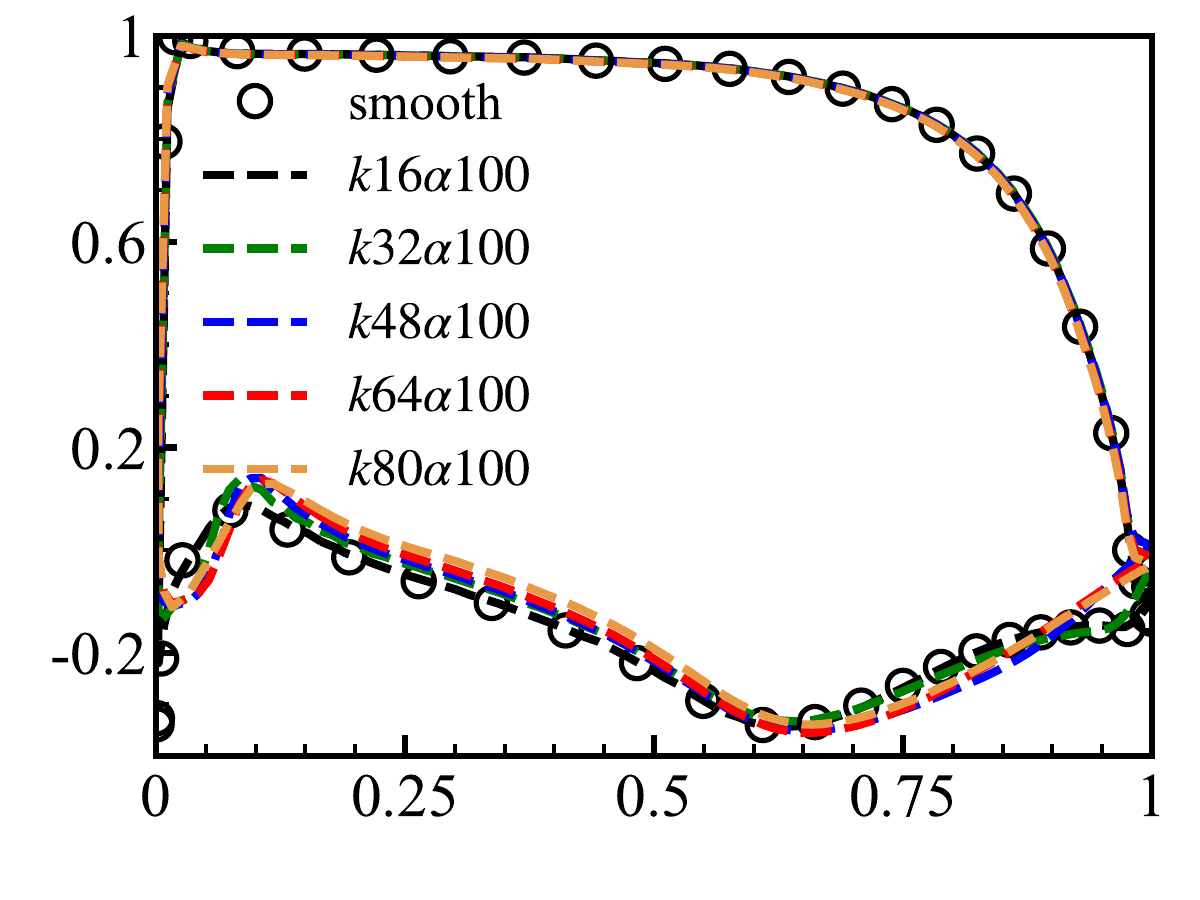}
        \put(-4,71){(\textit{a})}
        \put(-2,40){\scalebox{1.0}{$C_p$}}
        \begin{tikzpicture}[overlay, x=0.074\textwidth, y=0.074\textwidth]
            \draw[white] (0,0)--(0,1);
            \draw[white] (0,0)--(1,0);
            \draw[mycolor][line width = 1.5pt][densely dashed][-latex](2.3,1)--+(60:1.3);
            \node[mycolor] (A) at (3.3,2.1) {$k$};
            \draw[mycolor][line width = 1pt][->](3.6,1.98)--+(90:0.3);
            \draw[mycolor][line width = 1.5pt][dashed][rounded corners] (4.4,0.8) rectangle (5.9,1.8);
        \end{tikzpicture}
	\end{overpic}
    \hspace{0.02\textwidth}
    \begin{overpic}[width=0.46\textwidth]{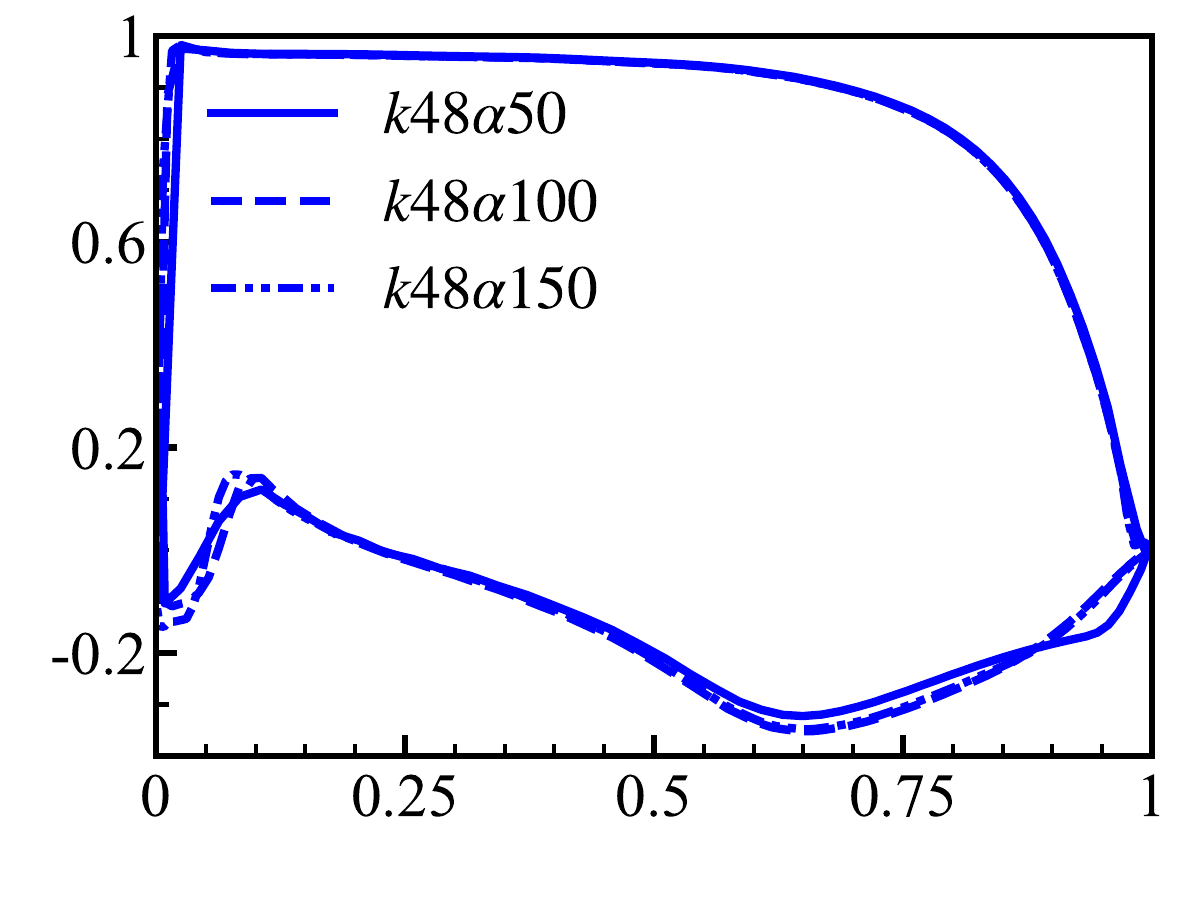}
    \put(-4,71){(\textit{b})}
	\end{overpic}\\
    \begin{overpic}[width=0.46\textwidth]{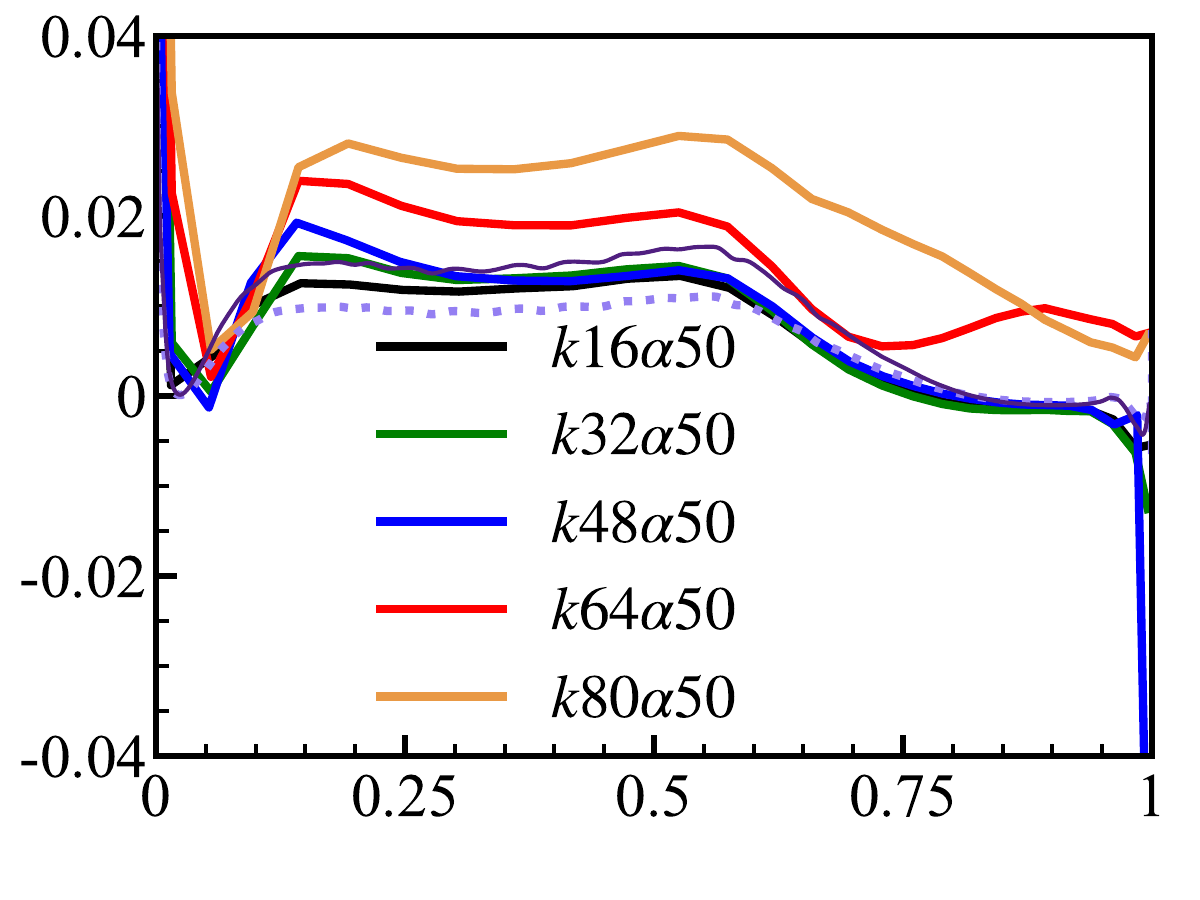}
        \put(-4,71){(\textit{c})}
        \put(50,0){\scalebox{1.0}{$x/C_{ax}$}}
		  \put(-5,40){\scalebox{1.0}{$C_d$}}
	\end{overpic}
    \hspace{0.02\textwidth}
    \begin{overpic}[width=0.46\textwidth]{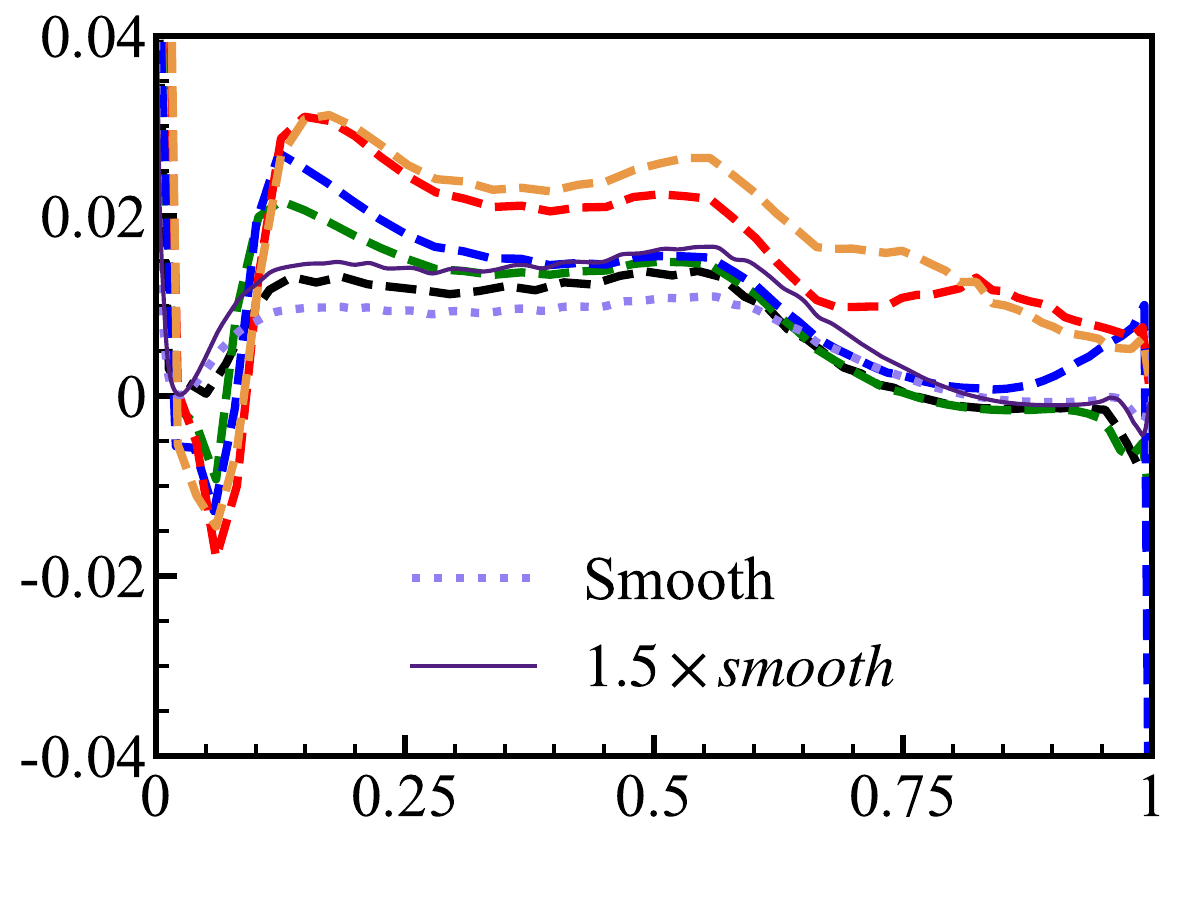}
		\put(-4,71){(\textit{d})}
        \put(50,0){\scalebox{1.0}{$x/C_{ax}$}}
        \begin{tikzpicture}[overlay, x=0.074\textwidth, y=0.074\textwidth]
            \draw[white] (0,0)--(0,1);
            \draw[white] (0,0)--(1,0);
            \draw[mycolor][line width = 1.5pt][densely dashed][-latex](2.1,2.8)--+(70:1.5);
            \node[mycolor] (A) at (2.8,4.1) {$k$};
            \draw[mycolor][line width = 1pt][->](3.1,3.98)--+(90:0.3);
            \draw[mycolor][line width = 1.5pt][dashed] (1.1,2.2) ellipse (0.35 and 0.8);
        \end{tikzpicture}
	\end{overpic}
	\caption{Pressure and drag coefficient distribution over the suction-side blade surface. (\textit{a, b}) Pressure coefficient for the $\alpha100$ cases and $k48$ cases, respectively. (\textit{c, d}) drag coefficient for the $\alpha50$ cases and $\alpha100$ cases, respectively. }
    \label{fig:Cp+Cd}
\end{figure*}

In order to quantify the roughness effect on the overall boundary layer flow, the spanwise- and time-averaged pressure coefficients $C_p$ of selected cases are shown in Figs.~\ref{fig:Cp+Cd}(\textit{a,b}). 
Note that the $C_p$ value is integrated over the surface along the streamwise interval of $\lambda_{\xi}$, i.e. over one roughness element, as suggested by Vadlamani et al.~\cite{Vadlamani2018}. 
In particular, the suction-side blade boundary layer shows a complex pressure distribution, including the leading edge (LE) region with a strong adverse pressure gradient (APG) from the stagnation point at $x/C_{ax}=0.0$ to the peak of pressure coefficient at $x/C_{ax}=0.1$, the favorable pressure gradient (FPG) regime from $x/C_{ax}=0.1$ to $x/C_{ax}=0.65$, and the APG region from $x/C_{ax}=0.65$ to the trailing edge.
Comparing the cases with different roughness heights in Fig.~\ref{fig:Cp+Cd}(\textit{a}), the roughness distribution has little influence on the mean pressure distribution in the pressure-side boundary layer. 
For the suction-side boundary layer, however, varying the roughness height causes a different pressure distribution, especially near the blade trailing edge. 
This corresponds to the changes to the trailing-edge separation observed in Fig.~\ref{fig:tke_fluc_linear_contour}. 
Furthermore, for the $k48$ cases shown in Fig.~\ref{fig:Cp+Cd}(\textit{b}), the different roughness wavenumber has a noticeable effect on trailing-edge separation and thus the corresponding pressure distribution, which again agrees well with Figs.~\ref{fig:tke_fluc_linear_contour}(\textit{g,h,i}). 

\begin{figure}[h]
	\centering
	\begin{overpic}[width=0.46\textwidth]{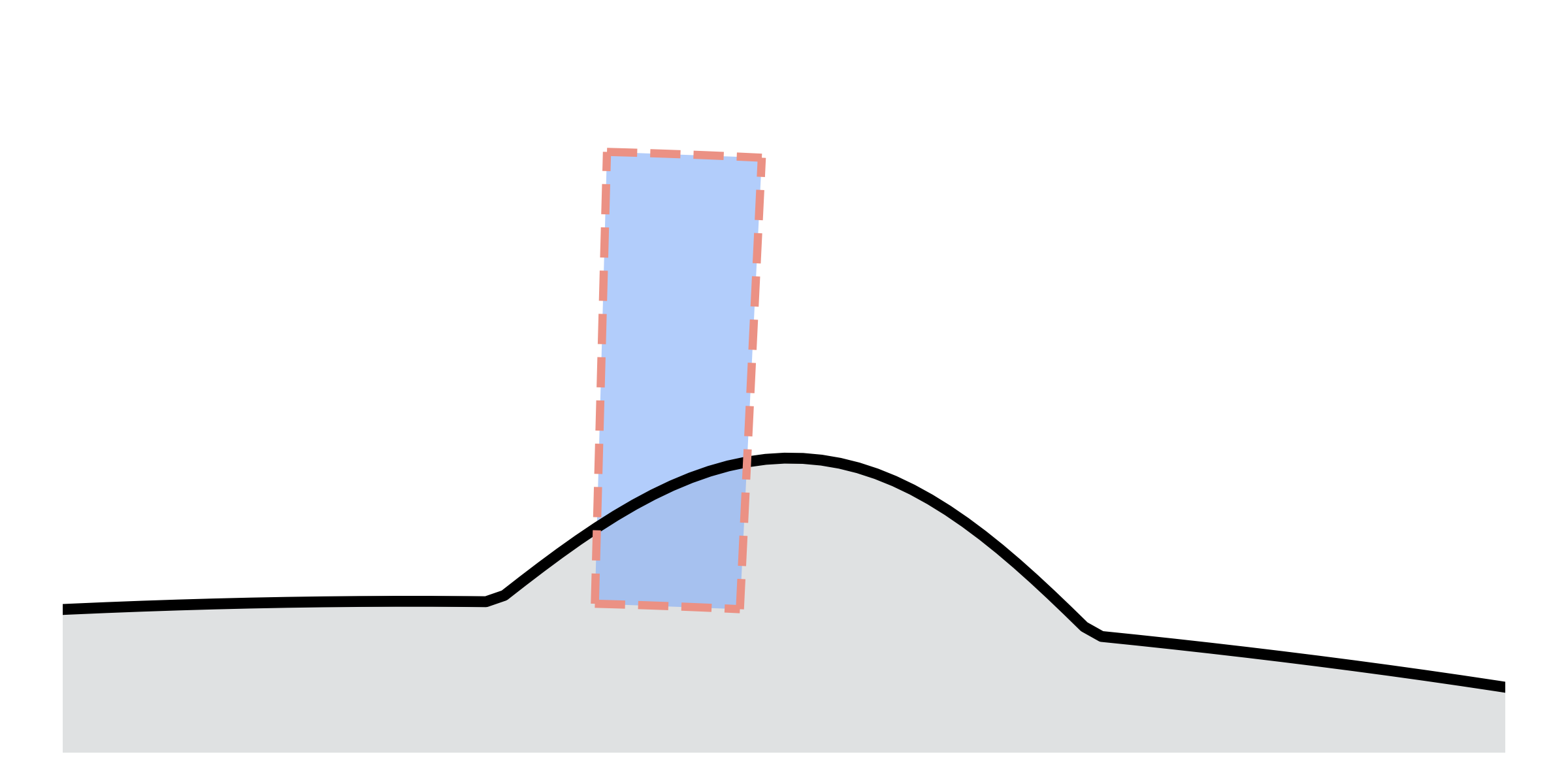}
            \begin{tikzpicture}[overlay, x=0.038\textwidth, y=0.038\textwidth]
                \draw[white] (0,0)--(0,1);
                \draw[white] (0,0)--(1,0);
                \draw [line width =1.5pt][latex-](4.5,3)--+(177:1.5);
                \node (D) at (2.8,3.1) {$v_n$};
                \draw [mycolor3][line width =1.5pt](5.9,4.9)--+(20:1);
                \node (A) at (8.5,5.3) {Control volume};
                \draw[mycolor][line width =1.5pt][dashed][->](4.7,1.1)--+(-3:1);
                \node[mycolor] (B) at (5.2,0.8) {$\bm{\xi}$};
                \draw[mycolor] [line width =1.5pt][dashed][->](4.5,2.2)--+(177:1);
                \node[mycolor] (C) at (3.2,2.2) {$\bm{n}$};
                \node (D) at (5.2,1.6) {$S_b$};
                \draw[mycolor] [line width =1.5pt][dashed][->](5.9,3)--+(-3:1);
                \node[mycolor] (E) at (7.1,2.9) {$\bm{n}$};
                \draw[mycolor] [line width =1.5pt][dashed][->](5.3,4.9)--+(87:1);
                \node[mycolor] (F) at (5.4,6.1) {$\bm{n}$};
            \end{tikzpicture}
	\end{overpic}\\[1em]
	\caption{Schematic visualization of the momentum balance method based control volumes.}
        \label{fig:Cf_sketch}
\end{figure}

The spanwise- and time-averaged drag coefficient $C_d$ is also calculated as an indicator for laminar-turbulent transition.
Note that in rough cases both the viscous and form drag need to be considered \cite{Joo2016}. 
In order to avoid computing derivatives along the rough surface, we introduce a control volume method based on the Navier-Stokes equation to compute the local drag \cite{Deyn2020} shown in Fig.~\ref{fig:Cf_sketch}, as 
\begin{equation}
    \tau_{w} = \left[ \int \rho v_n \mathbf{u} \, dS - \int \mathbf{n} \cdot (-p \bm{I} + \bm{\tau}) \, dS - \int \mathbf{n} \cdot \bm{R} \, dS \right] \cdot \bm{\xi} / S_{b}.
\end{equation}
Here $\bm{\xi}$ denotes the unit vector in the tangential direction, $\bm{n}$ denotes the unit outer normal vector, $S_b$ denotes the surface area of bottom surface of the control volume, $\bm{\tau}$ denotes the viscous stress, and $\bm{R}$ denotes the Reynolds stress. 
Based on the wall shear stress $\tau_w$, the drag coefficient is given by
\begin{equation}
    C_{d} = \frac{\tau_w}{(\rho / 2) U_{e}^{*2}}.
\end{equation}

The drag coefficients of the suction-side boundary layer in selected cases are shown in Fig.~\ref{fig:Cp+Cd} (\textit{c,d}). 
Considering the LE region ($x/C_{ax}<0.1$) first, for cases with low wavenumber $\alpha50$, the drag coefficient is generally positive for cases with different roughness heights. 
For comparison, for cases with higher wavenumbers $\alpha100$, except for the $k16$ cases with the lowest roughness elements, there are obvious negative drag coefficient regions, which correspond to the leading edge separation observed in Fig.~\ref{fig:tke_fluc_linear_contour}. 
For cases with $k>16$, the drag coefficient increases rapidly following the LE separation, significantly deviating from the smooth case, which is considered laminar.
Following the criterion by von Deyn et al.~\cite{Deyn2020}, the onset of laminar-turbulent transition can be defined as the point where the drag coefficient departs from the laminar value by a threshold of $50\%$. 
We can see that for cases with relatively high-amplitude of $k$, the $C_d$ value quickly reaches the transition onset, which aligns well with the observations in Fig.~\ref{fig:Q_vel}. 
Upon entering the FPG region ($0.1<x/C_{ax}<0.65$), the drag coefficient shows higher values for cases with larger roughness height.
Particularly, for cases with relatively low roughness amplitude ($k16$, $k32$ and $k48$), the drag coefficient presents a tendency for relaminarization, which is presumably due to the effect of the strong FPG. 
Finally, focusing on the APG region ($x/C_{ax}>0.65$), $C_d$ can be used to determine whether there is trailing-edge separation. 
Moreover, comparing to cases $k64$ and $k80$ in which the drag stays at a relatively high level, the $k48\alpha100$ cases, showing intermittent vortical structures in Fig.~\ref{fig:Q_vel} present a sudden increase of $C_d$ in the APG region.
This is inferred to be related to roughness-induced boundary layer transition. 

Based on the discussions on the overall flow above, we can see that the blade suction-side boundary layers with different surface roughness show extremely complex phenomena, including transition induced by LE separation, relaminarization in the FPG region, transition in the APG region, and also TE separation. 
Distinct from canonical flows, the complex flow phenomena, which obviously require further investigation, are affected by the surface curvature of the blade and also the pressure distribution across the vane, which is typical for turbomachinery flows.

\section{Roughness effects on suction side boundary layer}\label{sec:Results}

In the present section, we perform a detailed investigation on the flow mechanisms for the roughness effects on the suction-side boundary layer. 
Compared to the smooth case featuring the trailing-edge separation, the rough cases show complex behaviors induced by the surface roughness. 
Specifically, as discussed in Section~\ref{sec:Overview}, we divide the suction-side boundary layer into three regions: the LE region, the FPG region, and the APG region, aiming to shed light on how the varying surface roughness affects the flow behaviors in these different regions of the suction-side boundary layer. 

\subsection{Leading edge structures}\label{sec:vortex}
\begin{figure}[h!]
	\centering
	\begin{overpic}[width=0.24\textwidth]{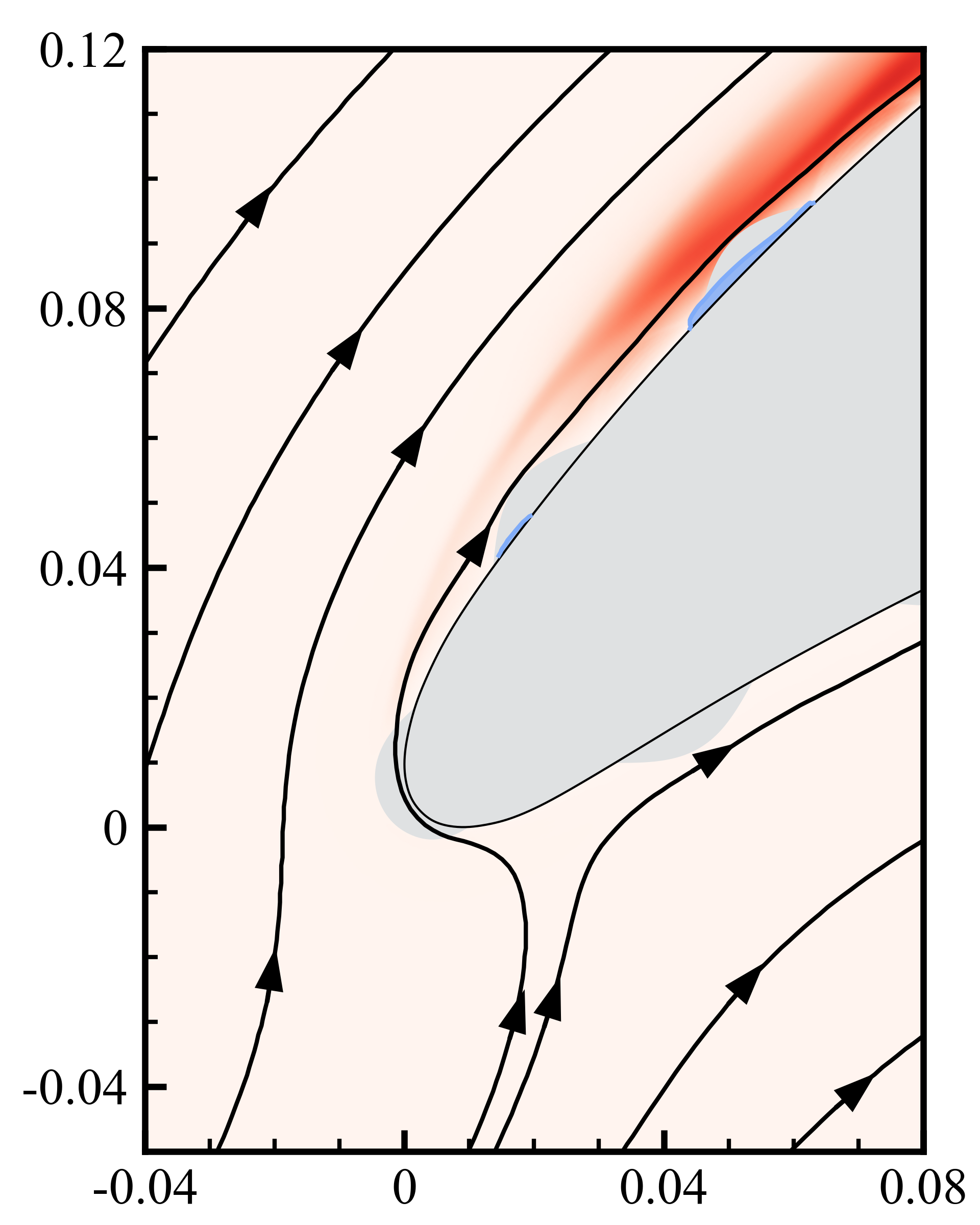}
		\put(-2,98){(\textit{a})}
            \put(44,-1){\scalebox{1.0}{$x$}}
            \put(0,50){\scalebox{1.0}{$y$}}
	\end{overpic}
	\begin{overpic}[width=0.24\textwidth]{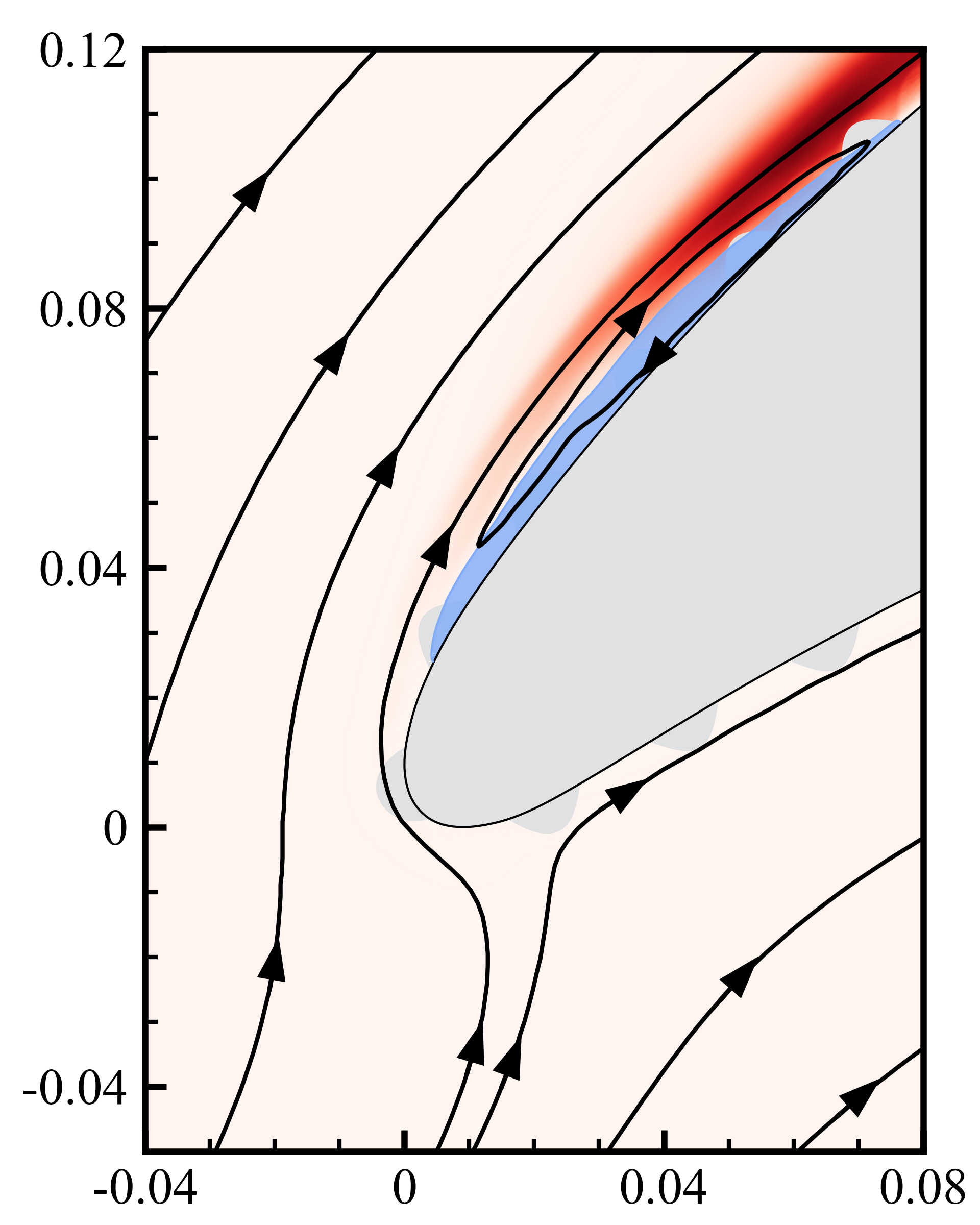}
		\put(-2,98){(\textit{b})}
            \put(44,-1){\scalebox{1.0}{$x$}}
	\end{overpic}\\[1em]
        \begin{overpic}[width=0.3\textwidth]{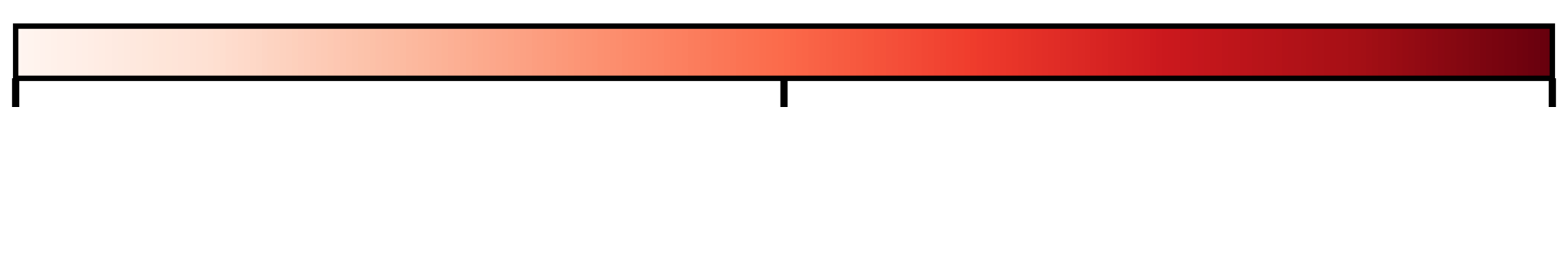}
            \put(0,4){\scalebox{1.0}{$0$}}
            \put(45,4){\scalebox{1.0}{$0.05$}}
            \put(95,4){\scalebox{1.0}{$0.1$}}
            \put(45,-2){\scalebox{1.0}{$TKE$}}
	\end{overpic}
	\caption{A zoom-in view of the time and spanwise averaged mean flow field at LE: (\textit{a}) case $k48\alpha50$; (\textit{b}) case $k48\alpha100$. Contours of the TKE are shown, with the gray-shaded area denoting the blade slice at $z=0$. The black lines with arrows indicate streamlines obtained from the mean flow, and the blue bubbles represent the reverse flow regions.}
	\label{fig:LE_streamline}
\end{figure}

In order to further analyze the leading-edge flow behaviors, taking case $k48\alpha50$ and case $k48\alpha100$ as examples, the time and spanwise averaged mean flow field are shown in Fig.~\ref{fig:LE_streamline}. 
For the $\alpha100$ case with higher effective streamwise slope, the flow exhibits a massive separation bubble that spans multiple roughness elements. 
This large-scale separation is accompanied by significant TKE enhancement, driven by the combined effects of the APG and surface roughness. 
In contrast, the reverse flow region in the $\alpha50$ case is confined to localized recirculations attached to individual roughness elements.
This observation about the effect of the roughness wavenumber on the LE separation is clearly consistent with Figs.~\ref{fig:tke_fluc_linear_contour} and \ref{fig:Cp+Cd}. 

\begin{figure}[h]
	\centering
	\begin{overpic}[width=0.46\textwidth]{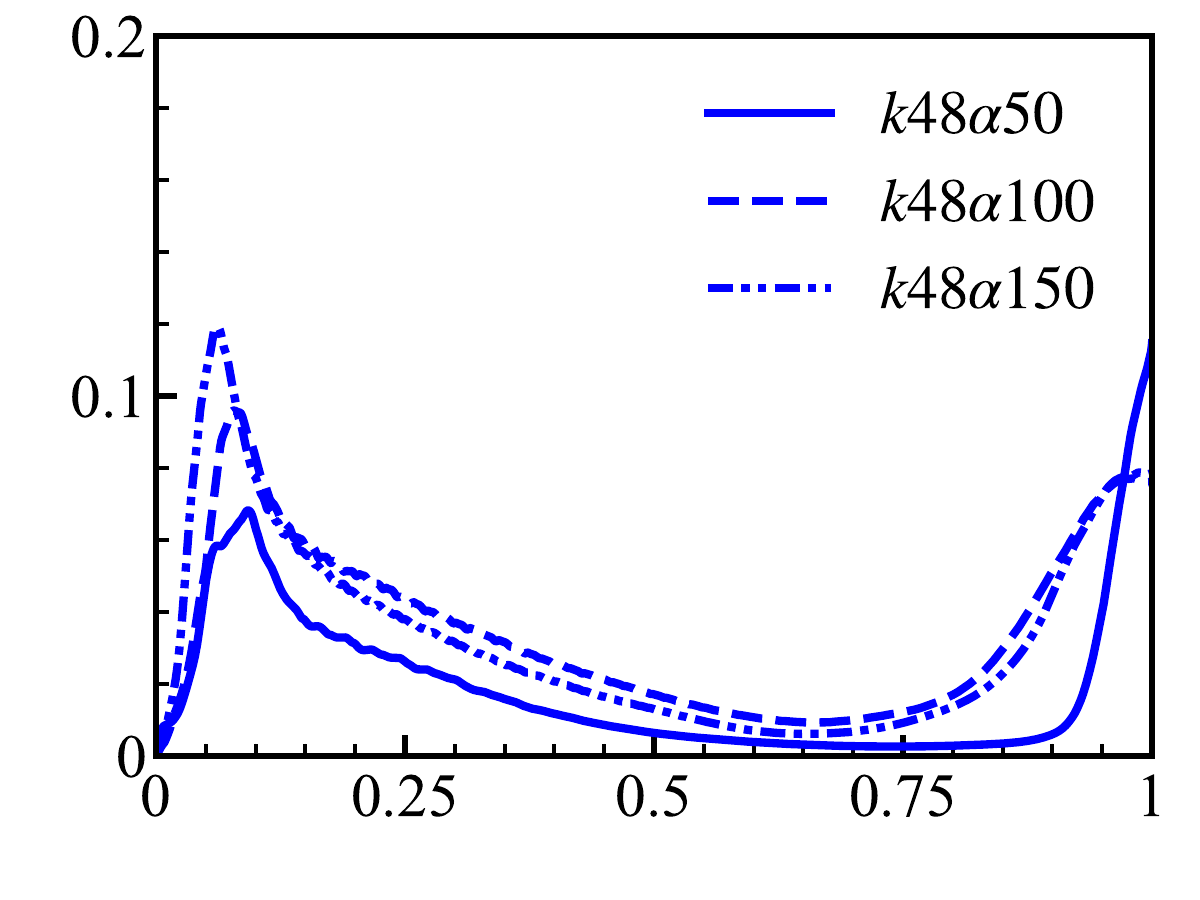}
        \put(-2,40){\scalebox{1}{\rotatebox[origin=c]{90}{$\mathrm{max}_{\eta}(tke)$}}}
		\put(50,3){\scalebox{1.0}{$x/C_{ax}$}}
	\end{overpic}
	\caption{The normal maximum of TKE on blade suction side compared with different roughness streamwise wavenumbers for $k48$.}
	\label{fig:tke_develop}
\end{figure}

The development of the fluctuations in the suction side boundary layer is further characterized by the wall-normal maximum of the turbulent kinetic energy as shown in Fig.~\ref{fig:tke_develop}. 
Regarding the effect of height, observations from Fig.~\ref{fig:tke_fluc_linear_contour} indicate that while increasing $k$ generally amplifies the fluctuation peak at the leading edge, this impact exhibits a saturation limit. 
Specifically, the peak magnitudes for cases $k48$, $k64$, and $k80$ remain nearly identical. 
This saturation suggests that, given the extremely thin boundary layer in the leading-edge region, roughness elements with $k > 48$ likely disturb the entire boundary layer, rendering further height increases less consequential. 
In terms of wavenumber, Fig.~\ref{fig:tke_develop} demonstrates that increasing the wavenumber $\alpha$ leads to an increase in the TKE peak. 
This enhancement is presumably attributed to the massive flow separation as visualized in Figs.~\ref{fig:tke_fluc_linear_contour} and \ref{fig:LE_streamline}, which significantly intensify the turbulent fluctuations. 

\subsection{FPG region}\label{sec:FPG}
\begin{figure*}[t!]
	\centering
	\begin{overpic}[width=0.8\textwidth]{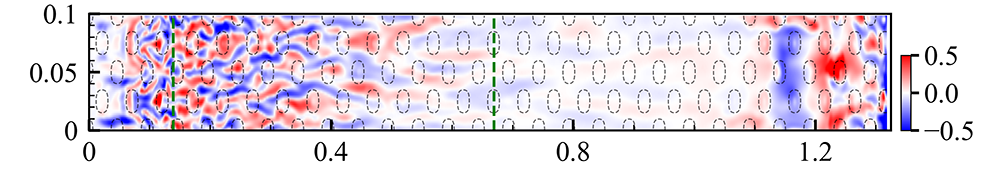}
		\put(-1,16){(\textit{a})}
		\put(0,9){\scalebox{1.0}{$z$}}
		\put(99,7){\scalebox{1.0}{$u_{\xi}^{\prime}$}}
	\end{overpic}\\
	\begin{overpic}[width=0.8\textwidth]{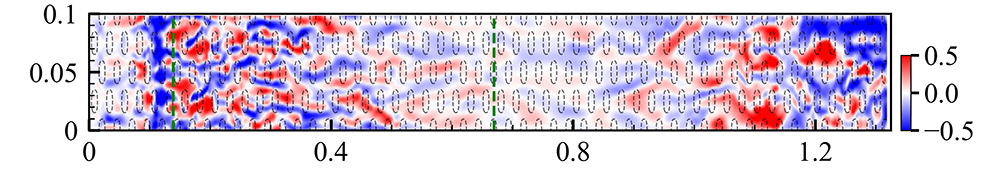}
		\put(-1,16){(\textit{b})}
		\put(0,9){\scalebox{1.0}{$z$}}
		\put(99,7){\scalebox{1.0}{$u_{\xi}^{\prime}$}}
	\end{overpic}\\
    \begin{overpic}[width=0.8\textwidth]{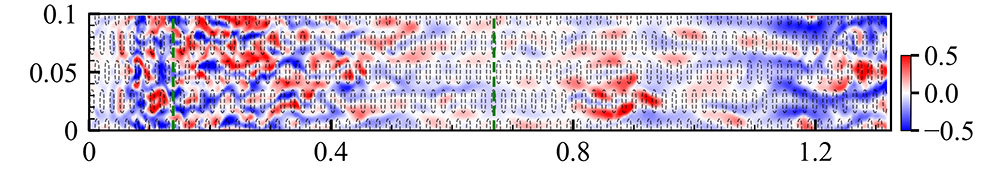}
        \put(-1,16){(\textit{c})}
        \put(50,-2){\scalebox{1.0}{$\xi$}}
        \put(0,9){\scalebox{1.0}{$z$}}
        \put(99,7){\scalebox{1.0}{$u_{\xi}^{\prime}$}}
	\end{overpic}\\
    \begin{overpic}[width=0.8\textwidth]{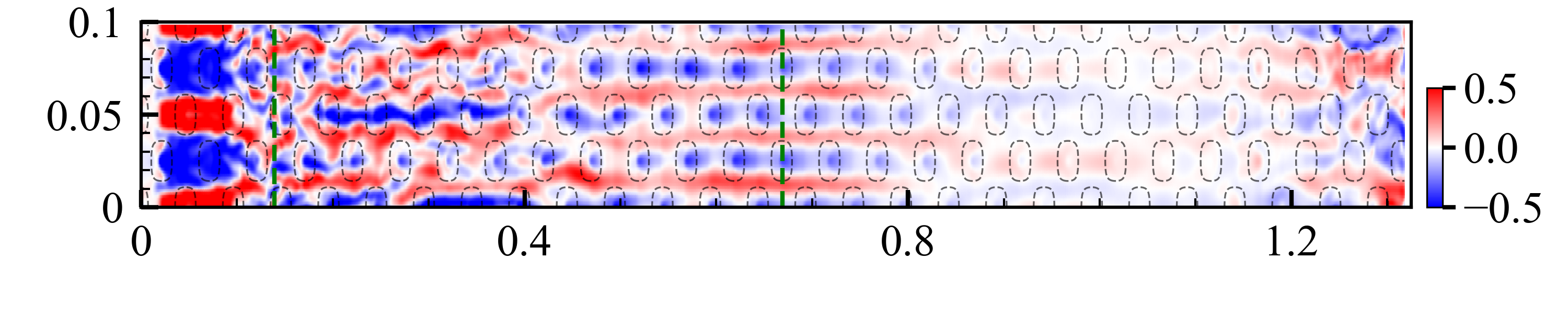}
		\put(-1,19){(\textit{d})}
		\put(0,12){\scalebox{1.0}{$z$}}
		\put(99,10){\scalebox{1.0}{$u_{dis}$}}
	\end{overpic}\\
	\begin{overpic}[width=0.8\textwidth]{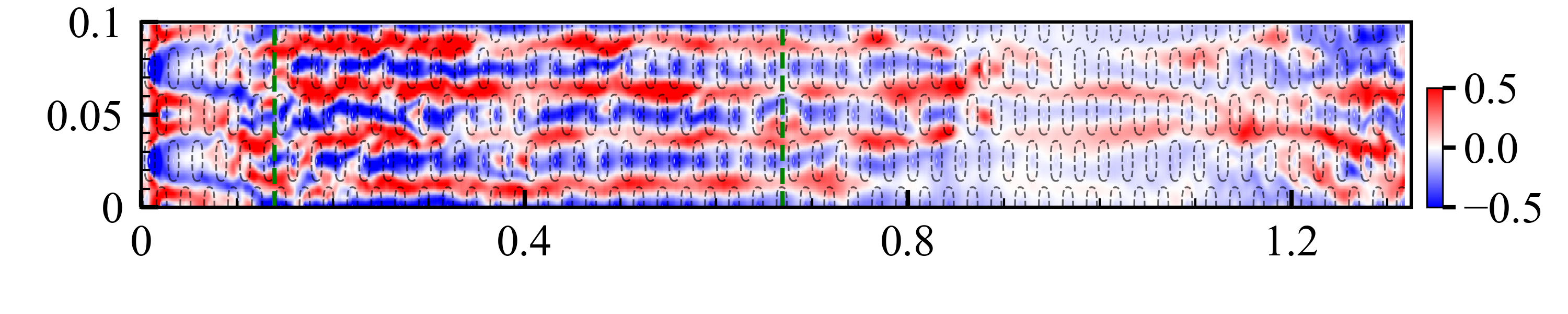}
		\put(-1,19){(\textit{e})}
		\put(0,12){\scalebox{1.0}{$z$}}
		\put(99,10){\scalebox{1.0}{$u_{dis}$}}
	\end{overpic}\\
    \begin{overpic}[width=0.8\textwidth]{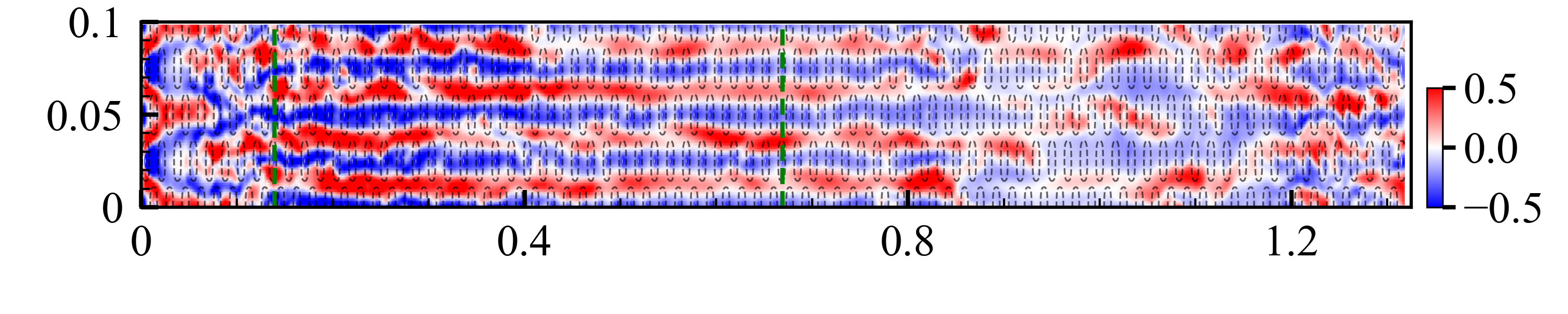}
		\put(-1,19){(\textit{f})}
        \put(50,0){\scalebox{1.0}{$\xi$}}
		\put(0,12){\scalebox{1.0}{$z$}}
		\put(99,10){\scalebox{1.0}{$u_{dis}$}}
	\end{overpic}\\
	\caption{The instantaneous tangential velocity fluctuations contours of $\xi - z$ plane cut at $\eta/k \approx 1$: (\textit{a}) $k48\alpha50$, turbulent; (\textit{b}) $k48\alpha100$, turbulent; (\textit{c}) $k48\alpha150$, turbulent; (\textit{d}) $k48\alpha50$, dispersive; (\textit{e}) $k48\alpha100$, dispersive; (\textit{f}) $k48\alpha150$, dispersive. The green dashed lines mark the positions of $x/C_{ax}=0.1$ and $x/C_{ax}=0.65$, respectively. The black dashed lines mark the location of the roughness element.
    These $\xi - z$ slices are located near the peak of the roughness element, however above the peak. }
	\label{fig:u_fluc+u_disp}
\end{figure*}

The blade suction-side boundary layer in the FPG region is affected by several factors, including the disturbances coming from the upstream LE region, the surface roughness in the local boundary layer, and the stabilizing effect of the FPG. 
One observation we can draw from flow visualizations in Figs.~\ref{fig:Q_vel} and \ref{fig:tke_fluc_linear_contour} and TKE plots in Fig.~\ref{fig:tke_develop} is that in most cases, the velocity fluctuations are suppressed by the stabilizing effects of the FPG, while in cases with high levels of roughness heights (the $k64$ and $k80$ cases), the FPG boundary layer remains highly disturbed.
The stabilizing effect of the FPG can be further shown by the contours of the instantaneous tangential turbulent fluctuating velocity in Fig.~\ref{fig:u_fluc+u_disp}(\textit{a, b, c}).
It can be seen that the turbulent velocity fluctuations are significantly reduced in the FPG region. 
Moreover, the cases $k48\alpha100$ and $k48\alpha150$ have stronger fluctuating velocity than case $\alpha50$. 

Compared to the turbulent fluctuations which show relatively chaotic behaviors in Fig.~\ref{fig:u_fluc+u_disp}(\textit{a, b, c}), the contours of the instantaneous  dispersive tangential velocity $u_{dis}$ shown in Fig.~\ref{fig:u_fluc+u_disp}(\textit{d, e, f}) are organized as streak-like patterns in the FPG region. 
The instantaneous dispersive velocity \cite{Deyn2020} $u_{dis}$ is defined as 
\begin{equation}
    u_{\mathrm{dis}}=u_{\xi}(x, y, z, t)-\bar{u}_{\xi}(x, y, t),
\end{equation}
where $\bar{u}_{\xi}(x, y, t)$ is obtained by taking spanwise averaging of the instantaneous velocity $u_{\xi}(x, y, z, t)$. 
Specifically, high-speed streaks are observed in the spanwise gap between the roughness elements, while low-speed streaks exist due to the blockage effects of the roughness elements.
Note that these streaks can be observed in all cases with different $\alpha$, even though the LE region appears to show different patterns, which presumably depends on whether there exists LE separation as discussed in Fig.~\ref{fig:LE_streamline}. 
Moreover, due to the larger slope and smaller spacing, the strength of streaks in the high wavenumber cases as shown in Figs.~\ref{fig:u_fluc+u_disp}(\textit{e,f}) is visibly stronger than in the low wavenumber case shown in Fig.~\ref{fig:u_fluc+u_disp}(\textit{d}). 
\begin{figure*}[t!]
	\centering
	\begin{overpic}[width=0.305\textwidth]{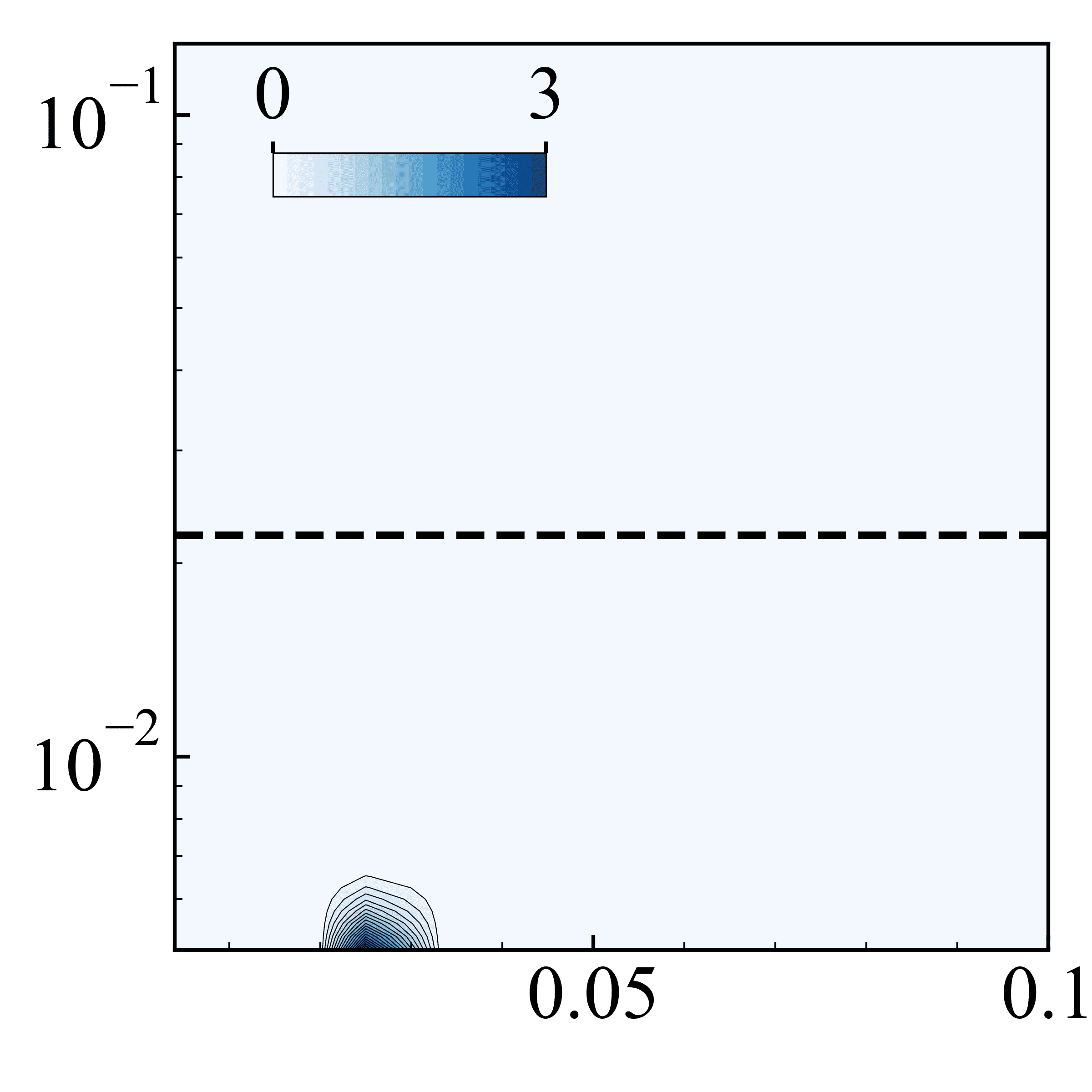}
		\put(-8,92){(\textit{a})}
		\put(-4,55){\scalebox{1.0}{$\eta$}}
		\put(55,82){\scalebox{1.0}{$k_z E_{\widetilde{u_{\xi}} \widetilde{u_{\xi}}}$}}
	\end{overpic}
	\hspace{0.02\textwidth}
	\begin{overpic}[width=0.305\textwidth]{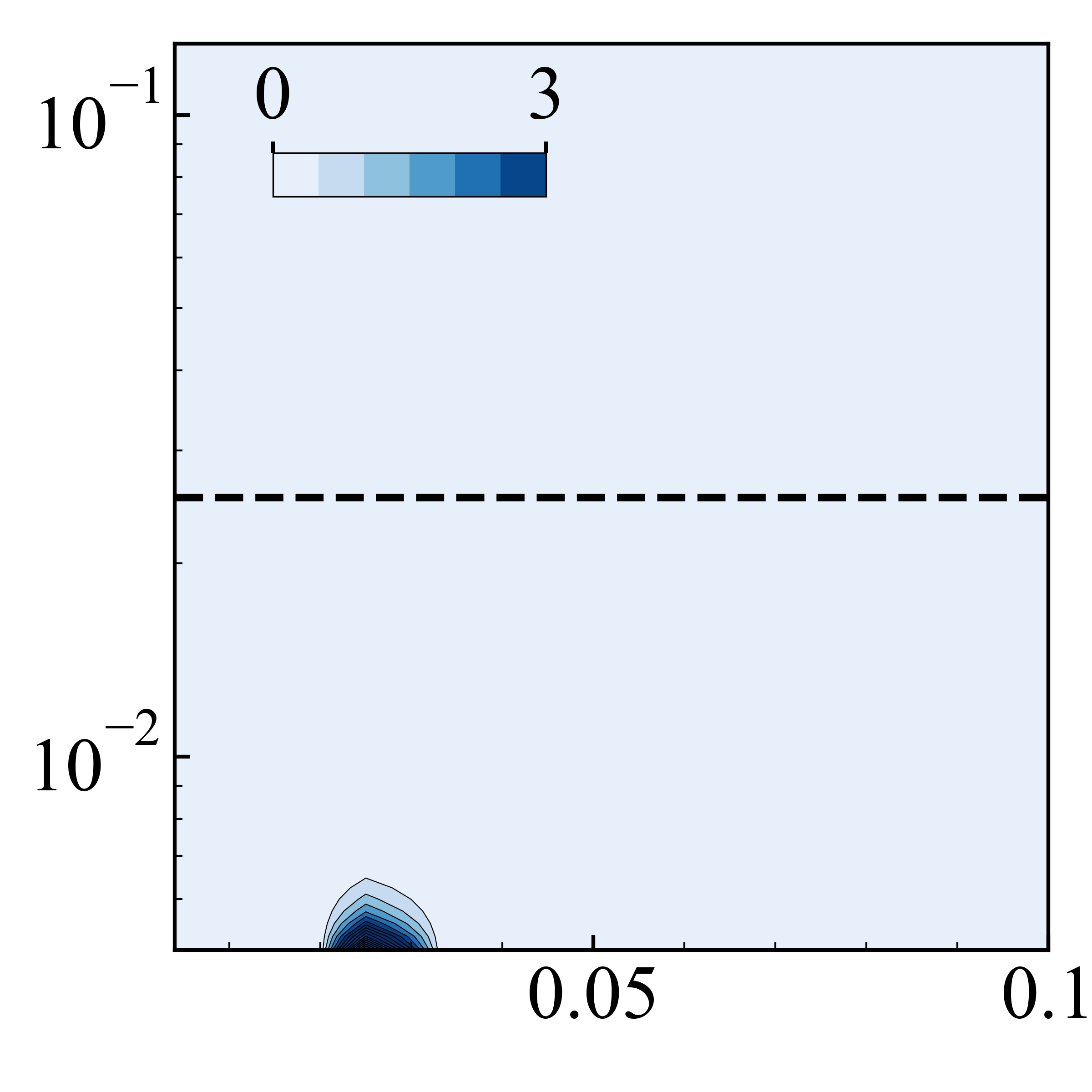}
		\put(-8,92){(\textit{b})}
		\put(55,82){\scalebox{1.0}{$k_z E_{\widetilde{u_{\xi}} \widetilde{u_{\xi}}}$}}
	\end{overpic}
	\hspace{0.02\textwidth}
	\begin{overpic}[width=0.305\textwidth]{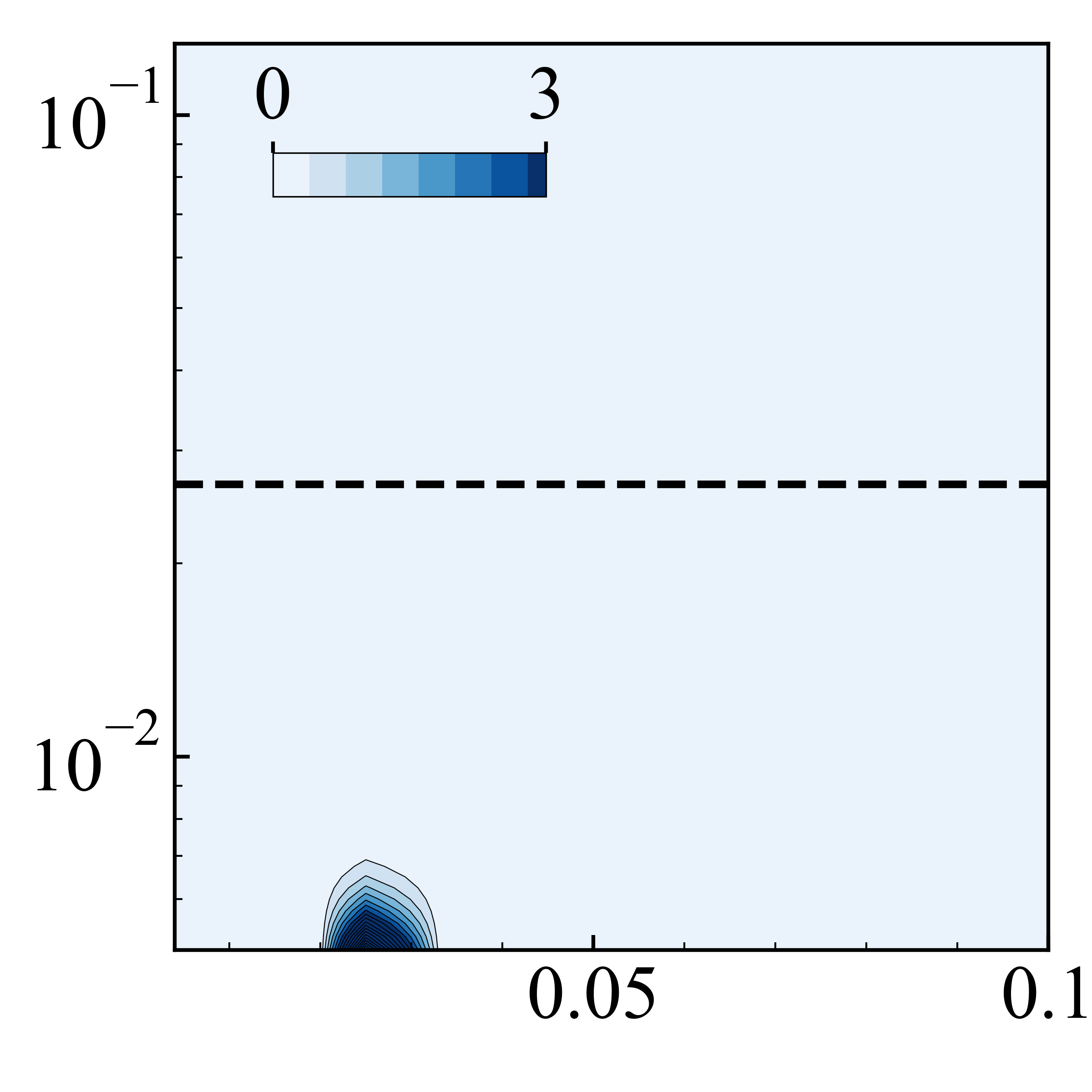}
		\put(-8,92){(\textit{c})}
		\put(55,82){\scalebox{1.0}{$k_z E_{\widetilde{u_{\xi}} \widetilde{u_{\xi}}}$}}
	\end{overpic}\\
	\begin{overpic}[width=0.305\textwidth]{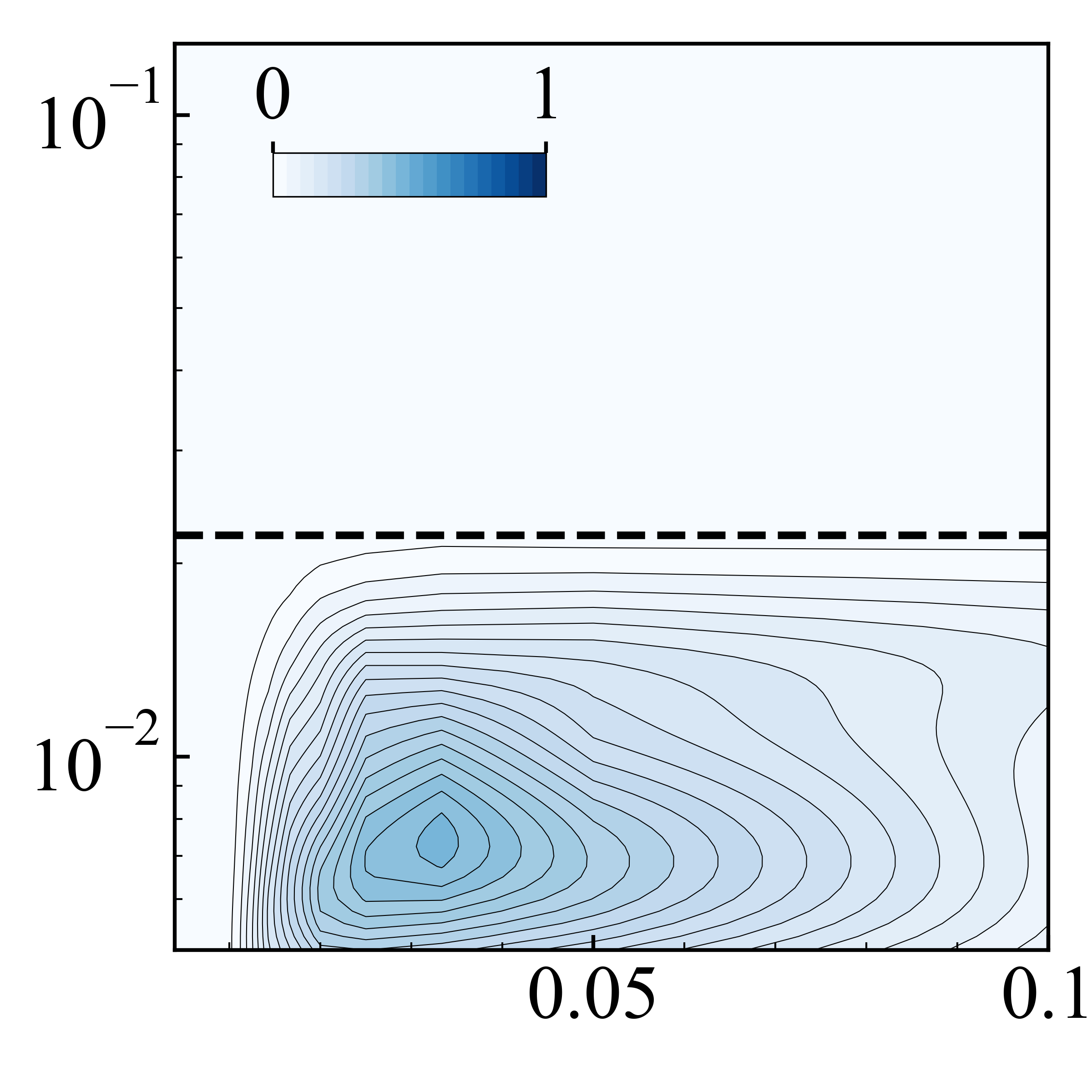}
		\put(-8,92){(\textit{d})}
		\put(-4,55){\scalebox{1.0}{$\eta$}}
		\put(52,-2){\scalebox{1.0}{$\lambda_z$}}
		\put(55,82){\scalebox{1.0}{$k_z E_{u_{\xi}^{\prime} u_{\xi}^{\prime}}$}}
	\end{overpic}
	\hspace{0.02\textwidth}
	\begin{overpic}[width=0.305\textwidth]{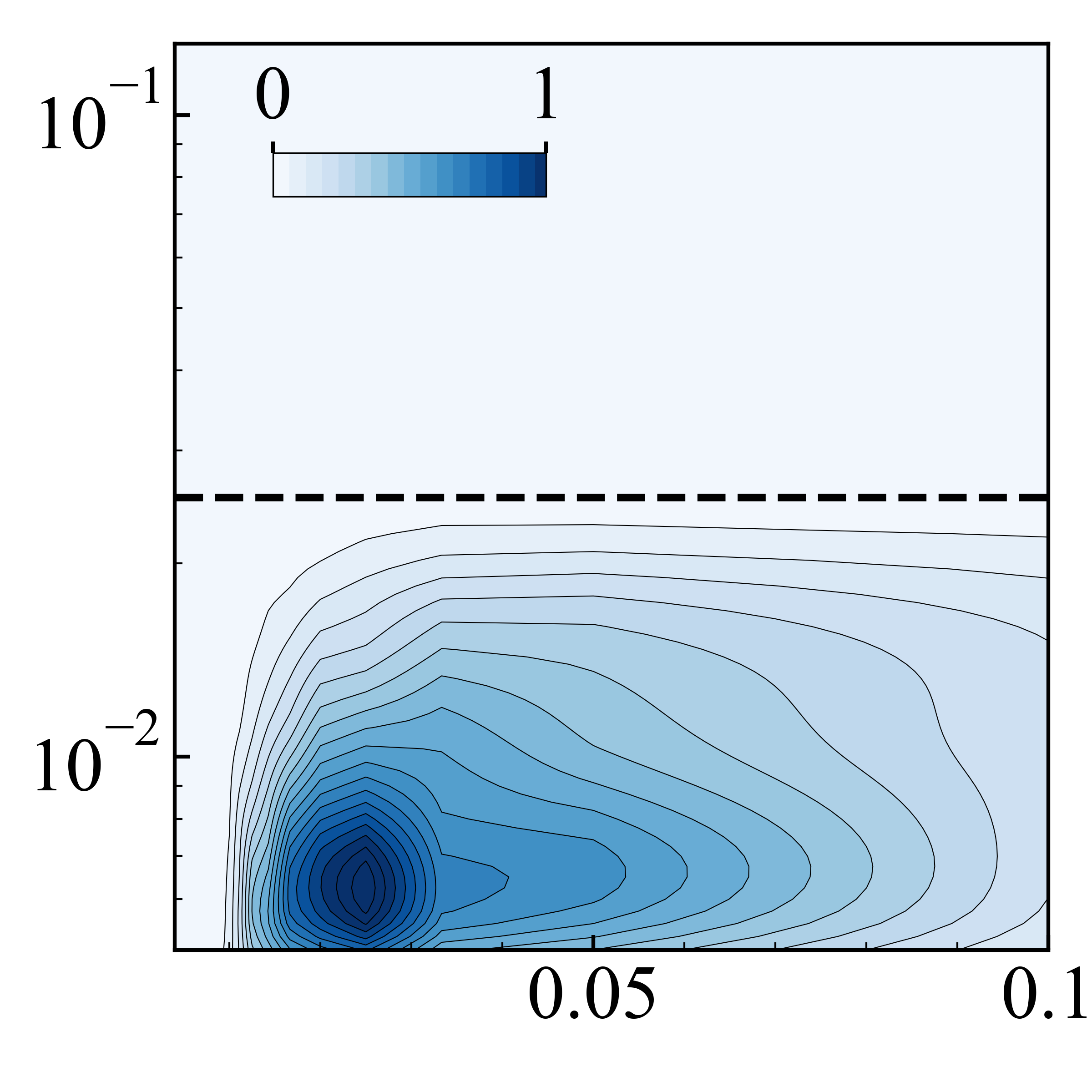}
		\put(-8,92){(\textit{e})}
		\put(52,-2){\scalebox{1.0}{$\lambda_z$}}
		\put(55,82){\scalebox{1.0}{$k_z E_{u_{\xi}^{\prime} u_{\xi}^{\prime}}$}}
	\end{overpic}
	\hspace{0.02\textwidth}
	\begin{overpic}[width=0.305\textwidth]{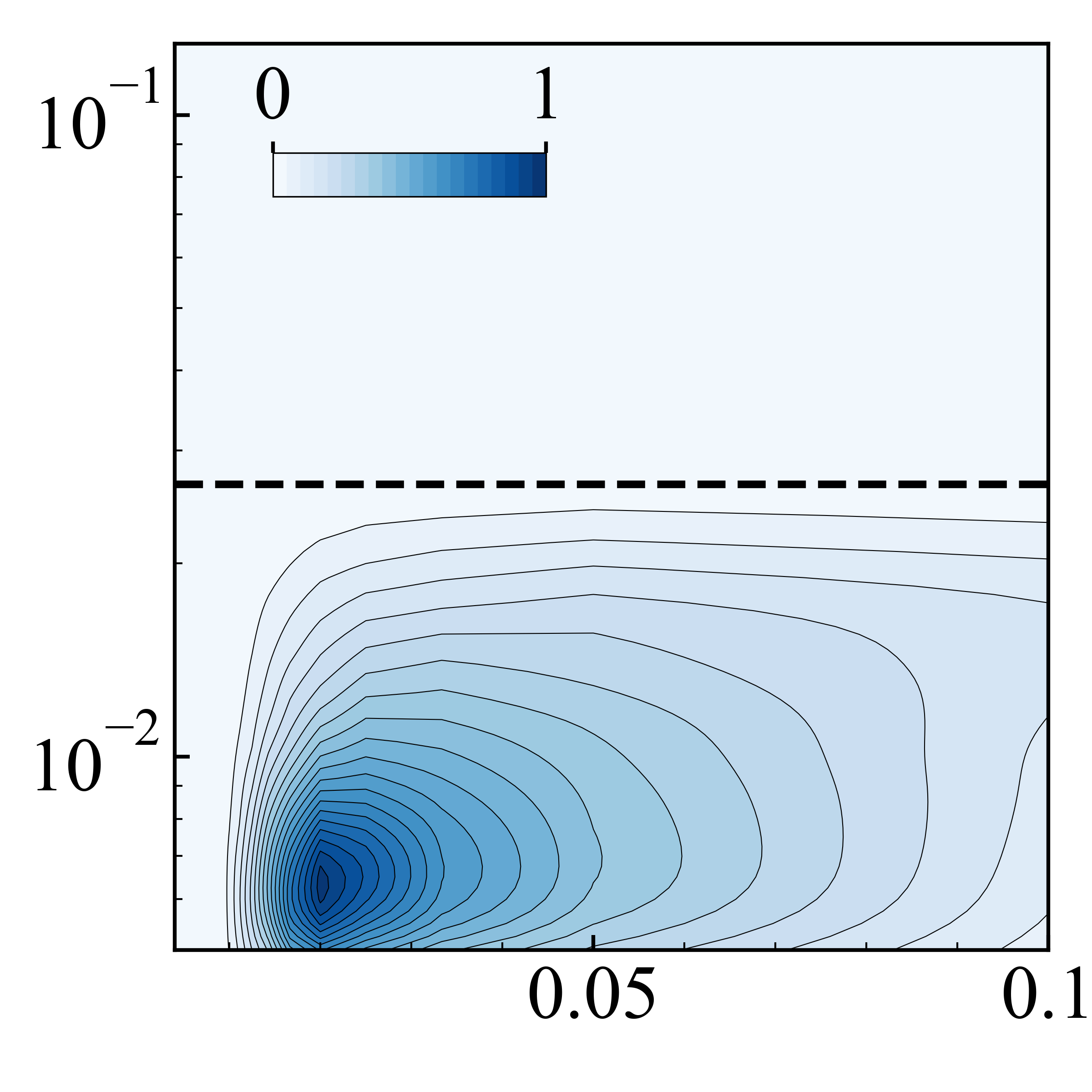}
		\put(-8,92){(\textit{f})}
		\put(52,-2){\scalebox{1.0}{$\lambda_z$}}
		\put(55,82){\scalebox{1.0}{$k_z E_{u_{\xi}^{\prime} u_{\xi}^{\prime}}$}}
	\end{overpic}\\[1em]
    \begin{overpic}[width=0.305\textwidth]{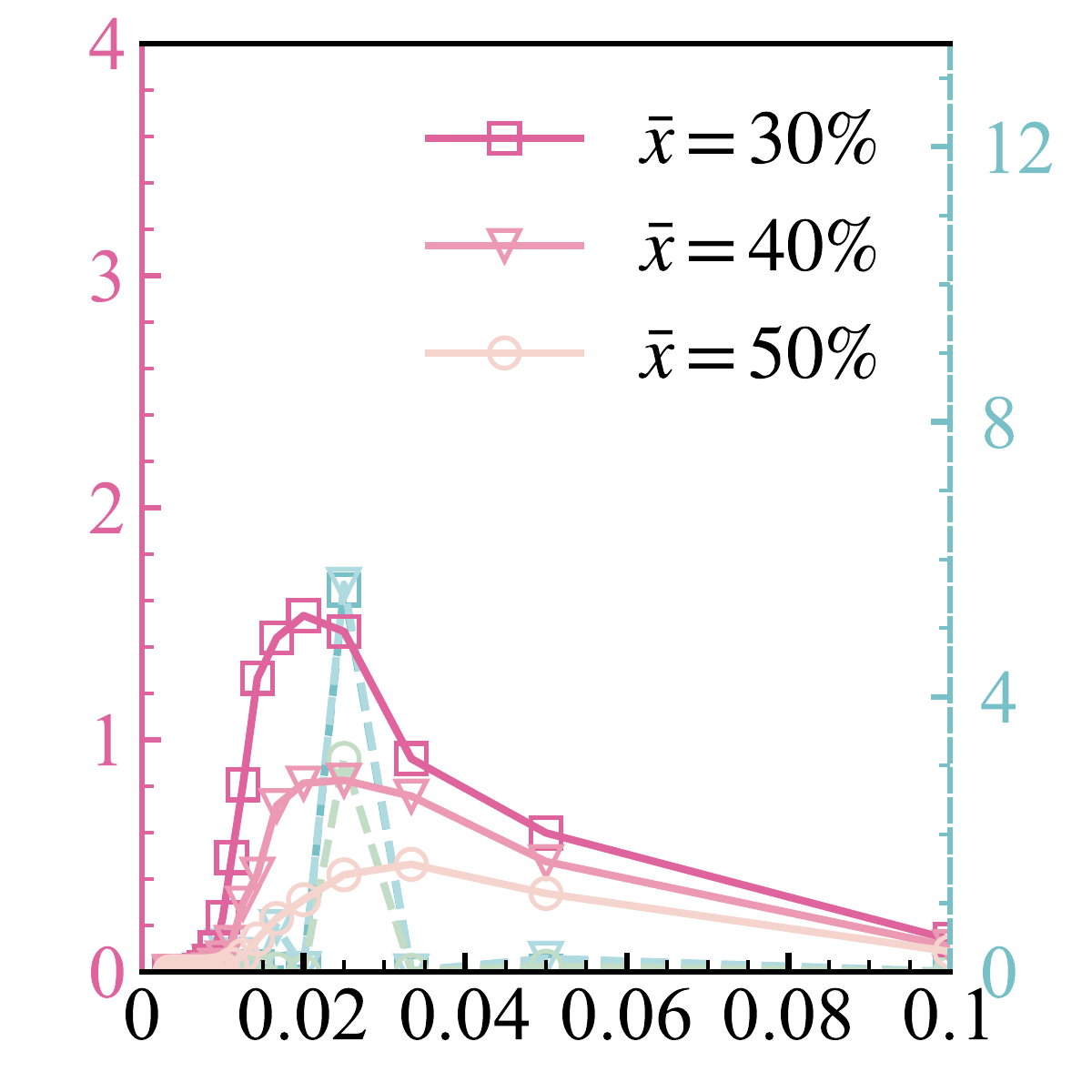}
		\put(-8,92){(\textit{g})}
        \put(-4,54){\scalebox{1}{\rotatebox[origin=c]{90}{$k_z E_{u_{\xi}^{\prime} u_{\xi}^{\prime}}$}}}
        \put(52,-2){\scalebox{1.0}{$\lambda_z$}}
	\end{overpic}
	\hspace{0.02\textwidth}
	\begin{overpic}[width=0.305\textwidth]{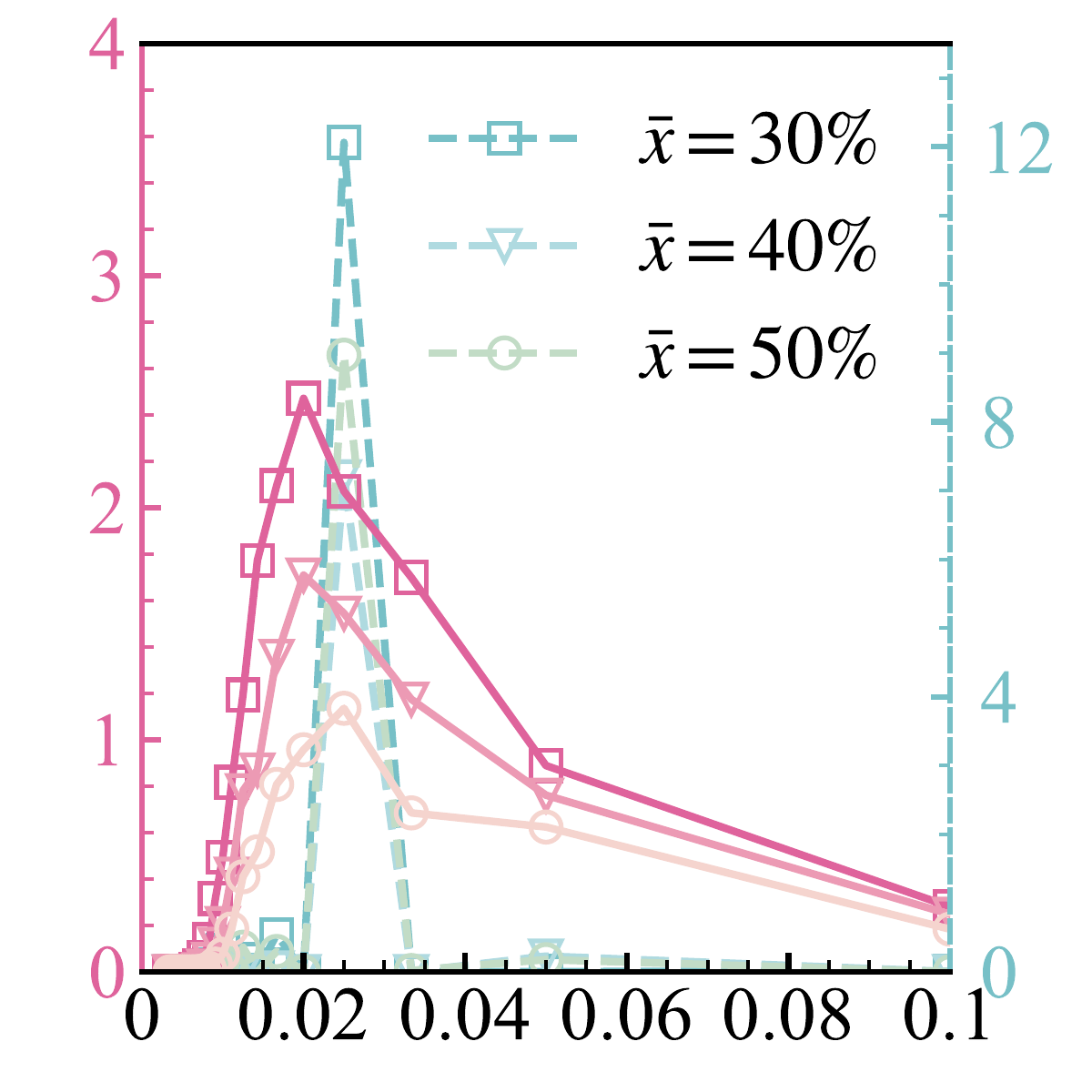}
		\put(-8,92){(\textit{h})}
		\put(52,-2){\scalebox{1.0}{$\lambda_z$}}
	\end{overpic}
	\hspace{0.02\textwidth}
	\begin{overpic}[width=0.305\textwidth]{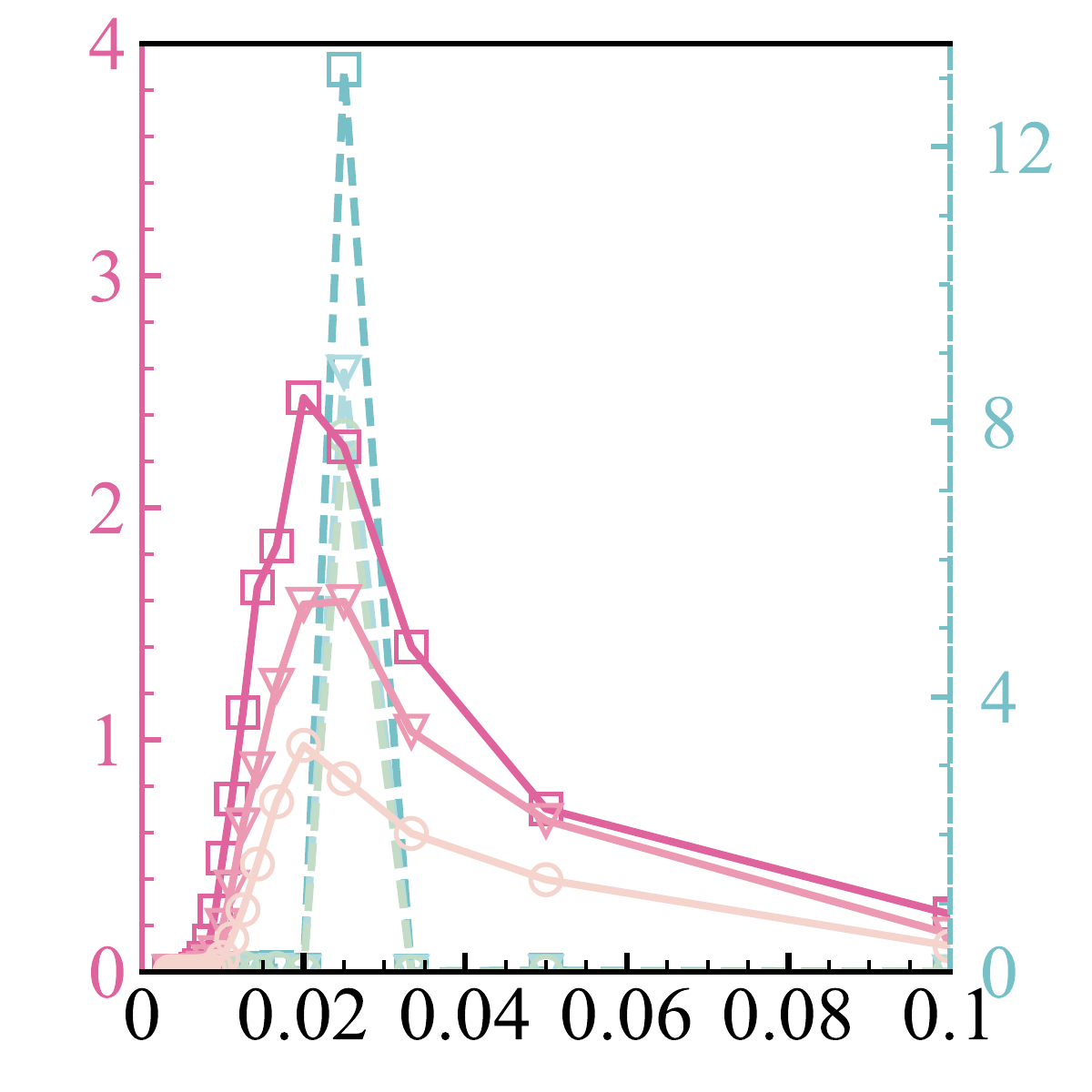}
		\put(-8,92){(\textit{i})}
        \put(96,54){\scalebox{1}{\rotatebox[origin=c]{90}{$k_z E_{\widetilde{u_{\xi}} \widetilde{u_{\xi}}}$}}}
		\put(52,-2){\scalebox{1.0}{$\lambda_z$}}
	\end{overpic}\\[1em]
	\caption{Contours of the one-dimensional pre-multiplied spanwise energy spectra of tangential velocity dispersive fluctuation: (\textit{a,b,c}) and turbulent fluctuation: (\textit{d,e,f}) at $x/C_{ax}=0.5$. (\textit{a,d}) $k48\alpha50$; (\textit{b,e}) $k48\alpha100$; (\textit{c,f}) $k48\alpha150$. Peak of one-dimensional pre-multiplied spanwise energy spectra of streamwise velocity fluctuation: (\textit{g}) $k48\alpha50$; (\textit{h}) $k48\alpha100$; (\textit{i}) $k48\alpha150$. Solid lines mean turbulent fluctuation, and dashed lines mean dispersive fluctuation.}
	\label{fig:spectra}
\end{figure*}

The disturbances in the FPG boundary layer can be further investigated based on the spectra of the dispersive and turbulent velocity. 
Specifically, the dispersive and turbulent tangential velocities are collected on the spanwise and wall-normal plane-cut at $x/C_{ax}=0.5$ for the $k48$ cases, and then their spectra in the spanwise wavenumber space are calculated and presented in Figs.~\ref{fig:spectra}(\textit{a,b,c,d,e,f}). 
Moreover, the pre-multiplied energy spectrum is also plotted at the wall-normal positions corresponding to the contour peak for different streamwise positions in Figs.~\ref{fig:spectra}(\textit{g,h,i}). 
For the dispersive velocity, the energy is mainly concentrated on the spanwise wavelengths $\lambda_{z}=0.025$. 
Note that the dispersive energy of cases $\alpha100$ and $\alpha150$ is significantly stronger than that of case $\alpha50$, which agrees with our observation about the streaks in Fig.~\ref{fig:u_fluc+u_disp}. 
Moreover, dispersive energy distributes mainly close to the wall, suggesting that the roughness elements are responsible for transferring kinetic energy from the mean flow to the dispersive parts. 
On the other hand, the turbulent kinetic energy clearly is distributed over a wider range of spanwise wavelengths and also wall-normal locations. 
This agrees with the observations from Fig.~\ref{fig:u_fluc+u_disp}, in which the tangential turbulent velocity shows a more chaotic behavior for all three cases. 
Nonetheless, the wavelengths corresponding to the peaks of the pre-multiplied spectra are also close to $\lambda_{z}=0.025$ as shown in Figs.~\ref{fig:spectra}(\textit{g,h,i}), suggesting modulation of the turbulent fluctuations by the dispersive velocity. 

\subsection{Transition in APG region}\label{sec:APG}
Based on observations from the flow visualizations in Figs.~\ref{fig:Q_vel} and \ref{fig:tke_fluc_linear_contour}, the suction-side boundary layer in the APG region eventually develops to a turbulent state in all cases.
Specifically, the boundary layer in cases with small amplitude roughness elements ($k<48$) stays laminar until the separation-induced transition near the blade trailing edge, while the APG boundary layer in cases with high roughness elements ($k>48$) seems to be packed with turbulent structures and thus suppresses the trailing edge separation.
For cases with intermediate roughness height ($k=48$), however, the streamwise wavenumber of the roughness is shown to have significant effects on the transitional behaviors in the APG boundary layer, which has been discussed in Figs.~\ref{fig:Q_vel}(\textit{g,h,i}).
Therefore, in this section, we focus on the $k48$ cases which show the most interesting transitional behaviors. 

\begin{figure*}[htbp!]
    \centering
    \begin{adjustbox}{valign=c}
		\begin{overpic}[width=0.7\textwidth]{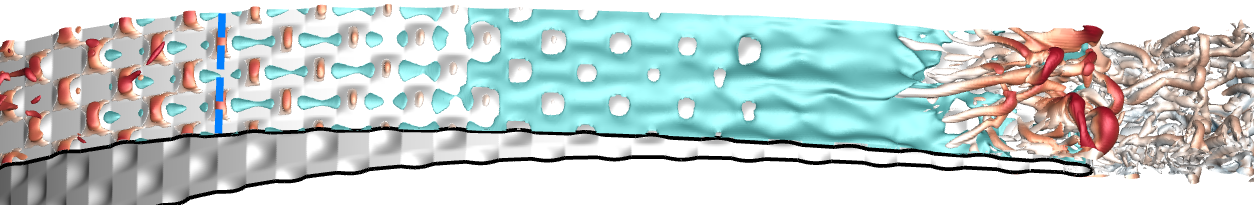}
        \put(-1,16){(\textit{a})}
		\end{overpic}
	\end{adjustbox}
	\hspace{0.03\textwidth}
	\begin{adjustbox}{valign=c}
		\begin{overpic}[width=0.24\textwidth]{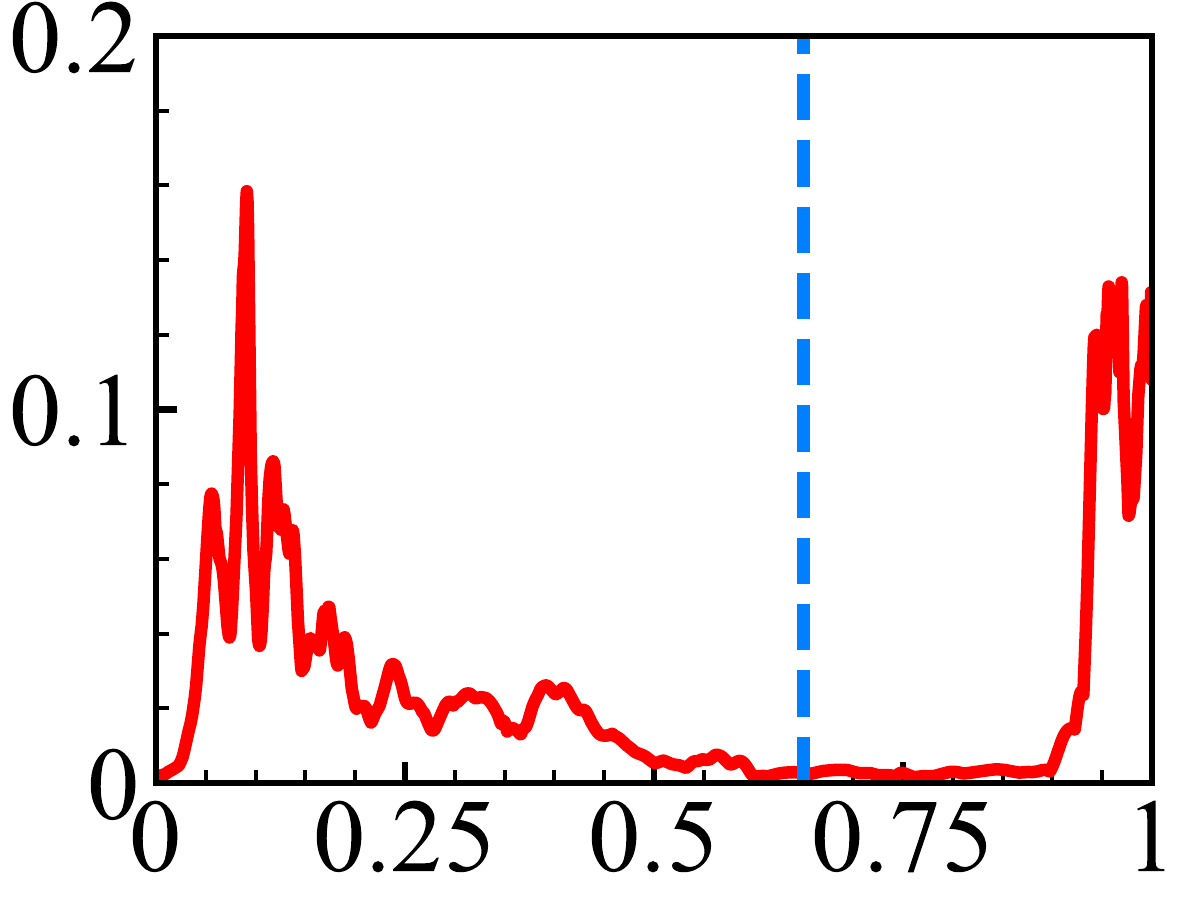}
			\put(45,-6){\scalebox{1.0}{$x/C_{ax}$}}
			\put(-10,40){\scalebox{1}{\rotatebox[origin=c]{90}{$\mathrm{max}_{\eta}(TKE)$}}}
		\end{overpic}
	\end{adjustbox}\\
    \begin{adjustbox}{valign=c}
        \begin{overpic}[width=0.7\textwidth]{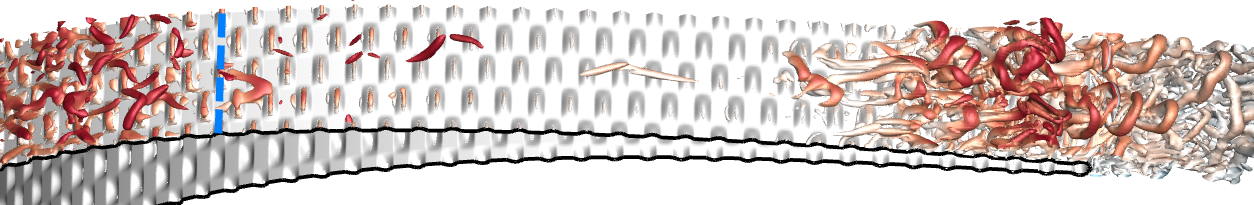}
            \put(0,15){(\textit{b})}
            \begin{tikzpicture}[overlay, x=0.074\textwidth, y=0.074\textwidth]
                \draw[white] (0,0)--(0,1);
                \draw[white] (0,0)--(1,0);
                \draw [magenta][line width = 1.5pt][densely dashed](1.8,0.9) ellipse (0.4 and 0.25);
                \put(70,-1){\scalebox{1.0}{$t_0$}}
            \end{tikzpicture}
        \end{overpic}
    \end{adjustbox}
    \hspace{0.03\textwidth}
    \begin{adjustbox}{valign=c}
        \begin{overpic}[width=0.24\textwidth]{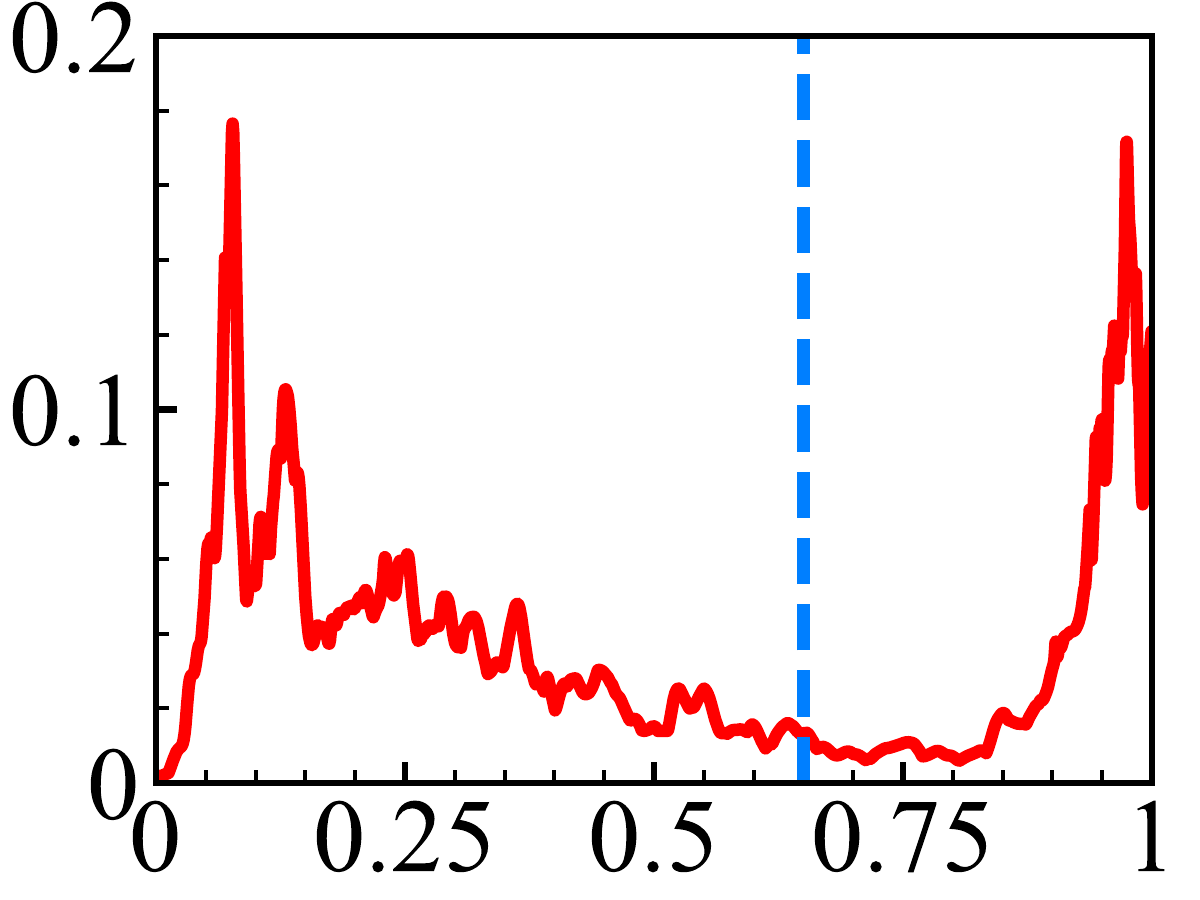}
            \put(-10,40){\scalebox{1}{\rotatebox[origin=c]{90}{$\mathrm{max}_{\eta}(tke)$}}}
        \end{overpic}
    \end{adjustbox} \\
    \begin{adjustbox}{valign=c}
        \begin{overpic}[width=0.7\textwidth]{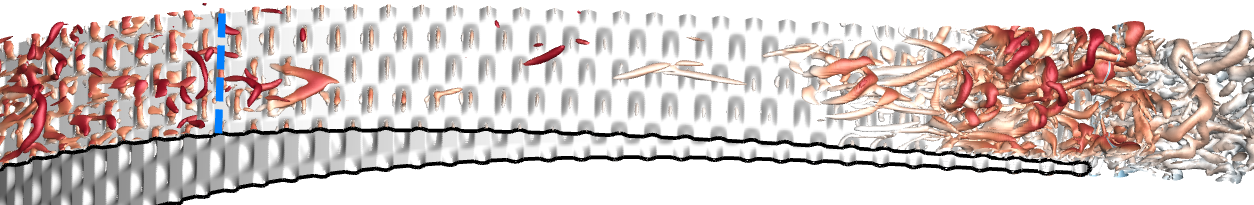}
            \put(0,15){(\textit{c})}
            \begin{tikzpicture}[overlay, x=0.074\textwidth, y=0.074\textwidth]
                \draw[white] (0,0)--(0,1);
                \draw[white] (0,0)--(1,0);
                \draw [magenta][line width = 1.5pt][densely dashed](2.1,0.9) ellipse (0.6 and 0.3);
                \put(70,-1){\scalebox{1.0}{$t_1$}}
            \end{tikzpicture}
        \end{overpic}
    \end{adjustbox}
    \hspace{0.03\textwidth}
    \begin{adjustbox}{valign=c}
        \begin{overpic}[width=0.24\textwidth]{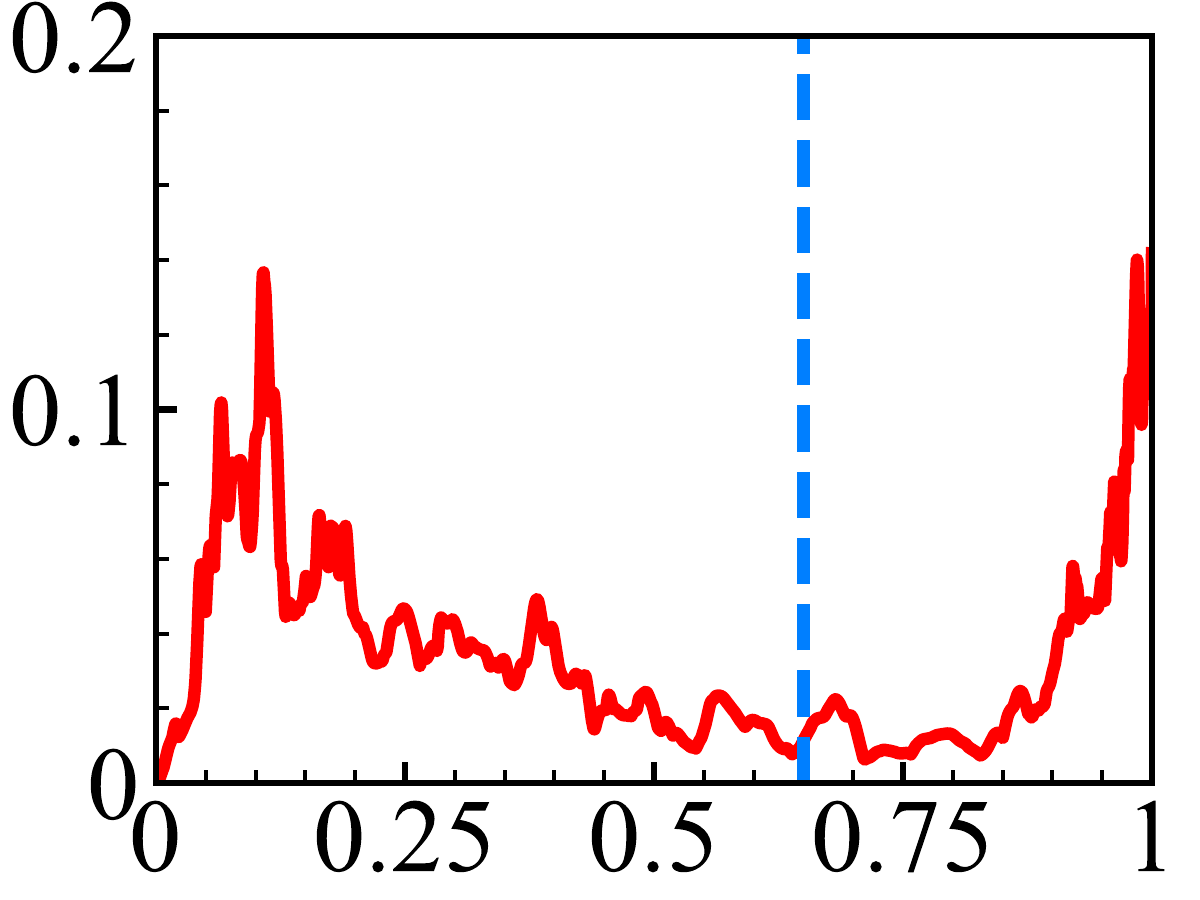}
            \put(-10,40){\scalebox{1}{\rotatebox[origin=c]{90}{$\mathrm{max}_{\eta}(tke)$}}}
            \begin{tikzpicture}[overlay, x=0.074\textwidth, y=0.074\textwidth]
                \draw[white] (0,0)--(0,1);
                \draw[white] (0,0)--(1,0);
                \draw [blue][line width = 1.5pt][<-](2.3,0.6)--+(70:1);
            \end{tikzpicture}
        \end{overpic}
    \end{adjustbox}\\
    \begin{adjustbox}{valign=c}
        \begin{overpic}[width=0.7\textwidth]{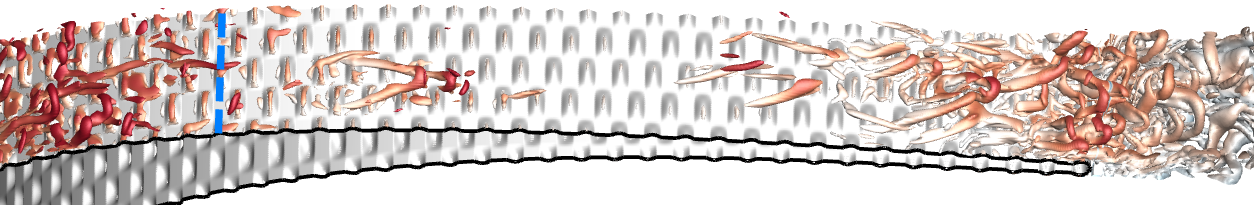}
            \put(0,15){(\textit{d})}
            \put(26,-2){PHV}
            \begin{tikzpicture}[overlay, x=0.074\textwidth, y=0.074\textwidth]
                \draw[white] (0,0)--(0,1);
                \draw[white] (0,0)--(1,0);
                \draw [magenta][line width = 1.5pt][densely dashed](3,0.9) ellipse (0.8 and 0.4);
                \draw[black][line width = 0.8pt](3.1,0.8)--+(-120:0.8);
                \put(70,-1){\scalebox{1.0}{$t_2$}}
            \end{tikzpicture}
        \end{overpic}
    \end{adjustbox}
    \hspace{0.03\textwidth}
    \begin{adjustbox}{valign=c}
        \begin{overpic}[width=0.24\textwidth]{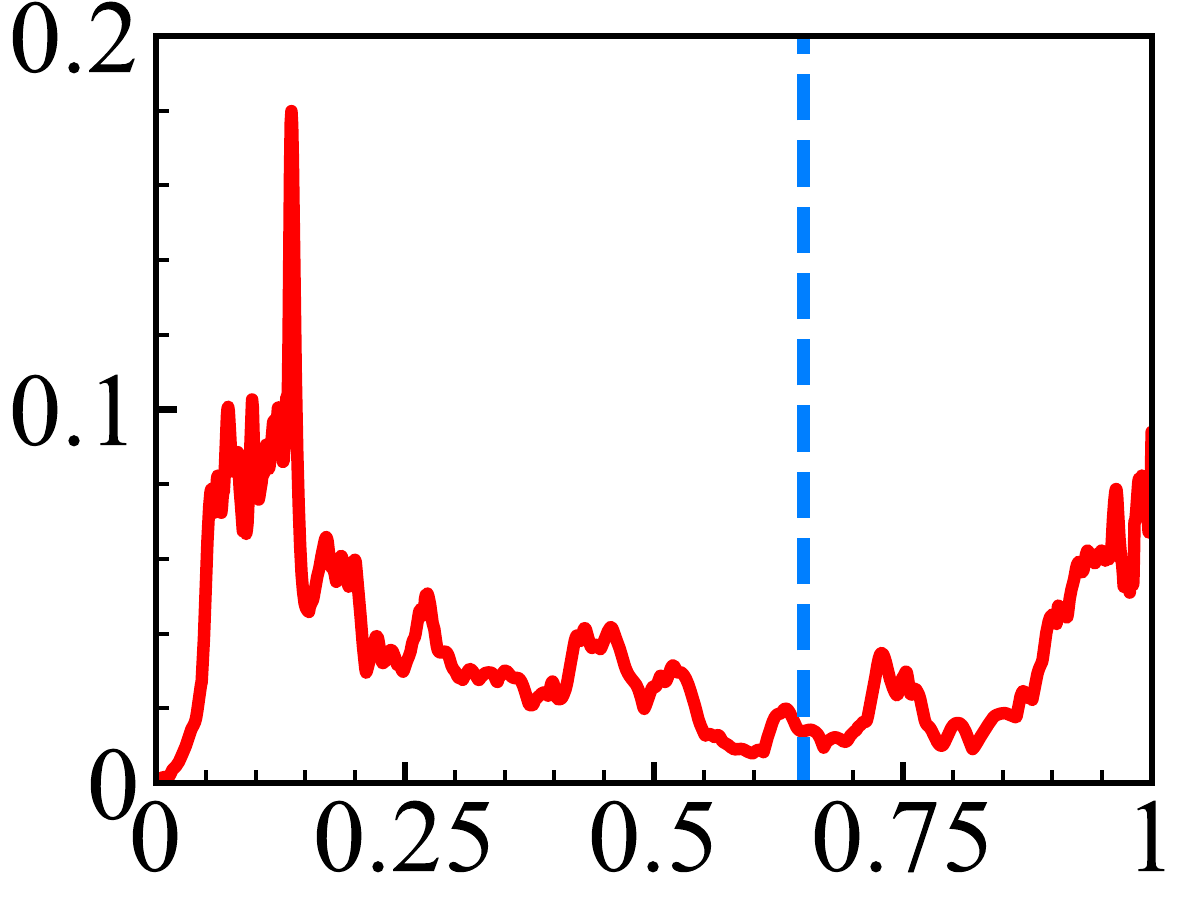}
            \put(-10,40){\scalebox{1}{\rotatebox[origin=c]{90}{$\mathrm{max}_{\eta}(tke)$}}}
            \begin{tikzpicture}[overlay, x=0.074\textwidth, y=0.074\textwidth]
                \draw[white] (0,0)--(0,1);
                \draw[white] (0,0)--(1,0);
                \draw [blue][line width = 1.5pt][<-](2.4,0.7)--+(70:1);
            \end{tikzpicture}
        \end{overpic}
    \end{adjustbox}\\
    \begin{adjustbox}{valign=c}
        \begin{overpic}[width=0.7\textwidth]{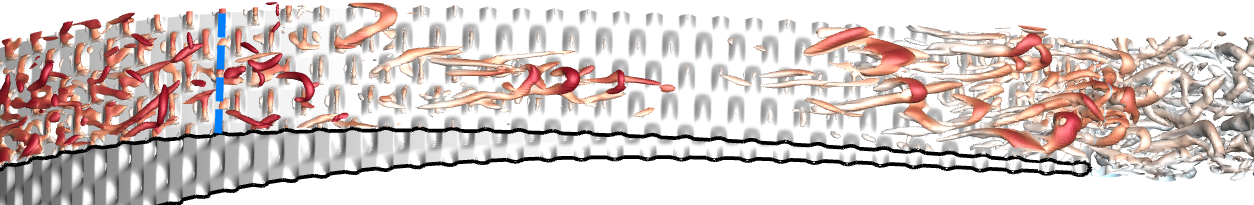}
            \put(0,15){(\textit{e})}
            \put(28,-2){SHV}
            \put(42,-2){PHV}
            \put(56,-2){DHV}
            \put(30,19){QSV}
            \begin{tikzpicture}[overlay, x=0.074\textwidth, y=0.074\textwidth]
                \draw[white] (0,0)--(0,1);
                \draw[white] (0,0)--(1,0);
                \draw [magenta][line width = 1.5pt][densely dashed](3.9,0.9) ellipse (1.2 and 0.5);
                \draw[black][line width = 0.8pt](3.9,0.8)--+(-150:1.2);
                \draw[black][line width = 0.8pt](4.2,0.8)--+(-90:0.7);
                \draw[black][line width = 0.8pt](4.6,0.8)--+(-30:1.2);
                \draw[black][line width = 0.8pt](3.1,1.2)--+(90:0.5);
                \put(70,-1){\scalebox{1.0}{$t_3$}}
            \end{tikzpicture}
        \end{overpic}
    \end{adjustbox}
    \hspace{0.03\textwidth}
    \begin{adjustbox}{valign=c}
        \begin{overpic}[width=0.24\textwidth]{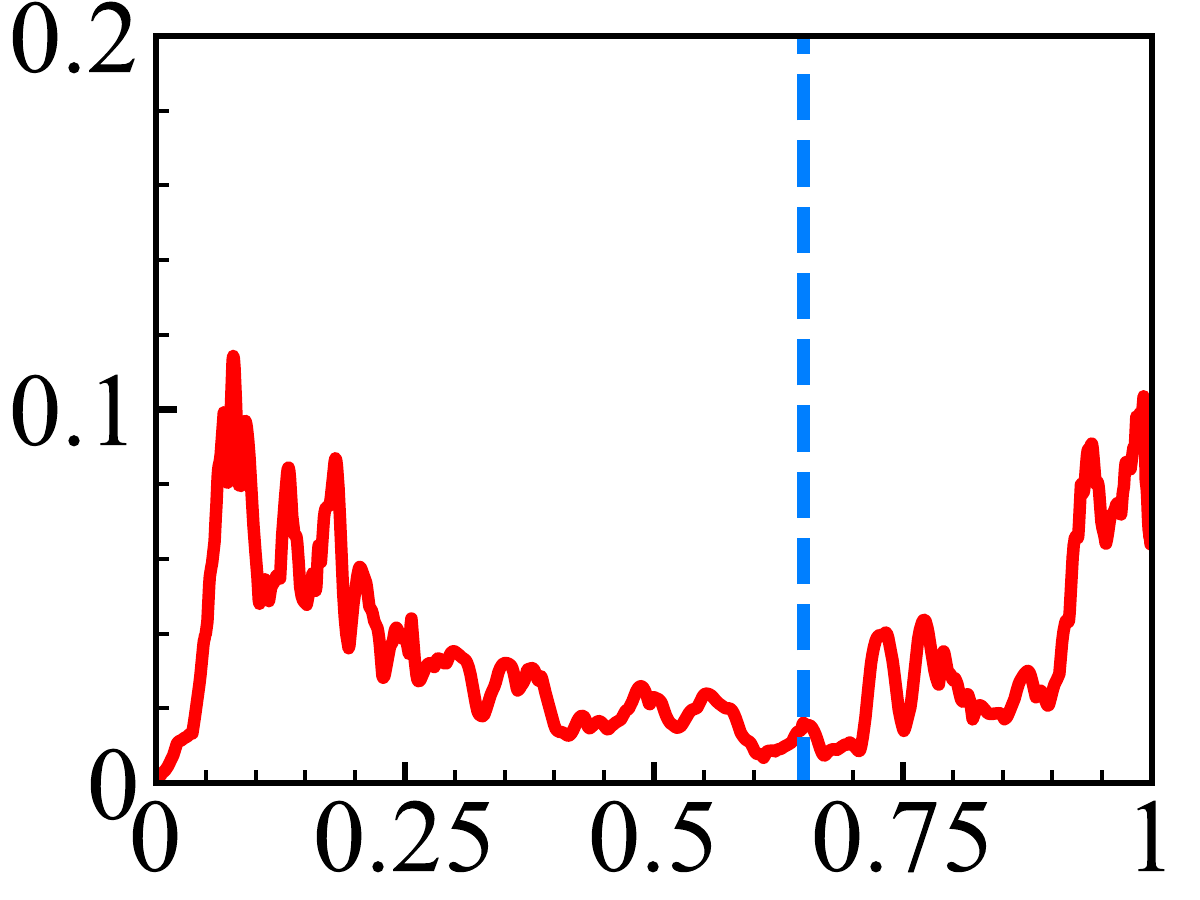}
            \put(-10,40){\scalebox{1}{\rotatebox[origin=c]{90}{$\mathrm{max}_{\eta}(tke)$}}}
            \begin{tikzpicture}[overlay, x=0.074\textwidth, y=0.074\textwidth]
                \draw[white] (0,0)--(0,1);
                \draw[white] (0,0)--(1,0);
                \draw [blue][line width = 1.5pt][<-](2.5,0.8)--+(70:1);
            \end{tikzpicture}
        \end{overpic}
    \end{adjustbox}\\
    \begin{adjustbox}{valign=c}
        \begin{overpic}[width=0.7\textwidth]{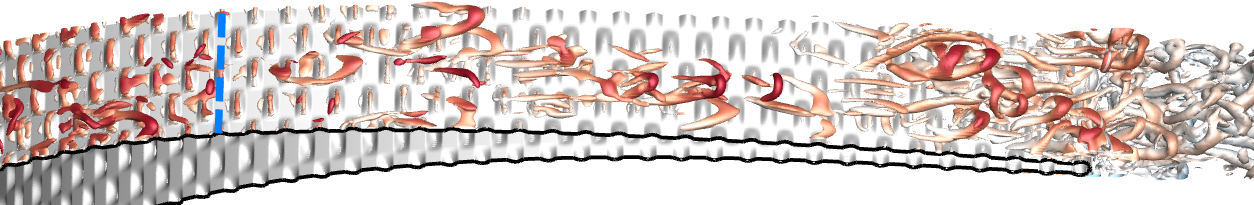}
            \put(0,15){(\textit{f})}
            \begin{tikzpicture}[overlay, x=0.074\textwidth, y=0.074\textwidth]
                \draw[white] (0,0)--(0,1);
                \draw[white] (0,0)--(1,0);
                \draw [magenta][line width = 1.5pt][densely dashed](4.8,0.9) ellipse (1.4 and 0.5);
                \put(70,-1){\scalebox{1.0}{$t_4$}}
            \end{tikzpicture}
        \end{overpic}
    \end{adjustbox}
    \hspace{0.03\textwidth}
    \begin{adjustbox}{valign=c}
        \begin{overpic}[width=0.24\textwidth]{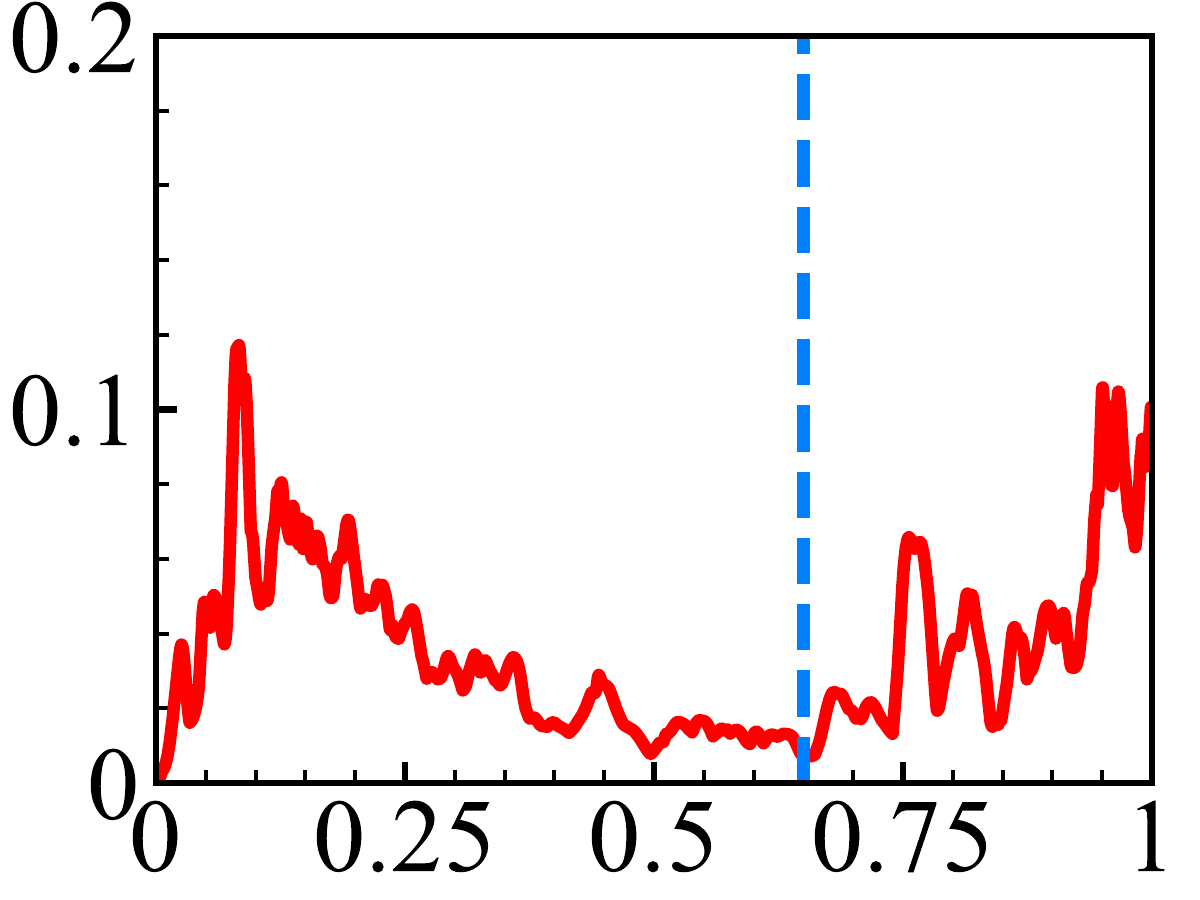}
            \put(-10,40){\scalebox{1}{\rotatebox[origin=c]{90}{$\mathrm{max}_{\eta}(tke)$}}}
            \begin{tikzpicture}[overlay, x=0.074\textwidth, y=0.074\textwidth]
                \draw[white] (0,0)--(0,1);
                \draw[white] (0,0)--(1,0);
                \draw [blue][line width = 1.5pt][<-](2.6,0.95)--+(70:1);
            \end{tikzpicture}
        \end{overpic}\end{adjustbox}\\
    \begin{adjustbox}{valign=c}
        \begin{overpic}[width=0.7\textwidth]{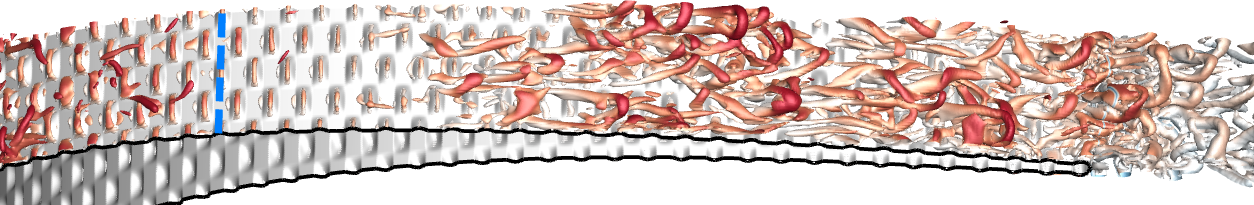}
            \put(0,15){(\textit{g})}
            \put(70,-1){\scalebox{1.0}{$t_5$}}
        \end{overpic}
    \end{adjustbox}
    \hspace{0.03\textwidth}
    \begin{adjustbox}{valign=c}
        \begin{overpic}[width=0.24\textwidth]{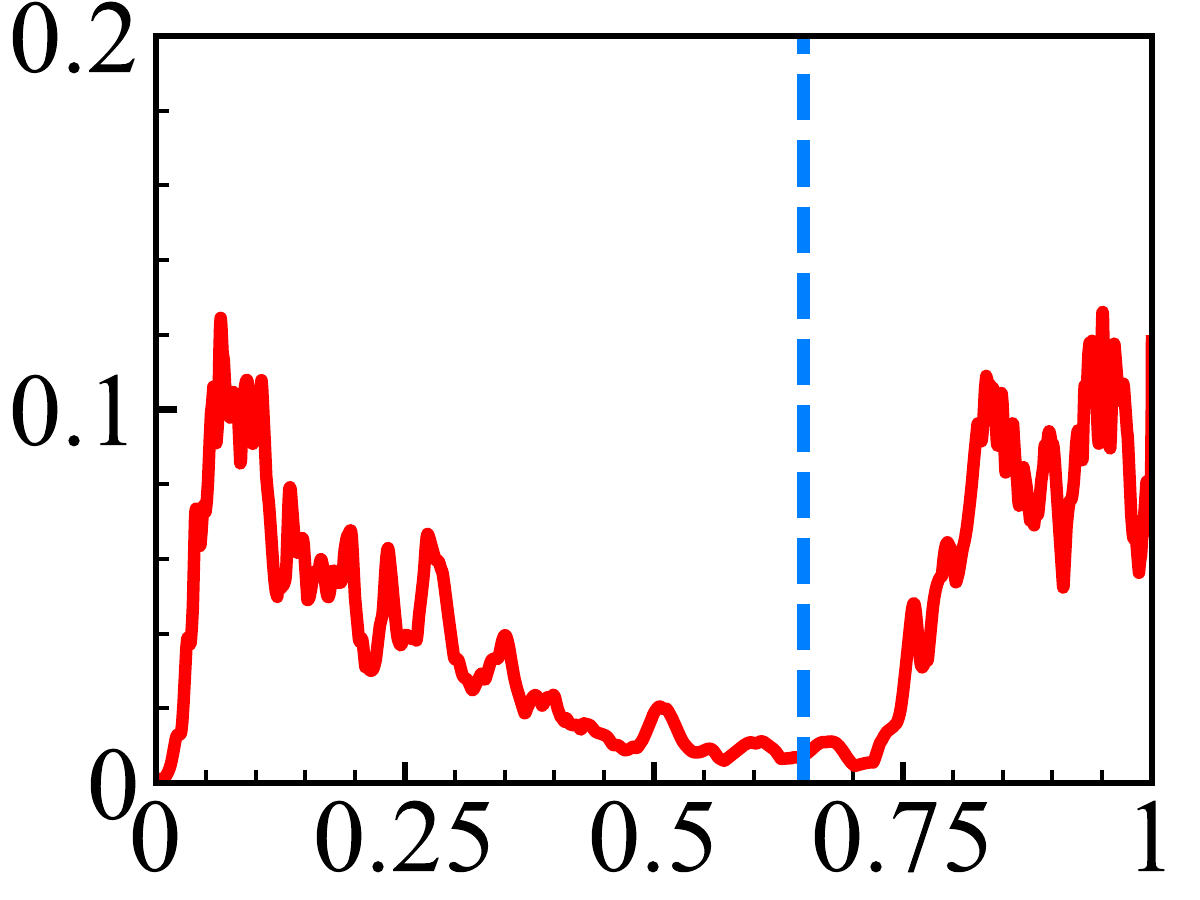}
            \put(-10,40){\scalebox{1}{\rotatebox[origin=c]{90}{$\mathrm{max}_{\eta}(tke)$}}}
            \put(45,-6){\scalebox{1.0}{$x/C_{ax}$}}
            \begin{tikzpicture}[overlay, x=0.074\textwidth, y=0.074\textwidth]
                \draw[white] (0,0)--(0,1);
                \draw[white] (0,0)--(1,0);
                \draw [blue][line width = 1.5pt][<-](2.7,1.45)--+(70:0.9);
            \end{tikzpicture}
        \end{overpic}\end{adjustbox}\\
    \caption{The vortical sturctures on the APG region for (\textit{a}) $k48\alpha 50$ and (\textit{b-g}) $k48\alpha 100$. The blue solid line indicate $x/C_{ax}=0.65$. Meanwhile, the instaneous normal maximum of tke on te suction-side surfacce is averaged in the spanwise direction and plotted against $x/C_{ax}$ for the corresponding snapshorts. PHV, primary hairpin vortex; SHV, secondary hairpin vortex; DHV, downstream hairpin vortex; QSV, quasi-streamwise vortices.}
    \label{fig:Q_k48s50+k48s100}
\end{figure*}

We further investigate the vortical structures in the APG boundary layer to understand the detailed transition behaviors for the $k48$ cases.
In Fig.~\ref{fig:Q_k48s50+k48s100}(\textit{a}), the instantaneous vortical structures for the $k48\alpha50$ case are identified by the Q-isosurface, accompanied by the flow separation region visualized via zero-velocity isosurfaces. 
Additionally, the distribution of the wall-normal maximum of TKE along the blade suction side is also presented for the corresponding instant. 
It can be seen that the APG boundary layer shows turbulent vortical structures only downstream of the separation region near the blade trailing edge.
Accordingly, the velocity fluctuation amplitudes remain small until they experience a sudden increase caused by the separation bubble.
This confirms that for this case with weak roughness effects and at relatively low Reynolds number, the boundary layer is dominated by the TE separation and the resulting separation-induced transition. 

Furthermore, we study the evolution of vortical structures for the $k48\alpha100$ case in a sequence of snapshots, along with the corresponding distribution of the wall-normal maximum TKE as shown in Fig.~\ref{fig:Q_k48s50+k48s100}(\textit{b-f}). 
Overall, the evolving structures observed here resemble those shown in previous transitional channel flows \cite{Zhou_1999,Zhao_2016} and flate plate boundary layer \cite{Sayadi_2013}.
Specifically, a $\Lambda$-shaped structure first appears as the boundary layer flow enters the APG region at $x/C_{ax}=0.65$, as shown in Fig.~\ref{fig:Q_k48s50+k48s100}(\textit{b}).
It is noted that the $\Lambda$-shaped structure causes the TKE to rapidly amplify, forming a local peak as shown by the blue arrow in Fig.~\ref{fig:Q_k48s50+k48s100}(\textit{c}).
Traveling downstream, the initial structure quickly evolves into a hairpin-like vortex as shown in Fig.~\ref{fig:Q_k48s50+k48s100}(\textit{d}), and the local TKE peak increases and moves downstream accordingly.
Furthermore, the primary hairpin vortex (PHV), once formed, induces the subsequent hairpin-like structures, forming a coherent packet of hairpins that propagate coherently \cite{Zhou_1999}.
The hairpin packets presented in Fig.~\ref{fig:Q_k48s50+k48s100}(\textit{e}) result in multiple TKE peaks, which keep convecting downstream while amplifying.
Moreover, the hairpins also generate quasi-streamwise vortices to the side of their legs. 
Besides the spanwise symmetric structures we have presented, there also exist asymmetric one-sided hairpins, such as `canes' shown in Fig.~\ref{fig:Q_k48s50+k48s100}(\textit{f}). 
From Fig.~\ref{fig:Q_k48s50+k48s100}(\textit{e}) to \ref{fig:Q_k48s50+k48s100}(\textit{g}), it can be observed that the vortical structures become increasingly chaotic, until breaking down into turbulence. 

\begin{figure*}[t!]
	\centering
	\begin{overpic}[width=0.9\textwidth]{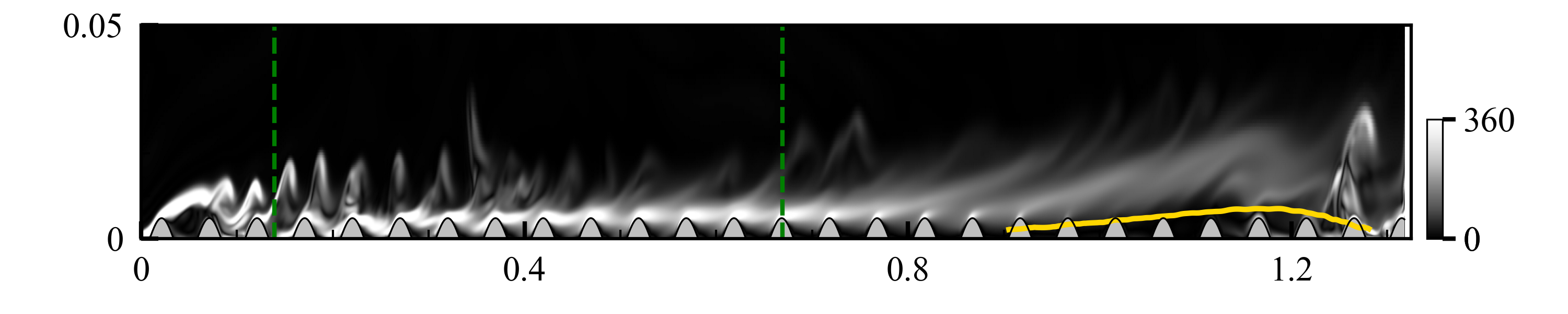}
		\put(0,18){(\textit{a})}
		\put(2,11){\scalebox{1.0}{$\eta$}}
        \put(97,8){\scalebox{1.0}{$|\omega|$}}
	\end{overpic}\\
	\begin{overpic}[width=0.9\textwidth]{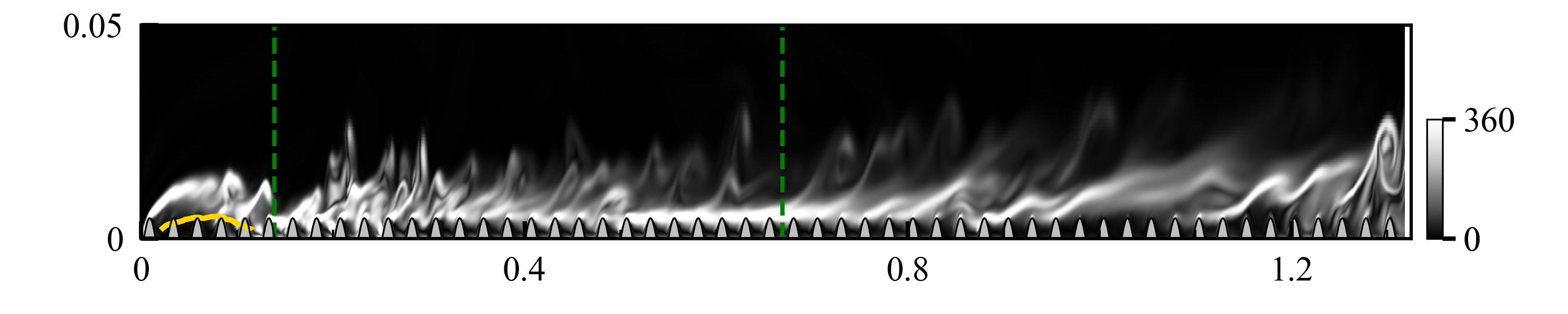}
		\put(0,18){(\textit{b})}
		\put(2,11){\scalebox{1.0}{$\eta$}}
		\put(50,-2){\scalebox{1.0}{$\xi$}}
        \put(97,8){\scalebox{1.0}{$|\omega|$}}
	\end{overpic}\\
	\caption{Instantaneous contours of vorticity magnitude on $\xi-\eta$ plane ($z=0.075$). The dashed green lines present $x/C_{ax}=0.1$ and $x/C_{ax}=0.65$. The yellow solid lines denote the zero-isoline of spanwise- and time-averaged tangential velocity. (\textit{a}) $k48\alpha50$, (\textit{b}) $k48\alpha100$. }
	\label{fig:omega_contour}
\end{figure*}

In roughness-induced transition, the Kelvin-Helmholtz (K-H) instability within the separated shear layer constitutes a ubiquitous mechanism \cite{Ma_Mahesh_2023,Wu2025}. 
To visually highlight the destabilized shear layer, we present instantaneous contours of vorticity magnitude in Fig.~\ref{fig:omega_contour}. 
A detached shear layer lifting away from the roughness elements is observed to show instability in the APG region from around $\xi\approx0.8$ for case $k48\alpha100$, consistent with observations by Vadlamani et al.~\cite{Vadlamani2018}. 
Conversely, for case $k48\alpha50$, the flow remains stable without transition, until the significant elevation and breakdown of the shear layer induced by the TE separation bubble as shown in  Fig.~\ref{fig:omega_contour}(a).

To summarize, for cases with small-amplitude roughness elements, disturbances remain weak throughout the APG region until transition is triggered by flow separation near the trailing edge. 
In contrast, cases with high-amplitude roughness elements exhibit earlier transition initiation, commencing during or even prior to the APG region. 
Notably, the streamwise wavenumber exerts negligible influence on transition location in high-roughness configurations. 
However, for medium-roughness cases $k48$, variations in streamwise wavenumber within a specific range profoundly alter both transition location and mechanism. 
When the wavenumber increases from $\alpha50$ to $\alpha100$, the transition path shifts from separation-induced instability to instability of roughness-induced elevated shear layers, the transition location advances considerably upstream.

\section{Conclusion}\label{sec:Conclusion}

In the present study, direct numerical simulations of a LPT with roughness elements distributed over the blade surface have been performed, and the roughness height and streamwise wavenumber are varied in a series of fifteen cases to present a systematic study on the complex boundary layer behaviors.
For cases with different surface roughness, various paths for transition are observed, including the transition induced by roughness elements in the LE region, transition triggered by TE separation, and also transition induced by shear layer instability in the APG region.

On one hand, the roughness height is indicated to be the dominating factor for suction-side boundary layer transition. Specifically, for cases with large roughness heights, such as the $k64$ and $k80$ cases, the roughness elements in the LE region induce wake structures and the shear layer elevated from the wall quickly breaks down into turbulence.
The turbulent fluctuations in these high-amplitude roughness cases sustain through the whole suction-side boundary layer, despite the stabilizing effect of the FPG region. 
For cases with relatively small roughness heights (the $k32$ and $k16$ cases), however, the disturbances induced by the LE roughness are suppressed in the FPG region, and the relaminarized boundary layer does not show transition until the TE separation.

The streamwise wavenumber of the distributed roughness, on the other hand, plays an important role in cases with intermediate roughness height, \emph{i.e.} the $k48$ cases in the present study.
The case with smaller wavenumber (the $k48\alpha50$ case, thus low-level effective slopes) relaminarizes in the FPG region and maintains a laminar mean flow, until boundary layer separation induces prompt breakdown into turbulence.
In contrast, the cases with larger wavenumbers (the $k48\alpha100$ and $k48\alpha150$ cases) show earlier transition triggered by a shear-layer instability, which manages to suppress the mean flow separation near the TE region.
This demonstrates that in this intermediate regime, the boundary layer is highly sensitive to the specific geometric topography beyond simple peak-to-valley height.

These findings pose a significant challenge for lower-fidelity modeling approaches, particularly RANS-based transition models commonly used in turbomachinery design. 
The present results suggest that to accurately predict losses and heat transfer in LPT flows, transition models must incorporate additional geometric descriptors, such as the streamwise spectral content or effective slope, rather than relying solely on roughness height. 
The current dataset can also serve as benchmarks for developing more non-local, topography-aware roughness models, or potentially data-driven closures, to account for these complex geometric sensitivities.

\begin{acknowledgment}
This work has been supported by the National Natural Science Foundation of China (Grant Nos.~92152202, 12432010 and 12588201).
\end{acknowledgment}

%
% ...

\bibliographystyle{asmems4}

\bibliography{zxw}

@string{AIAA  = {AIAA J.}}

@string{AIP   = {AIP Adv.}}

@string{AST   = {Aerosp. Sci. Technol.}}

@string{CF    = {Comp. Fluids}}

@string{FTaC  = {Flow Turbul. Combust.}}

@string{IJHFF = {Int. J. Heat Fluid Flow}}

@string{JCP   = {J. Comp. Phys.}}

@string{JFM   = {J. Fluid Mech.}}

@string{JoT   = {ASME J. Turbomach.}}

@string{PF    = {Phys. Fluids}}

@string{PRF   = {Phys. Rev. Fluids}}

@article{Bons2010review,
    author = {Bons, J. P.},
    title = "{A review of surface roughness effects in gas turbines}",
    journal = JoT,
    volume = {132},
    number = {2},
    year = {2010},
    note = {021004}
}

@article{Napoli2008,
    author = {Napoli, E. and Armenio, V. and De Marchis, M.},
    title = "{The effect of the slope of irregularly distributed roughness elements on turbulent wall-bounded flows}",
    journal = JFM,
    volume={613},
    pages = {385–394},
    year = {2008}
}

@article{Ma2022,
    author = {Ma, G. Z. and Xu, C. X. and Sung, H. J. and Huang, W. X.},
    title = "{Scaling of rough-wall turbulence in a transitionally rough regime}",
    journal = PF,
    volume = {34},
    number = {3},
    pages = {031701},
    year = {2022},
    month = {03}
}

@article{Chan2018, 
    author={Chan, L. and MacDonald, M. and Chung, D. and Hutchins, N. and Ooi, A.}, 
    title="{Secondary motion in turbulent pipe flow with three-dimensional roughness}", 
    journal=JFM,     
    volume={854}, 
    year={2018}, 
    pages={5–33}
}

@article{Vadlamani2018, 
    author={Vadlamani, N. R. and Tucker, P. G. and Durbin, P.}, 
    title="{Distributed roughness effects on transitional and turbulent boundary layers}", 
    journal=FTaC, 
    volume={100}, 
    number={3},
    year={2018}, 
    pages={627–649}
}

@article{Deyn2020, 
    author={von Deyn, L. H. and Forooghi, P. and Frohnapfel, B. and Schlatter, P. and Hanifi, A. and Henningson, D. S.}, 
    title="{Direct numerical simulations of bypass transition over distributed roughness}", 
    journal=AIAA, 
    volume={58}, 
    number={2},
    year={2020}, 
    pages={702–711}
}

@article{Wu2025, 
    author={Wu, H. and Yang, x. and Li, G. and Yin, Z.}, 
    title="{Interaction of freestream turbulence and surface roughness in separation-induced transition}", 
    journal=PRF, 
    volume={10}, 
    number={1}, 
    year={2025}, 
    pages={013903}
}

@article{Zhao2020, 
    author={Zhao, Y. and Sandberg, R. D.},
    title="{Bypass transition in boundary layers subject to strong pressure gradient and curvature effects}", 
    journal=JFM, 
    volume={888},  
    year={2020}, 
    pages={A4}
}

@inproceedings{Bammert1972,
    author = {Bammert, K. and Milsch, R.},
    title = "{Boundary layers on rough compressor blades}",
    volume = {ASME 1972 International Gas Turbine and Fluids Engineering Conference and Products Show},
    series = {Turbo Expo: Power for Land, Sea, and Air},
    pages = {V001T01A047},
    year = {1972},
    month = {03} 
}

@article{Bammert1980,
    author = {Bammert, K. and Sandstede, H.},
    title = "{Measurements of the boundary layer development along a turbine blade with rough surfaces}",
    journal = {J. Eng. Gas Turbines Power},
    volume = {102},
    number = {4},
    pages = {978-983},
    year = {1980}
}

@article{Kind1998,
    author = {Kind, R. J. and Serjak, P. J. and Abbott, M. W. P.},
    title = "{Measurements and prediction of the effects of surface roughness on profile losses and deviation in a turbine cascade}",
    journal = JoT,
    volume = {120},
    number = {1},
    pages = {20-27},
    year = {1998}
}

@article{Bogard1998,
    author = {Bogard, D. G. and Schmidt, D. L. and Tabbita, M.},
    title = "{Characterization and laboratory simulation of turbine airfoil surface roughness and associated heat transfer}",
    journal = JoT,
    volume = {120},
    number = {2},
    pages = {337-342},
    year = {1998}
}

@inproceedings{Boyle2003,
    author = {Boyle, R. J. and Senyitko, R. G.},
    title = "{Measurements and predictions of surface roughness effects on the turbine vane aerodynamics}",
    volume = {Volume 6: Turbo Expo 2003, Parts A and B},
    series = {Turbo Expo: Power for Land, Sea, and Air},
    pages = {291-303},
    year = {2003}
}

@article{Roberts2005,
    author = {Roberts, S. K. and Yaras, M. I.},
    title = "{Effects of surface-roughness geometry on separation-bubble transition}",
    journal = JoT,
    volume = {128},
    number = {2},
    pages = {349-356},
    year = {2005}
}

@article{Bons2002,
    author = {Bons, J. P. },
    title = {St and cf Augmentation for Real Turbine Roughness With Elevated Freestream Turbulence },
    journal = JoT,
    volume = {124},
    number = {4},
    pages = {632-644},
    year = {2002}
}

@inproceedings{Montis2010,
    author = {Montis, M. and Niehuis, R. and Fiala, A.},
    title = {Effect of Surface Roughness on Loss Behaviour, Aerodynamic Loading and Boundary Layer Development of a Low-Pressure Gas Turbine Airfoil},
    booktitle ={Proceedings of the ASME Turbo Expo},
    venue = {Glasgow, UK},
    eventdate = {June 14–18},
    type = {Vol.},
    number = {7},
    pages = {1535–1547},
    year = {2010}
}

@inproceedings{Montis2011,
    author = {Montis, M. and Niehuis, R. and Fiala, A.},
    title = {Aerodynamic Measurements on a Low Pressure Turbine Cascade With Different Levels of Distributed Roughness},
    booktitle ={Proceedings of the ASME Turbo Expo},
    venue = {Vancouver, British Columbia, Canada},
    eventdate = {June 6–10},
    type = {Vol.},
    number = {7},
    pages = {457–467.},
    year = {2011}
}

@article{Lorenz2012,
    author = {Lorenz, M. and Schulz, A. and Bauer, H.-J.},
    title = {Experimental Study of Surface Roughness Effects on a Turbine Airfoil in a Linear Cascade— Part II: Aerodynamic Losses},
    journal = JoT,
    volume = {134},
    number = {4},
    pages = {041007},
    year = {2012}
}

@article{Stripf2009,
    author = {Stripf, M. and Schulz, A. and Bauer, H.-J. and Wittig, S.},
    title = "{Extended models for transitional rough wall boundary layers with heat transfer—part I: Model formulations}",
    journal = JoT,
    volume = {131},
    number = {3},
    pages = {031016},
    year = {2009}
}

@article{Ge2015,
    author = {Ge, X. and Durbin, P. A.},
    title = "{An intermittency model for predicting roughness induced transition}",
    journal = IJHFF,
    volume = {54},
    pages = {55-64},
    year = {2015}
}

@article{Wei2017,
    author = {Wei, L. and Ge, X. and George, J. and Durbin, P.},
    title = "{Modeling of laminar-turbulent transition in boundary layers and rough turbine blades}",
    journal = JoT,
    volume = {139},
    number = {11},
    pages = {111009},
    year = {2017}
}

@inproceedings{Dassler2012,
    author = {Dassler, P. and Ko{\v{z}}ulovi{\'c}, D. and Fiala, A.},
    title = "{An approach for modelling the roughness-induced boundary layer transition using transport equations}",
    booktitle = {Europ. Congress on Comp. Methods in Appl. Sciences and Engineering, ECCOMAS 2012},
    year = {2012},
    address = {Vienna, Austria}
}

@article{Liu2020,
    author = {Liu, Z. and Zhao, Y. and Chen, S. and Yan, C. and Cai, F.},
    title = "{Predicting distributed roughness induced transition with a four-equation laminar kinetic energy transition model}",
    journal = AST,
    volume = {99},
    pages = {105736},
    year = {2020}
}

@inproceedings{Joo2016,
    author = {Joo, J. and Medic, G. and Sharma, O.},
    title = "{Large-eddy simulation investigation of impact of roughness on flow in a low-pressure turbine}",
    volume = {Volume 2C: Turbomachinery},
    series = {Turbo Expo: Power for Land, Sea, and Air},
    pages = {V02CT39A053},
    year = {2016}
}

@inproceedings{Hammer2018,
    author = {Hammer, F. and Sandham, N. D. and Sandberg, R. D.},
    title = "{Large eddy simulations of a low-pressure turbine: Roughness modeling and the effects on boundary layer transition and losses}",
    volume = {Volume 2B: Turbomachinery},
    series = {Turbo Expo: Power for Land, Sea, and Air},
    pages = {V02BT41A014},
    year = {2018}
}

@article{Wang2021,
    author = {Wang, M. and Lu, X. and Yang, C. and Zhao, S. and Zhang, Y.},
    title = "{Numerical investigation of distributed roughness effects on separated flow transition over a highly loaded compressor blade}",
    journal = PF,
    volume = {33},
    number = {11},
    pages = {114104},
    year = {2021},
    month = {11}
}

@article{Zeng2022,
    author = {Zeng, X. and Luo, J. and Cui, J.},
    title = "{Roughness effects on flow losses of a high-lift low-pressure turbine cascade}",
    journal = AIP,
    volume = {12},
    number = {1},
    page = {015316},
    year = {2022}
}

@article{Jelly2023,
    author = {Jelly, T. O. and Nardini, M. and Rosenzweig, M. and Leggett, J. and Marusic, I. and Sandberg, R. D.},
    title = "{High-fidelity computational study of roughness effects on high pressure turbine performance and heat transfer}",
    journal = IJHFF,
    volume = {101},
    pages = {109134},
    year = {2023}
}

@article{Nardini2023,
    author = {Nardini, M. and Kozul, M. and Jelly, T. O. and Sandberg, R. D.},
    title = "{Direct numerical simulation of transitional and turbulent flows over multi-scale surface roughness—Part I: Methodology and challenges}",
    journal = JoT,
    volume = {146},
    number = {3},
    pages = {031008},
    year = {2023}
}

@article{Jelly2025,
    author = {Jelly, T. O. and Nardini, M. and Sandberg, R. D. and Vitt, P. and Sluyter, G.},
    title = "{Effects of localized non-Gaussian roughness on high-pressure turbine aerothermal performance: Convective heat transfer, skin friction, and the Reynolds’ analogy}",
    journal = JoT,
    volume = {147},
    number = {5},
    pages = {051017},
    year = {2025}
}

@article{Nardini2023_2,
    author = {Nardini, M. and Jelly, T. O. and Kozul, M. and Sandberg, R. D. and Vitt, P. and Sluyter, G.},
    title = "{Direct numerical simulation of transitional and turbulent flows over multi-scale surface roughness—Part II: The Effect of roughness on the performance of a high-pressure turbine blade}",
    journal = JoT,
    volume = {146},
    number = {3},
    pages = {031009},
    year = {2023}
}

@article{Michelassi2015,
    author = {Michelassi, V. and Chen, L.-W. and Pichler, R. and Sandberg, R. D.},
    title = "{Compressible direct numerical simulation of low-pressure turbines—Part II: Effect of inflow disturbances}",
    journal = JoT,
    volume = {137},
    number = {7},
    pages = {071005},
    year = {2015}
}

@article{Demarchis2015,
    author = {M. {De Marchis} and B. Milici and E. Napoli},
    title = "{Numerical observations of turbulence structure modification in channel flow over 2D and 3D rough walls}",
    journal = IJHFF,
    volume = {56},
    pages = {108–123},
    year = {2015}
}

@article{Ciorciari2014,
    author = {Ciorciari, R. and Kirik, I. and Niehuis, R.},
    title = {Effects of unsteady wakes on the secondary flows in the linear T106 turbine cascade},
    journal = JoT,
    volume = {136},
    number = {9},
    pages = {091010},
    year = {2014}
}

@inproceedings{Tarada1993,
    author = {Tarada, F. and Suzuki, M.},
    title = "{External heat transfer enhancement to turbine blading due to surface roughness}",
    volume = {Volume 2},
    series = {Turbo Expo},
    pages = {V002T08A006},
    year = {1993}
}

@article{Schlanderer2017,
    author = {Schlanderer, S. C. and Weymouth, G. D. and Sandberg, R. D.},
    title = "{The boundary data immersion method for compressible flows with application to aeroacoustics}",
    journal = JCP,
    volume = {333},
    pages = {440-461},
    year = {2017},
}

@article{Sandberg2015,
    author = {Sandberg, R. D. and Michelassi, V. and Pichler, R. and Chen, L. and Johnstone, R.},
    title = "{Compressible direct numerical simulation of low-pressure turbines—Part I: Methodology}",
    journal = JoT,
    volume = {137},
    number = {5},
    pages = {051011},
    year = {2015}
}

@inproceedings{Hunt1988,
  author = {Hunt, J. C. R. and Wray, A. A. and Moin, P.},
  title = "{Eddies, streams, and convergence zones in turbulent flows}",
  booktitle = {Studying Turbulence Using Numerical Simulation Databases},
  volume    = {1},
  pages     = {193-208},
  series    = {2},
  year      = {1988}
}

@article{Kennedy2000,
    author = {Kennedy, C. A. and Carpenter, M. H. and Lewis, R. M.},
    title = "{Low-storage, explicit Runge–Kutta schemes for the compressible Navier–Stokes equations}",
    journal = {Appl. Numer. Maths.},
    volume = {35},
    number = {3},
    pages = {177-219},
    year = {2000}
}

@article{Deuse2020a,
    author = {Deuse, M. and Sandberg, R. D.},
    title = "{Implementation of a stable high-order overset grid method for high-fidelity simulations}",
    journal = CF,
    volume = {211},
    pages = {104449},
    year = {2020}
}

@article{Zhao2021,
    author = {Zhao, Y. and Sandberg, R. D.},
    title = "{High-fidelity simulations of a high-pressure turbine vane subject to large disturbances: Effect of exit Mach number on losses}",
    journal = JoT,
    volume = {143},
    number = {9},
    pages = {091002},
    year = {2021}
}

@article{Klein2003,
    author = {Klein, M. and Sadiki, A. and Janicka, J.},
    title = "{A digital filter based generation of inflow data for spatially developing direct numerical or large eddy simulations}",
    journal = JCP,
    volume = {186},
    number = {2},
    pages = {652-665},
    year = {2003}
}

@article{Sandberg2006,
    author = {Sandberg, R. D. and Sandham, N. D.},
    title = "{Nonreflecting zonal characteristic boundary condition for direct numerical simulation of aerodynamic sound}",
    journal = AIAA,
    volume = {44},
    number = {2},
    pages = {402-405},
    year = {2006}
}

@article{Stadtmuller2001,
    author={Stadtm{\"u}ller, P},
    title="{Investigation of wake-induced transition on the LP turbine cascade T106A-EIZ}",
    journal={DFG-Verbundprojekt Fo},
    volume={136},
    number={11},
    year={2001}
}

@article{Reynolds_Hussain_1972,
    author={Reynolds, W. C. and Hussain, A. K. M. F.},
    title="{The mechanics of an organized wave in turbulent shear flow. Part 3. Theoretical models and comparisons with experiments}",
    journal=JFM,
    volume={54}, 
    number={2},
    year={1972},
    pages={263–288}
}

@article{Zhao_2016,
    author={Zhao, Y. and Yang, Y. and Chen, S.},
    title="{Evolution of material surfaces in the temporal transition in channel flow}",
    journal=JFM,
    volume={793},
    year={2016},
    pages={840–876}
}

@article{Sayadi_2013,
    author={Sayadi, T. and Hamman, C. W. and Moin, P.},
    title="{Direct numerical simulation of complete H-type and K-type transitions with implications for the dynamics of turbulent boundary layers}",
    journal=JFM,
    volume={724},
    year={2013},
    pages={480–509}
}

@article{Zhou_1999,
    author={Zhou, J. and Adrian, R. J. and Balachandar, S. and Kendall, T. M.},
    title="{Mechanisms for generating coherent packets of hairpin vortices in channel flow}",
    journal=JFM,
    volume={387},
    year={1999},
    pages={353–396}
}

@article{Ma_Mahesh_2023,
    author={Ma, R. and Mahesh, K.}, 
    title="{Boundary layer transition due to distributed roughness: effect of roughness spacing}",
    journal=JFM,
    volume={977},
    year={2023}, 
    pages={A27}
}


\end{document}